\documentclass[aps,nofootinbib]{revtex4}
\usepackage{amsfonts}
\usepackage{amsmath}
\usepackage{amssymb}
\usepackage{graphicx}
\usepackage{srcltx}

\begin{document}

\title{Green's Functions for the Anderson model: the Atomic Approximation}
\author{M. E. Foglio}
\affiliation{Instituto de F\'{\i}sica Gleb Wataghin, Universidade Estadual de Campinas,
Bar\~ao Geraldo 13083-970 Campinas-SP, Brazil}
\author{T. Lobo and M. S. Figueira}
\affiliation{Instituto de F\'{\i}sica, Universidade Federal Fluminense, Av. Litor\^{a}nea
s/n, $24210-346$ Niter\'oi - Rio de Janeiro, Brazil.}

\begin{abstract}
We consider the cumulant expansion of the PAM employing the hybridization as
perturbation (Phys. Rev. B \textbf{50}, 17933 (1994)), and we obtain
formally exact one-electron Green's functions (GF). These GF contain
effective cumulants that are as difficult to calculate as the original GF,
and the Atomic Approximation consists in substituting the effective
cumulants by the ones that correspond to the atomic case, namely by taking a
conduction band of zeroth width and local hybridization. This approximation
has already been used for the case of infinite electronic repulsion $U$
(Phys. Rev. B \textbf{62}, 7882 (2000)), and here we extend the treatment to
the case of finite $U$. The method can also be applied to the single
impurity Anderson model (SIAM), and we give explicit expressions of the
approximate GF both for the PAM and the SIAM.
\end{abstract}

\date{\today}

\maketitle


\section{ \ \ \ \ Introduction}

In this work we discuss approximate Green's Functions (GF) for the Periodic
Anderson Model (PAM), obtained by starting from a formally exact expression
and approximating a component of this expression by the corresponding exact
solution of the atomic problem. We have already employed this technique in
the limit of infinite repulsion U of the localized electrons \cite%
{FoglioFeSi},\cite{FoglioFeSi2}, and here we shall extend the technique to
the case of a finite U. We call this technique the Atomic Approximation, not
to be confused with the atomic solution of the problem.

The Hamiltonian for the PAM is

\begin{align}
H & =\sum_{\mathbf{k},\sigma}E_{\mathbf{k},\sigma}C_{\mathbf{k},\sigma
}^{\dagger}C_{\mathbf{k},\sigma}+\sum_{j,\sigma}E_{\sigma\ }f_{j,\sigma
}^{\dagger}f_{j,\sigma}  \notag \\
& +U\sum_{j}n_{j,\sigma}n_{j,\overline{\sigma}}+H_{h},  \label{Eq1}
\end{align}

\noindent where the operators $C_{\mathbf{k},\sigma}^{\dagger}$ and $C_{%
\mathbf{k},\sigma}$ are the creation and destruction operators of conduction
band electrons (c-electrons) with wave vector $\mathbf{k}$, component of
spin $\sigma$ and energies $E_{\mathbf{k},\sigma}$. The $f_{j,\sigma}^{%
\dagger}$ and $f_{j,\sigma}$ are the corresponding operators for the $f$%
-electrons in the Wannier localized state at site j , with spin component $%
\sigma$ and site independent energy $E_{\sigma}$. The third term is the
Coulomb repulsion between the localized electrons at each site where $%
n_{j,\sigma}=f_{j,\sigma}^{\dagger}f_{j,\sigma}$ is the number of $f$%
-electrons with spin component $\sigma$ at site $j$ and the symbol $%
\overline{\sigma}$ denotes the spin component opposite to $\sigma$. The
fourth term $H_{h}$ describes the hybridization between the localized and
conduction electrons
\begin{equation}
H_{h}=\sum_{j,\mathbf{k},\sigma}(V_{j,\mathbf{k},\sigma}f_{j,\sigma}^{%
\dagger }C_{\mathbf{k},\sigma}+V_{j,\mathbf{k},\sigma}^{\ast}C_{\mathbf{k}%
,\sigma }^{\dagger}f_{j,\sigma}),  \label{Eq2}
\end{equation}
with a coupling strength given by

\begin{equation}
V_{j,\mathbf{k},\sigma}=\frac{1}{\sqrt{N_{s}}}V_{\sigma}(\mathbf{k})\exp{(i%
\mathbf{k}.\mathbf{R}_{j}),}  \label{Eq3}
\end{equation}

\noindent where $N_{s}$ is the number of sites in the system and $V_{\sigma
}(\mathbf{k})$ is independent of the wave vector $\mathbf{k}$ when the
mixing is purely local.

If we consider that the local repulsion between $f$-electrons is infinite ($%
U\rightarrow\infty$), so that the double occupancy at any site is zero, we
can employ Hubbard $X$ operators to make disappear the term proportional to $%
U$ in the Hamiltonian. To this purpose we consider first the definition of
the $X$ operators: the $X_{j,ba}$ transforms the state $\mid a>$ at site j
into the state $\mid b>$ at the same site, and we assume that $\mid a>$ and $%
\mid b>$ are eigenstates of the number of electrons. We say that $X_{j,ba}$
is of the Fermi type when $\mid a>$ and $\mid b>$ differ by an odd number of
Fermions, and that it is of the Bose type when they differ by an even number
of Fermions. By definition, two $X$-operators of the Fermi type at different
sites anti-commute, and commute when at least one of them is of the Bose
type. The algebra of these operators when they are at the same site is
defined by their product rule

\begin{equation}
X_{j,ab}\cdot X_{j,cd}=\delta_{b,c}\cdot X_{j,ad,}  \label{Hub2}
\end{equation}

\noindent and they are neither Fermions nor Bosons. For infinite $U$, the
only $f$-electron states at any site $j$ are the vacuum $\mid j,0>$ and the
two states $\mid j,\sigma>$ that have one electron with spin component $%
\sigma$, and the only Fermi type operators that we shall need in this case
are $X_{j,o\sigma}$ and their Hermitian conjugates $X_{j,\sigma
o}=X_{j,o\sigma }^{\dagger}$. Projecting $H$ into the subspace without
doubly occupied f-electron states we obtain the PAM Hamiltonian for infinite
$U$:

\begin{align}
H & =\sum_{\vec{k},\sigma}E_{\mathbf{k},\sigma}C_{\mathbf{k},\sigma
}^{\dagger}C_{\mathbf{k},\sigma}+\sum_{j\sigma}E_{j\sigma}X_{j,\sigma\sigma }
\notag \\
& +\sum_{j,\mathbf{k},\sigma}(V_{j,\mathbf{k},\sigma}X_{j,o\sigma}^{\dagger
}C_{\mathbf{k}\sigma}+V_{j,\mathbf{k},\sigma}^{\ast}C_{\mathbf{k}\sigma
}^{\dagger}X_{j,o\sigma})  \label{Hub9}
\end{align}

\noindent where $X_{j,\sigma\sigma}=X_{j,o\sigma}^{\dagger}X_{j,o\sigma} $
is the projector into the state $\mid j,\sigma>.$ The identity relation in
the reduced space of the localized states at site $j$ is

\begin{equation}  \label{Hub8}
X_{j , o o} + X_{j, \sigma\sigma} + X_{j , \overline{\sigma} \overline{\sigma%
}} = I
\end{equation}

\noindent and its statistical average gives the conservation of probability
in that space of states.

The generalization of Eq.~(\ref{Hub9}) to the case of several configurations
with a rather arbitrary choice of states is

\begin{equation}
H=\sum_{\vec{k},\sigma}E_{\mathbf{k},\sigma}C_{\mathbf{k},\sigma}^{\dagger
}C_{\mathbf{k},\sigma}+\sum_{ja}E_{ja}X_{j,aa}+H_{h}=H{_{o}+}H_{h},
\label{Modelo5}
\end{equation}

\noindent where\footnote{%
For Eq. (\ref{Eq1}) with finite $U$ there are four states $\left\vert
0\right\rangle $, $\left\vert +\right\rangle $, $\left\vert -\right\rangle $
,$\left\vert d\right\rangle =\left\vert +-\right\rangle $, and we have $%
f_{\sigma}=X_{0\sigma}+\sigma X_{\bar{\sigma}d}$. ( $\sigma=1$ corresponds
to $+$ and $\sigma=-1$ to $-$). When we write Eq. \ref{Eq2} employing $%
X_{0\sigma}$ and $X_{\bar{\sigma}d}$ it appears $\sigma X_{\bar{\sigma}d}$
rather than $X_{\bar{\sigma}d}$, and we have to write the $V_{j,ba,\mathbf{k}%
,\sigma}$ in Eq. \ref{Hibr} in the following way: $V_{j,\bar{\sigma}d,%
\mathbf{k},\sigma}=\sigma V_{j,0\sigma,\mathbf{k},\sigma }=\sigma V_{j,%
\mathbf{k},\sigma}$.}

\begin{equation}
H_{h}=\sum_{jba,\vec{k}\sigma}\left( V_{jba,\mathbf{k}\sigma}X_{j,ba}^{%
\dagger}C_{\mathbf{k}\sigma}+V_{jba,\mathbf{k}\sigma}^{\ast}C_{\mathbf{k}%
\sigma}^{\dagger}X_{j,ba}\right) .  \label{Hibr}
\end{equation}

The $a$ and $b$ summations are over all the states $\mid a>$ and $\mid b>$
that we want to include in the model, and the only restriction is that any
hybridization constant must vanish unless state $\mid a>$ has just one
electron more than the state $\mid b>$: this last condition is necessary to
satisfy the conservation of electrons. In this general case, the energies $%
E_{j,a}$ include all the Coulomb repulsions of the type described by the
third term in Eq.~(\ref{Eq1}).

To abbreviate, we can write Eq.~(\ref{Hibr}) in the interaction picture in
the more compact form:

\begin{equation}
H_{h}(\tau)=\sum_{l,l^{\prime}}V(l,l^{\prime})Y(l)Y(l^{\prime}).
\label{Hibri1}
\end{equation}
where

\begin{equation}
Y(l)\equiv Y_{\gamma}(\tau)=\exp{(\tau}H{_{o})}Y_{\gamma}\exp{(-\tau}H{_{o})}
\label{Representacao6}
\end{equation}

\noindent is the operator $Y_{\gamma}$ in the interaction picture (the
subindex $\gamma$ is discussed in more detail after Eq. (\ref{Fourier2})).
The only non-zero coupling coefficients $V(l,l^{\prime})$ are those that
correspond to the correct combination of indices $l$ and $l^{\prime}$ in
Eq.~(\ref{Hibr}) and a factor $1/2$ is not necessary in Eq.~(\ref{Hibri1})
if we choose to retain only terms in which $Y(l)$ corresponds to the $f$%
-electrons and $Y(l^{\prime})$ to the conduction electron (to achieve this
ordering in the second term of the parenthesis in Eq.~(\ref{Hibr}) one must
anti commute two Fermi type operators, and the corresponding minus sign is
absorbed into a redefined hybridization constant).

As we are interested in the Grand Canonical Ensemble of electrons, we should
replace the total Hamiltonian $H$ by

\begin{equation}
\mathcal{H}=H-\mu\left\{ \sum_{\mathbf{k},\sigma}C_{\mathbf{k},\sigma
}^{\dagger}C_{\mathbf{k},\sigma}+\sum_{ja}\nu_{a}X_{j,aa}\right\} =\mathcal{H%
}_{o}+H_{h}  \label{Modelo7}
\end{equation}

\noindent where $X_{j , a a}$ is the occupation number operator of state $%
\mid a >$ at site $j$, and $\nu_{a}$ is the number of electrons in that
state. This transformation is easily performed by changing the energies $%
E_{j , a}$ of all ionic states $\mid a >$ into

\begin{equation}
\varepsilon_{j,a}=E_{j,a}-\mu\nu_{a}  \label{Eq1.13}
\end{equation}

\noindent and the energies $E_{\mathbf{k},\sigma}$ of the conduction
electrons into

\begin{equation}
\varepsilon\left( \mathbf{k},\sigma\right) =E_{\mathbf{k},\sigma}-\mu
\label{Energcond}
\end{equation}

\section{The Green's functions in imaginary frequency for several particles.}

\label{S2}

In this work we shall consider Green's Functions ((GF) of imaginary times of
both conduction electron operators $C_{\mathbf{k}\sigma}$ and Hubbard
operators $X_{j,ba}$, and a general GF can be written employing the $%
Y_{\gamma}$ operators:

\begin{equation}
\mathcal{G}(\gamma_{1},\tau_{1};\cdots;\gamma_{n},\tau_{n})=\left\langle
\left( \hat{Y}(\gamma_{1},\tau_{1})\cdots\hat{Y}(\gamma_{n},\tau_{n})\right)
_{+}\right\rangle _{\mathcal{H}},  \label{Fourier1}
\end{equation}

\noindent where

\begin{equation}  \label{Fourier2}
\hat Y(\gamma,\tau)=\exp{(\tau\mathcal{H})}Y_{\gamma}\exp{(-\tau\mathcal{H})}
\end{equation}

\noindent is defined for $\beta\geq\tau\geq0$. Besides the Fermi-like
operators $Y_{\gamma}$ that appear in $H_{h}$, we shall also consider
Bose-like Hubbard operators that do not change the number of electrons. At
this point it is necessary to be more specific about the argument $\gamma$
of the operators $Y_{\gamma}$ in Eqs.~(\ref{Representacao6}) and (\ref%
{Fourier2}). When the corresponding $Y_{\gamma}$ is a Fermi type $X_{j,ba}$,
we use $\gamma=(f;j,\alpha,u)$, with $u=-$, and the single index $\alpha$
identifies the transition $\mid a>$\hspace{0.2cm}$\rightarrow$\hspace{0.2cm}$%
\mid b>$, with the same restriction stated after Eq.~(\ref{Hibr}), namely
that state $\mid a>$ has just one electron more than the state $\mid b>$.
The inverse transition (operator $X_{j,ba}^{\dagger}$) is described by the
same $\alpha$ but with $u=+$. The $j$ identifies the site, $\tau$ is the
imaginary time (cf. Eq.~(\ref{Representacao6}) and Eq.~(\ref{Fourier2})) ,
and $f$ is only used when necessary to avoid confusion. When $Y_{\gamma}$ is
$C_{\mathbf{k}\sigma}$ we use $\gamma=(c;\mathbf{k},\sigma,u)$ with $u=-$
and change to $u=+$ for $C_{\mathbf{k}\sigma}^{\dagger}$. It is not
necessary to assign a $u$ parameter to the Bose-type operators, but to unify
the notation we shall keep the $u$ and put always $u=1$ for these operators.
The only restriction on the two states $\mid a>$ and $\mid b>$ of the
transition $\alpha=(b,a)$ for Bose type operators, is that they should have
the same number of electrons.

One can not use Feynman type expansions for the GF in Eq. (\ref{Fourier1}),
because the Hubbard operators are not Fermi operators, and we shall use a
cumulant expansion \cite{FFM} that is an extension of the one derived by
Hubbard \cite{Hubbard5} for his model. The diagrammatic expansion of the GF
is obtained employing the Theorem 3.3 from Reference \cite{FFM}, that
expresses the GF as the sum of the contributions of all the topologically
distinct and vacuum free graphs, drawn according to Rule 3.4 of that
reference. The\ corresponding contributions are calculated with Rule 3.6 of
\cite{FFM}, and in this section we shall summarize some details of these GF
calculation.

To avoid repeating the same term in Eq.~(\ref{Hibri1}) we assumed that $%
V(l,l^{\prime})$ is non-zero only when the first index corresponds to an $X$%
-operator. These coefficients do not depend on $\tau$ or $\tau^{\prime}$,
and to abbreviate it is convenient to introduce $v(j,\alpha,\mathbf{k}%
,\sigma,u)$ in Eq.~(\ref{Hibr}) :

\begin{equation}
\begin{array}{ccc}
v(j,\alpha,\mathbf{k},\sigma,+) & = & V(f;j,\alpha,+;c;\mathbf{k}%
,\sigma,-)=V_{j,ba,\mathbf{k},\sigma}, \\
v(j,\alpha,\mathbf{k},\sigma,-) & = & -V(f;j,\alpha,-;c;\mathbf{k}%
,\sigma,+)=V_{j,ba,\mathbf{k},\sigma}^{\ast}.%
\end{array}
\label{Ger3}
\end{equation}
The minus sign that should multiply into $V_{j,ba,\vec{k},\sigma}^{\ast}$,
because we anti-commuted two Fermi-type operators from Eq.~(\ref{Hibr}) in
the corresponding terms of Eq. (\ref{Hibri1}), will be absorbed in the rules
for the sign of the graph contributions when $\xi=0$ (cf. Appendix~\ref{ApA})%
\footnote{%
Note also that a factor $\left( -1\right) ^{n}$ appears in the perturbation
expansion contribution of any graph of order $n$, i.e. with $n$ internal
edges (cf. the cumulant expansion for the Ising model in Ref. \cite{Wortis74}%
, where this sign has been included in the interaction constant in its Eq.
(2)) \ We have\ then added a factor $\left( -1\right) $ to every internal
edge, and therefore this extra factors would only change the sign of a
graph's contribution when it is of odd order This sign appears explicitly in
the expansion of the PAM in \cite{FFM} (cf. Eqs. (3.8),(3.11) of that
reference) but it was left out from the diagrams contribution by an
oversight. Note that this sign does not depend on the Fermionic character of
the $X$ operators\label{myfoot3}.}.

To Fourier transform with respect to time the GF of Eq. (\ref{Fourier1}), it
is essential that they obey the boundary condition

\begin{align}
& \left\langle \left( \hat Y(\gamma_{1},\tau_{1})\cdots\hat Y(\gamma_{j},
\tau_{j}=\beta)\cdots\hat Y(\gamma_{n},\tau_{n})\right) _{+} \right\rangle _{%
\mathcal{H}} =  \notag \\
& \pm\left\langle \left( \hat Y(\gamma_{1},\tau_{1})\cdots\hat Y(\gamma
_{j},\tau_{j}=0)\cdots\hat Y(\gamma_{n},\tau_{n})\right) _{+}\right\rangle _{%
\mathcal{H}}  \label{Fourier3}
\end{align}

\noindent with respect to all the operators $\hat{Y}(\gamma_{1},\tau
_{1})\cdots\hat{Y}(\gamma_{n},\tau_{n})$, where the $-(+)$ corresponds to
Fermi-like (Bose-like) operators $\hat{Y}(\gamma_{j},\tau_{j})$.

When Eq.~(\ref{Fourier3}) is satisfied for all the variables and $H$ does
not depend on $\tau$, we can treat the GF as periodic (anti periodic) with
period $\beta$ in $\tau$, for all Bose-like (Fermi-like) operators $%
Y(\gamma,\tau)$, and we then write

\begin{align}
& \left\langle \left( \hat Y(\gamma_{1},\tau_{1})\cdots\hat Y(\gamma
_{n},\tau_{n})\right) _{+}\right\rangle _{\mathcal{H}} =  \notag \\
& \beta^{-\frac n2}\sum_{\omega_{1}\cdots\omega_{n}}\left\langle \left( \hat
Y(\gamma_{1},\omega_{1})\cdots\hat Y(\gamma_{n},\omega_{n})\right)
_{+}\right\rangle _{\mathcal{H}}  \notag \\
& \times\exp\left[ -i(\omega_{1}\tau_{1}+\cdots+\omega_{n}\tau_{n})\right]
\label{Fourier4}
\end{align}

The frequencies $\omega_{j}$ are different for the two type of operators $%
Y_{\gamma}$:

\begin{equation}
\omega_{j}=\frac{\pi\nu_{j}}{\beta}\qquad\text{where}\qquad\left\{
\begin{array}{ll}
\nu_{j}=0,\mp2,\mp4\cdots & \mbox{Bose-like} \\
\nu_{j}=1,\mp3,\mp5\cdots & \mbox{Fermi-like}.%
\end{array}
\right.  \label{Fourier5}
\end{equation}

The notation of the Fourier coefficients in Eq.~(\ref{Fourier4}) is purely
symbolic, because the $\tau$-ordering $(\ldots)_{+}$ has no meaning there.

\subsection{Rules for reciprocal space and imaginary frequencies}

\label{S2.0}

To Fourier transform the spatial dependence one has to remember that the $c$%
-operators are already in reciprocal space, so it is only necessary to
transform the $f$-operators. For a GF with $r$ operators of the $f$-type
(Fermi-like or Bose-like) and $n-r$ operators of the $c$-type we write in an
abbreviated notation

\widetext
\begin{align}
& \left\langle \left( \hat{Y}(f,\tau;1)\cdots\hat{Y}(f,\tau;r)\hat{Y}%
(c,\tau;r+1)\cdots\hat{Y}(c,\tau;n)\right) _{+}\right\rangle _{\mathcal{H}}=
\notag \\
& \beta^{-\frac{n}{2}}N_{s}^{-\frac{r}{2}}\sum_{\mathbf{k}_{1}\cdots \mathbf{%
k}_{r}}\sum_{\omega_{1}\cdots\omega_{n}}\exp[-i(\mathbf{k}_{1}u_{1}\mathbf{R}%
_{1}+\cdots+\mathbf{k}_{r}u_{r}\mathbf{R}_{r})-i(\omega_{1}\tau_{1}+\cdots+%
\omega_{n}\tau_{n})]  \notag \\
& \times\left\langle \left( \hat{Y}(f,\omega;1)\cdots\hat{Y}(f,\omega ;r)%
\hat{Y}(c,\omega;r+1)\cdots\hat{Y}(c,\omega;n)\right) _{+}\right\rangle _{%
\mathcal{H}},  \label{Fourier6}
\end{align}

\noindent where $R_{s}$ is the position of site $j_{s},$ $\hat{Y}(f,\tau;s)=%
\hat{Y}(f;j_{s},\alpha_{s},u_{s},\tau_{s})$, $\hat{Y}(c,\tau;s)=\hat{Y}(c;%
\mathbf{k}_{s},\sigma_{s},u_{s},\tau_{s})$, and we substitute the $\tau_{s}$
by $\omega_{s}$ in $\hat{Y}(f,\omega;s)$ and $\hat{Y}(c,\omega;s)$, as well
as $j_{s}$ by $\mathbf{k}_{s}$ in $\hat {Y}(f,\omega;s)$. With the same
notation, the inverse relation is then

\begin{align}
& \left\langle \left( \hat{Y}(f,\omega;1)\cdots\hat{Y}(f,\omega;r)\hat {Y}%
(c,\omega;r+1)\cdots\hat{Y}(c,\omega;n)\right) _{+}\right\rangle _{\mathcal{H%
}}=  \notag \\
& \beta^{-\frac{n}{2}}N_{s}^{-\frac{r}{2}}\sum_{j_{1}\cdots j_{r}}\int
_{0}^{\beta}d\tau_{1}\cdots\int_{0}^{\beta}d\tau_{n}\exp[+i(\mathbf{k}%
_{1}u_{1}\cdot\mathbf{R}_{1}+\cdots+\mathbf{k}_{r}u_{r}\cdot\mathbf{R}%
_{r})+i(\omega_{1}\tau_{1}+\cdots+\omega_{n}\tau_{n})]  \notag \\
& \times\left\langle \left( \hat{Y}(f,\tau;1)\cdots\hat{Y}(f,\tau;r)\hat {Y}%
(c,\tau;r+1)\cdots\hat{Y}(c,\tau;n)\right) _{+}\right\rangle _{\mathcal{H}}.
\label{Fourier7}
\end{align}

The present definition is slightly different from Hubbard's \cite{Hubbard5},
because we include the parameter $u=\pm1$ into the spatial part of the
exponential\footnote{%
The parameter $u$ was defined after Eq. (\ref{Fourier2}), as well as in the
sentences after Rule 3.4 in reference \cite{FFM}. The $u$ was convenient to
organize our calculation, but we did not use it in the temporal part of the
exponential because it was not particularly useful there.\label{myfoot4a}}
in Eqs.~(\ref{Fourier6}) and (\ref{Fourier7}).

From the invariance under time translation (i.e. $\mathcal{H}$ does not
depend on $\tau$) one can show that the GF in Eq.~(\ref{Fourier7}) vanishes
unless

\begin{equation}  \label{Fourier8}
\omega_{1}+\omega_{2}+\cdots+\omega_{n}=0.
\end{equation}

To prove the corresponding property for the wave vectors $\mathbf{k}_{j}$ in
Eq.~(\ref{Fourier7}), it is necessary to transform first the $c$-operators
into the Wannier representation

\begin{equation}
C_{j\sigma}^{\dagger}=\frac{1}{\sqrt{N_{s}}}\sum_{\mathbf{k}}\exp (-i\mathbf{%
k}\cdot\mathbf{R}_{j})C_{\mathbf{k}\sigma}^{\dagger}.  \label{Fourier9}
\end{equation}

Substituting Eq.~(\ref{Fourier9}) (or its Hermitian conjugate for $%
C_{j\sigma }$ ) into the GF in the r.h.s. of Eq.~(\ref{Fourier7}), and
employing the invariance under lattice translation, one finds that the GF in
Eq.~(\ref{Fourier7}) vanishes unless

\begin{equation}
\mathbf{k}_{1}u_{1}+\mathbf{k}_{2}u_{2}+\cdots+\mathbf{k}_{n}u_{n}=0.
\label{Fourier10}
\end{equation}
It is clear that the relations in Eqs.~(\ref{Fourier6}-\ref{Fourier7}) can
also be employed for the corresponding cumulant averages. When $H_{h}=0$
many $Y_{\gamma}$ are statistically independent, and the only cumulants left
in the imaginary-time and real-space expansion (cf. Rule 3.6 in reference
\cite{FFM}) must either have all their $Y_{\gamma}$ of the $f$-type and at
the same site, or else have all of the $c$-type with the same $\mathbf{k}$
(and same $\sigma$ when $\mathcal{H}_{o}$ is spin independent). Because of
the invariance of the system under lattice translations, the local cumulants
that appear in Rule 3.6a in \cite{FFM} are independent of the site position
and it is not necessary to take their spatial Fourier transform; on the
other hand, the $Y_{\gamma}$ of the $C$-electron cumulants of Rule 3.6 b' in
\cite{FFM} have been already transformed. From these two facts it follows
that to obtain the Fourier transformed version (in reciprocal space and
imaginary time) of Rule 3.6 in \cite{FFM} it would be sufficient to apply
only the transformation from time to frequency (cf. Eq.~(\ref{Fourier4})) to
the cumulants in that rule.

To set the notation we write

\begin{align}
& \left\langle \left( Y(f;j,\alpha_{p},u_{p},\tau_{p})\cdots Y(f;j,\alpha
_{1},u_{1},\tau_{1})\right) _{+}\right\rangle _{c} =  \notag \\
& \beta^{-\frac p2}\sum_{\omega_{1}\cdots\omega_{p}}\exp[-i(\omega_{1}%
\tau_{1}+\cdots+\omega_{p}\tau_{p})]  \notag \\
& \left\langle \left( Y(j,\alpha_{p},u_{p},\omega_{p})\cdots Y(j,\alpha
_{1},u_{1},\omega_{1})\right) _{+}\right\rangle _{c}  \label{Fourier11}
\end{align}

and

\begin{align}
& \left\langle \left( Y(c;\mathbf{k},\sigma_{2},-u_{2},\tau_{2})Y(c;\mathbf{k%
},\sigma_{1},u_{1},\tau_{1})\right) _{+}\right\rangle _{c}=  \notag \\
& \beta^{-1}\sum_{\omega_{1}\omega_{2}}\exp[-i(\omega_{1}\tau_{1}+\omega
_{2}\tau_{2})]  \notag \\
& \left\langle \left( C(\mathbf{k},\sigma_{2},-u_{2},\omega_{2})C(\mathbf{k}%
,\sigma_{1},u_{1},\omega_{1})\right) _{+}\right\rangle _{c}
\label{Fourier12}
\end{align}

Note that the invariance under time translation guarantees that Eq.~(\ref%
{Fourier8}) would be satisfied for the frequency dependent cumulants of
Eqs.~(\ref{Fourier11}) and (\ref{Fourier12}). To proceed with the
transformation of Rule 3.6 in \cite{FFM}, we use the prescriptions
summarized above to express the GF in the r.h.s. of Eq.~(\ref{Fourier7}) as
a sum of terms, each corresponding to the contribution of some graph. In
each term one introduces Eqs.~(\ref{Fourier11}) and (\ref{Fourier12}) and
then performs explicitly all the integrations over $\tau$ and all the
non-restricted summations over the sites $j$.

In each integration over $\tau$ there are two possibilities: the $\tau$
corresponds either to an external operator or else to an internal line. When
the $\tau_{j}$ corresponds to an external operator $Y(\gamma_{j},\tau_{j})$,
the Eq.~(\ref{Fourier7}) provides the integration, and the integrand has two
factors: one $\exp(i\omega_{j}\tau_{j})$ from Eq.~(\ref{Fourier7}) and
another $\exp(-i\omega_{s}\tau_{j})$ from applying Eqs.~(\ref{Fourier11})
and (\ref{Fourier12}) to the cumulant of Rule 3.6 that contains the external
operator $Y(\gamma_{j},\tau_{j})$. As both $\omega_{j}$ and $\omega_{s}$ are
of the same type (cf. Eq.~(\ref{Fourier5})), the integral vanishes unless $%
\omega_{j}=\omega_{s}$, and from the sum over all the $\omega_{s}$ in Eqs.~(%
\ref{Fourier11}) and (\ref{Fourier12}) only the external frequency $%
\omega_{j}$ remains.

When the $\tau_{s}$ belongs to an internal line, the integration comes from
the perturbation expansion (cf. Eq.~(\ref{Expansao2})), and the integrand is
$\exp[-i(\omega_{s}+\omega_{s}^{\prime})\tau_{s}]$\thinspace where $\omega
_{s}$ and $\omega_{s}^{\prime}$ come from expanding with Eq.~(\ref{Fourier11}%
) or Eq.~(\ref{Fourier12}) the two cumulants of Rule 3.6 in \cite{FFM} that
contain the $C$-operator and the $X$-operator of the internal line. The
integration is again zero unless $\omega_{s}+\omega_{s}^{\prime}=0$, and one
can then associate only one of these two frequencies to the internal line in
the transformed rules.

In Eq.~(\ref{Fourier7}) we have applied the spatial Fourier transformation
to the external $X$-operators, which together with \ items (3a) and (2d) of
Rules 3.5-3.6 from \cite{FFM} \ imply a sum over all the sites in the
lattice. It is then convenient to write explicitly the dependence with $%
R_{j} $ of the coupling constants of Eq.~(\ref{Ger3}):

\begin{equation}
v(j,\alpha,\mathbf{k},\sigma,u)=V(\alpha,\mathbf{k},\sigma,u)N_{s}^{-\frac {1%
}{2}}\exp(iu\mathbf{k}.\mathbf{R}_{j}),  \label{Fourier13}
\end{equation}
and one then obtains the following Rule.

\textbf{Rule 3.7}

To calculate the contribution of any diagram obtained from Rule 3.4 of \cite%
{FFM}

\begin{description}
\item[1.\ ] Assign to each internal line a momentum $\mathbf{k}_{s}$, a
frequency $\omega_{s}$, a spin $\sigma_{s}$ and an index $u_{s}$. Assign
dummy labels $\alpha_{s}$ and $\pm u_{s}$ to the $X$-operators at the FV
side of the internal line, and dummy labels $\mathbf{k}_{s}$, $\sigma_{s}$
and $\mp u_{s}$ to the $C$-operators at the CV side. Use $+u_{s}$ and $%
+\omega_{s}$ at the side of the edge to which points the arrow (cf. item iv
of Rule 3.4 in \cite{FFM}) and $-u_{s}$ and $-\omega_{s}$ to the opposite
side.

Assign to the external lines the labels of the corresponding external
operators, namely the momentum $\mathbf{k}_{s}$, frequency $\omega_{s}$,
index $u_{s}$ and also the transition $\alpha_{s}=(b_{s},a_{s})$ for $X$%
-operators and the spin component $\sigma_{s}$ for the $C$-operators (we use
always $+u_{r}$ and $+\omega_{r}$ for the external lines).

\item[2.\ ] Form the product of the following factors:

\item[ (a)] For each FV with lines $s=1,2,\cdots,p$ running to that vertex
(both internal and external) the factor\footnote{%
To simplify the notation we use $\Delta\left( x\right) =0$ when $x\neq0$ but
$\Delta\left( 0\right) =1$.\label{myfoot4b}}
\end{description}

\begin{align}
& N_{s}\Delta(\pm u_{p}\mathbf{k}_{p}\pm\cdots\pm u_{2}\mathbf{k}_{2}\pm
u_{1}\mathbf{k}_{1})  \notag \\
& \times\left\langle \left( X(j,\alpha_{p},\pm u_{p},\pm\omega_{p})\cdots
X(j,\alpha_{1},\pm u_{1},\omega_{1})\right) _{+}\right\rangle _{c},
\label{Fourier14}
\end{align}

\begin{description}
\item[ \ \ \ \ ] where $\mathbf{k}_{s}$, $\omega_{s}$, $\alpha_{s}$ and $%
u_{s}$ are the momentum, frequency transition, \mbox{$\alpha _s=(b_s,a_s)$}
and index $u_{s}$ labels of the $X$-operators associated to line s (always $%
+u_{s}$ and $+\omega_{s}$ for the external lines).

\item[ (b) ] For each CV a factor
\end{description}

\begin{equation}
\left\langle \left( C(\mathbf{k}_{1}^{\prime},\sigma_{1}^{\prime},-u_{1}^{%
\prime},-\omega_{1}^{\prime})\ C(\mathbf{k}_{1},\sigma_{1},u_{1},\omega_{1})%
\right) _{+}\right\rangle _{c},  \label{Fourier15}
\end{equation}

\begin{description}
\item[ \ \ \ \ ] where $\mathbf{k}_{1}$, $\sigma_{1}$, $u_{1}$ and $\omega
_{1}$ are the parameters of the edge with the arrow pointing towards the CV.
As we discussed before, this cumulant vanishes unless $\mathbf{k}_{1}=%
\mathbf{k}_{1}^{\prime}$, $u_{1}=u_{1}^{\prime}$ and $\omega_{1}=%
\omega_{1}^{\prime}$. When the Bloch states $\mid\mathbf{k},\sigma>$ are
eigenstates of $\mathcal{H}_{o}$, we have also $\sigma_{1}=\sigma_{1}^{%
\prime }$ and the factor above (cf. footnote \ref{myfoot6} in Appendix \ref%
{ApE}) is equal to

\begin{align}
& \frac{1}{i\omega_{1}+u_{1}\varepsilon(\mathbf{k}_{1},\sigma_{1})}  \notag
\\
& \times\delta(\mathbf{k}_{1},\mathbf{k}_{1}^{\prime})\delta(u_{1},u_{1}^{%
\prime})\delta(\sigma_{1},\sigma_{1}^{\prime})\delta(\omega_{1},\omega_{1}^{%
\prime})  \label{Fourier16}
\end{align}

where the parameters with sub index 1 correspond again to the edge with the
arrow pointing towards the CV (when the outgoing line is external with given
$u$ and $\omega$, we put $-u_{1}^{\prime}=u$ and $-\omega_{1}^{\prime}=%
\omega)$.

\item[ (c)] A factor $\left( -1\right) \ V(\alpha,\vec{k},\sigma,\pm u)$ for
each internal line$^{\text{\ref{myfoot3}}}$ with labels $\alpha$, $\pm u$ at
the FV site and labels $\vec{k}$, $\sigma$ and $\mp u$ at the CV side, as
written in \cite{FFM} (cf. Eq.~(\ref{Fourier13}))\footnote{%
Note that in Reference \cite{FFM} we use $V(\alpha,k,\sigma,\pm u)$ in
reciprocal space (rule 3.7, item 2.c) and $v(\alpha,k,\sigma,\pm u)$ in real
space (rule 3.5, item 2.c)\label{myfoot4}}.

\item[ (d)] A factor $\pm1$ determined by the rules in Appendix C.

\item[ (e)] A factor $1/g$ determined by the rules in Appendix D.

\item[ (f)] A factor $1/\sqrt{N_{s}}$ for each external line running to a FV.

\item[3.\ ] Sum the resulting product with respect to

\item[ (a)] The momenta $\mathbf{k}_{s}$, the frequencies $\omega_{s}$ and
the indices $u_{s}$ of all the internal edges. Divide each sum over momenta
into $\sqrt{N_{s}}$.

\item[ (b)] The labels $\alpha_{s}$ of the $X$-operators at the FV side of
all internal lines.

\item[ (c)] The label $\sigma_{{s}}$ of the $C$-operators at the CV side of
all internal lines.$\bullet$
\end{description}

Two points should be stressed: i) The frequencies of each local cumulant in
2.a satisfy Eq.~(\ref{Fourier8}), thus reducing by one the number of
frequency summations at each FV. ii) The rules are also valid for vacuum
graphs, and are employed to calculate the GPF with the Linked Cluster
Theorem.

\subsection{Rules for real space and imaginary frequencies (Valid for
impurities)}

\label{S2.1}

We shall transform Fourier the imaginary times of the diagrammatic expansion
calculated with Rule 3.6 in \cite{FFM}, but leave the real space description
of the local sites unchanged.

We employ the Rule 3.4 in \cite{FFM} for drawing the nth-order graphs for
the cumulant expansion. The following relations give the Fourier transforms,
following the same definitions employed in \cite{FFM}

.

\begin{align}
& \left\langle \left( \hat{Y}(f,\tau;1)\cdots\hat{Y}(f,\tau;r)\hat{Y}%
(c,\tau;r+1)\cdots\hat{Y}(c,\tau;n)\right) _{+}\right\rangle _{\mathcal{H}}=
\notag \\
& \beta^{-\frac{n}{2}}\sum_{\omega_{1}\cdots\omega_{n}}\exp[%
-i(\omega_{1}\tau_{1}+\cdots+\omega_{n}\tau_{n})]  \notag \\
& \times\left\langle \left( \hat{Y}(f,\omega;1)\cdots\hat{Y}(f,\omega ;r)%
\hat{Y}(c,\omega;r+1)\cdots\hat{Y}(c,\omega;n)\right) _{+}\right\rangle _{%
\mathcal{H}},  \label{Fourier6a}
\end{align}

\noindent where $\hat{Y}(f,\tau;s)=\hat{Y}(f;j_{s},\alpha_{s},u_{s},%
\tau_{s}) $ , $\hat{Y}(c,\tau;s)=\hat{Y}(c;\vec{k}_{s},\sigma_{s},u_{s},%
\tau_{s})$, and we substitute the $\tau$ by $\omega$ in $\hat{Y}(f,\omega;s)$
and $\hat {Y}(c,\omega;s)$, but keep the $j_{s}$ in $\hat{Y}(f,\omega;s)$
because here we do not transform the GF into reciprocal space. With the same
notation, the inverse relation is then

\begin{align}
& \left\langle \left( \hat{Y}(f,\omega;1)\cdots\hat{Y}(f,\omega;r)\hat {Y}%
(c,\omega;r+1)\cdots\hat{Y}(c,\omega;n)\right) _{+}\right\rangle _{\mathcal{H%
}}=  \notag \\
& \beta^{-\frac{n}{2}}\int_{0}^{\beta}d\tau_{1}\cdots\int_{0}^{\beta}d%
\tau_{n}\exp[i(\omega_{1}\tau_{1}+\cdots+\omega_{n}\tau_{n})]  \notag \\
& \times\left\langle \left( \hat{Y}(f,\tau;1)\cdots\hat{Y}(f,\tau;r)\hat {Y}%
(c,\tau;r+1)\cdots\hat{Y}(f,\tau;n)\right) _{+}\right\rangle _{\mathcal{H}}.
\label{Fourier7a}
\end{align}

From the invariance under time translation (i.e. $\mathcal{H}$ does not
depend on $\tau)$ one can show again that the GF in Eq.~(\ref{Fourier7a})
vanishes unless Eq. (\ref{Fourier8}) is satisfied (i.e. $\omega_{1}+%
\omega_{2}+\cdots+\omega_{n}=0$).

It is clear that the relations in Eqs.~(\ref{Fourier6a}-\ref{Fourier7a}) can
also be employed for the corresponding cumulant averages. When $H_{h}=0$
many $Y_{\gamma}$ are statistically independent, and the only cumulants left
in Rule 3.6 must either have all their $Y_{\gamma}$ of the $f$-type and at
the same site, or else have all their $Y_{\gamma}$ of the $c$-type with the
same $\mathbf{k}$ (and same $\sigma$ when $\mathcal{H}_{o}$ is spin
independent).We shall then apply the transformation from time to frequency
(cf. Eq.~(3.29) in \cite{FFM}) to the cumulants in that rule. To set the
notation we write

\begin{align}
& \left\langle \left( Y(f;j,\alpha_{p},u_{p},\tau_{p})\cdots Y(f;j,\alpha
_{1},u_{1},\tau_{1})\right) _{+}\right\rangle _{c}=  \notag \\
& \beta^{-\frac{p}{2}}\sum_{\omega_{1}\cdots\omega_{p}}\exp[%
-i(\omega_{1}\tau_{1}+\cdots+\omega_{p}\tau_{p})]  \notag \\
& \left\langle \left( X(j,\alpha_{p},u_{p},\omega_{p})\cdots X(j,\alpha
_{1},u_{1},\omega_{1})\right) _{+}\right\rangle _{c}  \label{Fourier11a}
\end{align}

and

\begin{align}
& \left\langle \left( Y(c;\vec{k},\sigma_{2},-u_{2},\tau_{2})Y(c;\vec {k}%
,\sigma_{1},u_{1},\tau_{1})\right) _{+}\right\rangle _{c}=  \notag \\
& \beta^{-1}\sum_{\omega_{1}\omega_{2}}\exp[-i(\omega_{1}\tau_{1}+\omega
_{2}\tau_{2})]  \notag \\
& \left\langle \left( C(\vec{k},\sigma_{2},-u_{2},\omega_{2})C(\vec {k}%
,\sigma_{1},u_{1},\omega_{1})\right) _{+}\right\rangle _{c}
\label{Fourier12a}
\end{align}

Note that the invariance under time translation guarantees that Eq.~(\ref%
{Fourier8}) would be satisfied for the frequency dependent cumulants of
Eqs.~(\ref{Fourier11a}) and (\ref{Fourier12a}). To proceed with the
transformation of Rule 3.6, we use the prescription given in Rule 3.4 to
express the GF in the r.h.s. of Eq.~(\ref{Fourier7a}) as a sum of terms,
each corresponding to some graph. In the contribution of each graph one
introduces Eqs.~(\ref{Fourier11a}) and (\ref{Fourier12a}) and then performs
explicitly all the integrations over $\tau\ $while the non-restricted
summations over the sites $j$ remain expressed formally. In each integration
over $\tau$ there are two possibilities: the $\tau$ corresponds either to an
external operator or else to an internal line. When the $\tau_{j}$
corresponds to an external operator $Y(\gamma_{j},\tau_{j})$, the Eq.~(\ref%
{Fourier7a}) provides the integration, and the integrand has two factors:
one $\exp(i\omega_{j}\tau _{j})$ from Eq.~(\ref{Fourier7a}) and another $%
\exp(-i\omega_{s}\tau_{j})$ from applying Eqs.~(\ref{Fourier11a}) and (\ref%
{Fourier12a}) to the cumulant of Rule 3.6 that contains the external
operator $Y(\gamma_{j},\tau_{j})$. As both $\omega_{j}$ and $\omega_{s}$ are
of the same type (cf. Eq.~(3.30) in \cite{FFM}), the integral vanishes
unless $\omega_{j}=\omega_{s}$, and from the sum over all the $\omega_{s}$
in Eqs.~(\ref{Fourier11a}) and (\ref{Fourier12a}) only the external
frequency $\omega_{j}$ remains. When the $\tau_{s}$ belongs to an internal
line, the integration comes from the perturbation expansion (cf. Eq.~(3.11)
in \cite{FFM}), and the integrand is $\exp[-i(\omega_{s}+\omega_{s}^{%
\prime})\tau_{s}]$\thinspace where $\omega _{s}$ and $\omega_{s}^{\prime}$
come from expanding with Eq.~(\ref{Fourier11a}) or Eq.~(\ref{Fourier12a})
the two cumulants of Rule 3.6 that contain the $C$-operator and the $X$%
-operator of the internal line. The integration is again zero unless $%
\omega_{s}+\omega_{s}^{\prime}=0$, and one can then associate only one of
these two frequencies to the internal line in the transformed rules. Before
explicitly giving the rules for the GF calculation, it is convenient to
recall the definition Eq. (\ref{Ger3}) of the coefficients $v(j,\alpha,\vec{k%
},\sigma,u)$:
\begin{equation*}
\begin{array}{ccc}
v(j,\alpha,\vec{k},\sigma,+) & = & V(f;j,\alpha,+;c;\vec{k},\sigma
,-)=V_{j,ba,\vec{k},\sigma}, \\
v(j,\alpha,\vec{k},\sigma,-) & = & -V(f;j,\alpha,-;c;\vec{k},\sigma
,+)=V_{j,ba,\vec{k},\sigma}^{\ast}.%
\end{array}%
\end{equation*}
The minus sign that should appear with $V_{j,ba,\vec{k},\sigma}^{\ast}$
because we anti-commuted two Fermi-type operators from Eq.~(2.8) in \cite%
{FFM}, will be absorbed in the rules for the sign of the graph contributions
when $\xi=0$ (cf. Appendix~\ref{ApA}). We can now give the rules without
further discussion.

\textbf{Rule 3.7a} To calculate the contribution of a diagram obtained from
Rule 3.4

\begin{description}
\item[1.\ ] Assign to each FV a site label $j_{s}$. To each internal line a
momentum $\vec{k}_{s}$, a frequency $\omega_{s}$ and an index $\pm u_{s}$.
Assign dummy labels $\alpha_{s}$ and $\pm u_{s}$ to the $X$-operators at the
FV side of the internal line, and dummy labels $\mathbf{k}_{s}$, $\sigma_{s}$
and $\mp u_{s}$ to the $C$-operators at the CV side. Use $+u_{s}$ and $%
+\omega_{s}$ at the side of the edge to which points the arrow (cf. item iv
of Rule 3.4) and $-u_{s}$ and $-\omega_{s}$ to the opposite side. Assign to
the external lines the labels of the corresponding external operators,
namely the frequency $\omega_{s}$, index $u_{s}$ and either the site $j_{s}$
and transition $\alpha_{s}=(b_{s},a_{s})$ for $X$-operators or the momentum $%
\mathbf{k}_{s}$ and the spin component $\sigma_{s}$ for the $C$-operators
(we use always $+u_{r}$ and $+\omega_{r}$ for the external lines).

\item[2.\ ] Form the product of the following factors:

\item[ (a)] For each FV with lines $s=1,2,\cdots,p$ running to that vertex
(both internal and external) the factor
\end{description}

\begin{equation}
\left\langle \left( X(j_{p},\alpha_{p},\pm u_{p},\pm\omega_{p})\cdots
X(j_{1},\alpha_{1},\pm u_{1},\omega_{1})\right) _{+}\right\rangle _{c},
\label{Fourier14a}
\end{equation}

\begin{description}
\item[ \ \ \ \ ] where $j_{s}$, $\omega_{s}$, $\alpha_{s}$ and $u_{s}$ are
the site, frequency transition, $\alpha_{s}=\left( b_{s},a_{s}\right) $ and
index $u_{s}$ labels of the $X$-operators associated to line s (always $%
+u_{s}$ and $+\omega_{s}$ for the external lines).

\item[ (b) ] For each CV a factor
\end{description}

\begin{equation}
\left\langle \left( C(\mathbf{k}_{1}^{\prime},\sigma_{1}^{\prime},-u_{1}^{%
\prime},-\omega_{1}^{\prime})C(\mathbf{k}_{1},\sigma_{1},u_{1},\omega_{1})%
\right) _{+}\right\rangle _{c},  \label{Fourier15a}
\end{equation}

\begin{description}
\item[ \ \ \ \ ] where $\mathbf{k}_{1}$, $\sigma_{1}$, $u_{1}$ and $\omega
_{1}$ are the parameters of the edge with the arrow pointing towards the CV.
As we discussed before, this cumulant vanishes unless $\mathbf{k}_{1}=%
\mathbf{k}_{1}^{\prime}$, $u_{1}=u_{1}^{\prime}$ and $\omega_{1}=%
\omega_{1}^{\prime}$. When the Bloch states $\mid\mathbf{k},\sigma>$ are
eigenstates of $\mathcal{H}_{o}$, we have also $\sigma_{1}=\sigma_{1}^{%
\prime }$ and the factor above (cf. footnote \ref{myfoot6} in Appendix \ref%
{ApE}) is equal to
\begin{align}
& \frac{1}{i\omega_{1}+u_{1}\epsilon(\mathbf{k}_{1},\sigma_{1})}  \notag \\
& \times\delta(\mathbf{k}_{1},\mathbf{k}_{1}^{\prime})\delta(u_{1},u_{1}^{%
\prime})\delta(\sigma_{1},\sigma_{1}^{\prime})\delta(\omega_{1},\omega_{1}^{%
\prime})  \label{Fourier16a}
\end{align}
where the parameters with sub index 1 correspond again to the edge with the
arrow pointing towards the CV (when the outgoing line is external with given
$u$ and $\omega$, we put $-u_{1}^{\prime}=u$ and $-\omega_{1}^{\prime}=%
\omega)$.

\item[ (c)] A factor $\left( -1\right) \ v(j,\alpha,\mathbf{k},\sigma,\pm u)
$ for each internal edge$^{\text{\ref{myfoot3}}}$ joining a FV at site $j$
with labels $\alpha$, $\pm u$ to a CV with labels $\mathbf{k}$, $\sigma$ and
$\mp u$.

\item[ (d)] A $\delta(j_{s},j_{i})$ for each external line $X$-operator at
site $j_{i}$ running to an FV site labeled with $j_{s}.$ The labels $j_{s}$
are dummy labels, but the Kroenecker deltas in the present item take care of
fixing its value when there is an external line running to a FV.

\item[ (e)] A factor $\pm1$ determined by the rules in Appendix C.

\item[ (f)] A factor $1/g$ determined by the rules in Appendix D.

\item[3.\ ] Sum the resulting product with respect to

\item[ (a)] The site labels $j_{s}$ of all the FV.

\item[ (b)] The momenta $\mathbf{k}_{s}$, the frequencies $\omega_{s}$ and
the indices $u_{s}$ of all the internal edges.

\item[ (c)] The labels $\alpha_{s}$ of the $X$-operators at the FV side of
all internal lines.

\item[ (d)] The label $\sigma_{{s}}$ of the $C$-operators at the CV side of
all internal lines.$\bullet$
\end{description}

Two points should be stressed: i) The frequencies of each local cumulant in
2.a satisfy Eq.~(\ref{Fourier8}), thus reducing by one the number of
frequency summations at each FV. ii) The rules are also valid for vacuum
graphs, and are employed to calculate the GCP with the Linked Cluster
Theorem.

\bigskip

\section{The effective cumulant\label{S4}}

The general GF in reciprocal space and imaginary frequency that we shall use
is
\begin{align}
& \mathcal{G}^{ff}(\mathbf{k},\alpha,u,\omega;\mathbf{k}^{\prime},\alpha^{%
\prime},u^{\prime},\omega^{\prime})=\   \notag \\
& =\frac{1}{\beta N_{s}}\sum_{j,\text{ }j^{\prime}}\exp\left[ i\left( u%
\mathbf{k}\cdot\mathbf{R}+u^{\prime}\mathbf{k}^{\prime}\cdot\mathbf{R}%
^{\prime}\right) \right] \int_{0}^{\beta}d\tau\int_{0}^{\beta}d\tau^{\prime
}\exp\left[ i\left( \omega\tau+\omega^{\prime}\tau^{\prime}\right) \right]
\left\langle \left( \widehat{Y}\left( f;j,\alpha,u,\tau\right) \widehat {Y}%
\left( f;j^{\prime},\alpha^{\prime},u^{\prime},\tau^{\prime}\right) \right)
_{+}\right\rangle _{\mathcal{H}}  \label{Eq3.1}
\end{align}

where $\mathbf{R}^{\prime}$ is the position of the site $j^{\prime}$, and in
particular we need the transform of $\left\langle \left( \widehat {X}%
_{j,\alpha}(\tau_{1})X_{j^{\prime},\alpha^{\prime}}^{\dagger}(0)\right)
_{+}\right\rangle _{\mathcal{H}}$, i.e.: $\mathcal{G}^{ff}(\mathbf{k}%
,\alpha,u=-,\omega;\mathbf{k}^{\prime},\alpha^{\prime},u^{\prime}=+,\omega^{%
\prime})$ . Employing Eqs. ( \ref{Fourier8},\ref{Fourier10}) and the
conservation of the number of electrons, we can abbreviate$^{\text{\ref%
{myfoot4b}}}$%
\begin{equation}
\mathcal{G}^{ff}(\mathbf{k},\alpha,u=-,\omega;\mathbf{k}^{\prime},\alpha^{%
\prime},u^{\prime},\omega^{\prime})=\mathcal{G}_{\alpha\alpha
^{\prime}}^{ff}(\mathbf{k},i\omega_{j})\ \Delta\left( u+u^{\prime}\right)
\Delta\left( u\mathbf{k}+u^{\prime}\mathbf{k}^{\prime}\right) \Delta\left(
\omega_{j}+\omega_{j}^{\prime}\right)  \label{Eq3.1.1}
\end{equation}
where $\omega$ and $\omega^{\prime}$ are given by Eq. (\ref{Fourier5}).

In the calculation with the usual Fermi or Bose operators, the one-particle
propagator of the f-electron is given by a sum of diagrams of the type shown%
\footnote{%
But note that the usual meaning of vertices and edges is exchanged with that
employed in the cumulant expansion.} in figure \ref{GffCHA}a but with each
local vertex corresponding to the sum of all \textquotedblleft
proper\textquotedblright\ (or irreducible) diagrams \cite{FetterW,LuttingerW}%
. The same result is found in the cumulant expansion of the Hubbard model
for $d\rightarrow\infty$ \cite{Metzner,CracoG} when the electron hopping is
employed as perturbation. The vertices then represent an \textquotedblleft
effective cumulant\textquotedblright\ $M_{2,\sigma}^{eff}(z_{n})$, that is
independent of $\mathbf{k}$ because only diagrams of a special type
contribute to this quantity for $d\rightarrow\infty$.

In the cumulant expansion of the Anderson lattice \cite{FFM} we employ the
hybridization rather than the hopping as a perturbation, and the exact
solution of the conduction electrons problem in the absence of hybridization
is part of the zeroth order Hamiltonian. For this reason it became necessary
to extend Metzner's derivation \cite{Metzner} to the Anderson lattice for $%
U\rightarrow\infty$, and we have shown\cite{FFdi} that the same type of
results obtained by Metzner are also valid for this model. These results had
been used to obtain the exact GF employed in \cite{Foglio}, but the
expression of the exact GF is valid for all dimensions and it has been used
to study \textrm{FeSi }\cite{FoglioFeSi,FoglioFeSi2}. As with the Feynman
diagrams, one can rearrange for $U\rightarrow\infty$ all the diagrams that
contribute to the exact $\mathcal{G}_{\sigma}^{ff}(\mathbf{k},z_{n})$, by
defining an effective cumulant $M_{2,\sigma}^{eff}(z_{n})$ that is given by
all the diagrams of $\mathcal{G}_{\sigma}^{ff}(\mathbf{k},z_{n})$ that can
not be separated by cutting a single edge (usually called \textquotedblleft
proper\textquotedblright\ or \textquotedblleft
irreducible\textquotedblright\ diagrams). The exact GF $\mathcal{G}%
_{\sigma}^{ff}(\mathbf{k},z_{n})$ is then given by the family of diagrams in
figure \ref{GffCHA}a, but with the effective cumulant $M_{2,\sigma}^{eff}(%
\mathbf{k,}z_{n})$ in place of the bare cumulant $M_{2,%
\sigma}^{0}(z_{n})=-D_{\sigma}^{0}/(z_{n}-\varepsilon_{f})$ at all the
filled vertices.

For finite $U$ one has sixteen exact GF\ $\mathcal{G}_{\alpha\alpha^{%
\prime}}^{ff}(\mathbf{k},z_{n})$ that define a \ $4\times4$ matrix, but when
the Hamiltonian commutes with the spin it can be split into two independent $%
2\times2$ matrices, one for each spin component. We name $\mathbf{G}^{ff}(%
\mathbf{k},z_{n})$ these two matrices, one for each spin component $\sigma$,
and\ when there is no possibility of confusion we do not write explicitly
the $\sigma$. In Appendix \ref{ApD} we obtain the expression of $\mathbf{G}%
^{ff}(\mathbf{k},z_{n})$ as a function of an effective cumulant matrix $%
\left\{ \mathbf{M}\right\}
_{_{\alpha,\alpha^{\prime}}}=M_{\alpha\alpha^{\prime}}^{eff}(\mathbf{k},z,u)$
(cf. \ref{ApD23}). Similar results are obtained in real space for $\mathcal{G%
}_{\alpha\alpha^{\prime}}^{ff}(\mathbf{j},z_{n})$\ as a function of an
effective cumulant matrix $\left\{ \mathbf{M}\right\}
_{_{\alpha,\alpha^{\prime}}}=M_{\alpha \alpha^{\prime}}^{eff}(\mathbf{j}%
,z,u) $ (cf. \ref{ApD43}). As before, these effective cumulants define in
each case two independent $2\times2$ matrices $\mathbf{M}$

The derivation given in Appendix \ref{ApD} is rather general and can be
extended to any number of transitions $\alpha$, but in this work we shall
only consider the case of $U\rightarrow\infty$ (only one transition per spin
$\alpha=(0,\sigma)$) and the case of finite $U$ (two transitions per spin $%
\alpha=(0,\sigma),(\bar{\sigma},d)$).

\section{The exact Green's functions from the cumulant expansion}

\subsubsection{The exact formal Green's functions for the PAM\label{S02_4}}

The contribution to\ $\mathcal{G}_{\alpha \alpha ^{\prime }}^{ff}(\mathbf{k}%
,z)$ (cf. Eq. (\ref{Eq3.1.1})) of the term in the series that has $n+1$
effective cumulants is given by (cf. Eq. (\ref{ApD25}) in Appendix \ref{ApD}%
)
\begin{equation}
\left\{ \left( \mathbf{M\cdot W}\right) ^{n}\cdot \mathbf{M}\right\}
_{\alpha \alpha ^{\prime }}=\left\{ \mathbf{M\cdot }\left( \mathbf{W\cdot M}%
\right) ^{n}\right\} _{\alpha \alpha ^{\prime }},  \label{Eq3.12}
\end{equation}%
with
\begin{equation}
\left\{ \mathbf{W}\right\} _{\alpha ^{\prime }\alpha }\equiv W_{\alpha
^{\prime }\alpha }\left( \mathbf{k},\sigma ,z\right) =V^{\ast }\left( \alpha
^{\prime },\mathbf{k},\sigma \right) \ V\left( \alpha ,\mathbf{k},\sigma
\right) \ \mathcal{G}_{c,\sigma }^{0}\left( \mathbf{k},z\right) ,
\label{Eq3.13}
\end{equation}%
where we introduced the conduction electron free GF (cf. footnote \ref%
{myfoot6} inAppendix \ref{ApE})
\begin{equation}
\mathcal{G}_{c,\sigma }^{0}\left( \mathbf{k},z_{n}\right) =\frac{-1}{%
z_{n}-\varepsilon \left( \mathbf{k},\sigma \right) },  \label{Eq3.14}
\end{equation}%
and we have used $V(\alpha ^{\prime },\mathbf{k},\sigma ,+)=V\left( \alpha
^{\prime },\mathbf{k},\sigma \right) $ and $V(\alpha ,\mathbf{k},\sigma
,-)=V^{\ast }\left( \alpha ,\mathbf{k},\sigma \right) $ (cf. Eqs(\ref{Ger3})
and (\ref{Fourier13})). To abbreviate we define
\begin{equation}
A_{\alpha \alpha ^{\prime }}\left( \mathbf{k},\sigma ,z\right) \equiv \left(
\mathbf{W\cdot M}\right) _{\alpha \alpha ^{\prime }}=\sum_{\alpha
_{1}}W_{\alpha \alpha _{1}}\mathbf{\left( \mathbf{k},\sigma ,z\right) \ }%
M_{\alpha _{1}\alpha ^{\prime }}\left( \mathbf{k},\sigma ,z\right) ,
\label{Eq3.15}
\end{equation}%
and introduce it in the series for the exact GF:
\begin{align}
\mathcal{G}_{\alpha \alpha ^{\prime }}^{ff}& =\left\{ \mathbf{M+M\cdot
W\cdot M+M\cdot }\left( \mathbf{W\cdot M}\right) ^{2}+\ldots \right\}
_{\alpha \alpha ^{\prime }}=  \notag \\
& =\left\{ \mathbf{M+M\cdot A+M\cdot }\left( \mathbf{A}\right) ^{2}+\ldots
\right\} _{\alpha \alpha ^{\prime }}=\left\{ \mathbf{M\cdot }\left( \mathbf{%
I+A+A}^{2}+\ldots \right) \right\} _{\alpha \alpha ^{\prime }}.
\label{Eq3.16}
\end{align}%
We now use
\begin{equation}
\mathbf{S}\mathbf{=}\left( \mathbf{I+A+A}^{2}+\ldots \right) =\mathbf{%
I+A\cdot S=}\left( \mathbf{I-A}\right) ^{-1},  \label{Eq3.17}
\end{equation}%
so that
\begin{equation}
\mathcal{G}_{\alpha \alpha ^{\prime }}^{ff}\left( \mathbf{k},\sigma
,z\right) =\left\{ \mathbf{M\cdot }\left( \mathbf{I-A}\right) ^{-1}\right\}
_{\alpha \alpha ^{\prime }},  \label{Eq3.18}
\end{equation}%
which are the components of the matrix $\mathbf{G}^{ff}\left( \mathbf{k}%
,\sigma ,z\right) $:
\begin{equation}
\mathbf{G}^{ff}=\mathbf{M\cdot }\left( \mathbf{I-A}\right) ^{-1}=\mathbf{%
M\cdot }\left( \mathbf{I-W\cdot M}\right) ^{-1}.  \label{Eq3.19}
\end{equation}

We can also express $\mathbf{M}\left( \mathbf{k},\sigma ,z\right) $ as a
function of $\mathbf{G}^{ff}\left( \mathbf{k},\sigma ,z\right) $: we use Eq.
(\ref{Eq3.19}) to write
\begin{equation}
\mathbf{M=G}^{ff}\cdot \left( \mathbf{I-A}\right) =\mathbf{G}^{ff}-\mathbf{G}%
^{ff}\cdot \mathbf{W\cdot M,}  \label{Eq3.20}
\end{equation}%
so that
\begin{equation}
\left( \mathbf{I+G}^{ff}\cdot \mathbf{W}\right) \cdot \mathbf{M=G}^{ff},
\label{Eq3.21}
\end{equation}%
and therefore
\begin{equation}
\mathbf{M=}\left( \mathbf{I+G}^{ff}\cdot \mathbf{W}\right) ^{-1}\cdot
\mathbf{G}^{ff}.  \label{Eq3.22}
\end{equation}%
The calculation of the effective cumulant $\mathbf{M}$ is as difficult as
that of the exact GF $\mathbf{G}^{ff}$, and to obtain an approximate GF we
shall substitute the exact $\mathbf{M}$ by one that corresponds to an
exactly soluble model that is similar to the system of interest. To this
purpose we shall employ the same Anderson model, but for a conduction band
that has zero width as well as local hybridization (i.e. $\varepsilon \left(
\mathbf{k},\sigma \right) =\varepsilon _{o}\left( \sigma \right) $ and $%
V\left( \alpha ^{\prime },\mathbf{k},\sigma \right) =V\left( \alpha ^{\prime
},\sigma \right) $) . This model has the same cumulant graphs that appear in
the system of interest, but its GF $\mathbf{G}^{ff,at}(z)$ can be calculated
exactly. To find the approximate effective cumulant $\mathbf{M}^{ap}\left(
z\right) $, we then substitute $\mathbf{G}^{ff}\left( \mathbf{k},\sigma
,z\right) $ in Eq. (\ref{Eq3.22}) by $\mathbf{G}^{ff,at}(z)$ and obtain
\begin{equation}
\mathbf{M}^{ap}\left( z\right) \mathbf{=}\left( \mathbf{I+G}^{ff,at}\left(
z\right) \cdot \mathbf{W}^{at}\right) ^{-1}\cdot \mathbf{G}^{ff,at}\left(
z\right) .  \label{Eq3.23}
\end{equation}%
This $\mathbf{M}^{ap}$ is independent of $\mathbf{k}$ because both $\mathbf{G%
}^{ff,at}\left( z\right) $ and
\begin{equation*}
\left\{ \mathbf{W}^{at}(z)\right\} _{\alpha ,\alpha ^{\prime }}=-\frac{%
V^{\ast }\left( \alpha ^{\prime },\sigma \right) \ V\left( \alpha ,\sigma
\right) }{z-\varepsilon _{o}\left( \sigma \right) }\
\end{equation*}%
are also independent of $\mathbf{k}$. The approximate GF is\ then obtained
by substituting $\mathbf{M}$ in Eq. (\ref{Eq3.19}) by $\mathbf{M}^{ap}\left(
z\right) $ from Eq. (\ref{Eq3.23}).

\begin{figure}[ptb]
\begin{center}
\includegraphics[width=6.9647in]{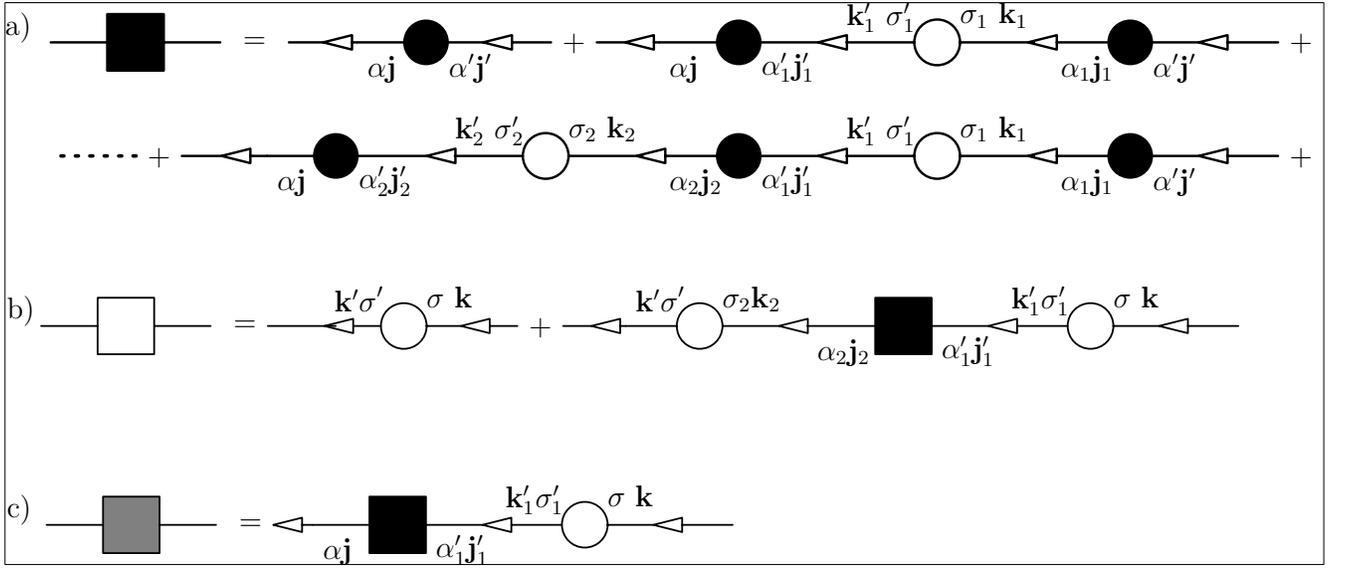}
\end{center}
\caption{Cumulant expansion diagrams of the PAM, that give the one-electron
GF in the Chain Approximation (CHA) \protect\cite{FFM}. The filled circles
(vertices) corresponds to the $f$-electron cumulants and the empty ones to
those of the $c$-electrons. The lines (edges) joining two vertices represent
the perturbation (hybridization) (cf. Rules 3.1 and 3.2 in reference
\protect\cite{FFM}). a) Diagrams for the $f$-electron GF $G_{j,\protect\alpha%
;j^{\prime},\protect\alpha^{\prime}}^{ff}(i\protect\omega_{n}=z_{n})$ in the
CHA, represented by the filled square to the left. b) Diagrams for the $c$
electron GF $G_{\mathbf{k}^{\mathbf{\prime}}\mathbf{,}\protect\sigma%
^{\prime};\mathbf{k,}\protect\sigma}^{cc}(z_{n})$ in the CHA, represented by
the square symbol to the left. c) Diagrams for the $f$-$c$ electron GF $G_{%
\mathbf{j};\mathbf{k,}\protect\sigma}^{fc}(z_{n})$ in the CHA, represented
by the square symbol to the left.}
\label{GffCHA}
\end{figure}

\subsubsection{The exact formal Green's functions for the impurity Anderson
model\label{S02_5}}

In the impurity case, only the Fourier transform of the imaginary time is
necessary (cf. Eq. (\ref{Eq3.1})), and to abbreviate $\left\langle \left(
Y\left( f;\mathbf{j,}\alpha,u=-,\omega\right) \ \ Y\left( f;\mathbf{j}%
^{\prime}\mathbf{,}\alpha^{\prime},u^{\prime},\omega^{\prime}\right) \right)
_{+}\right\rangle =\mathcal{G}^{ff}(\mathbf{j},\alpha,u=-,\omega ;\mathbf{j}%
^{\prime},\alpha^{\prime},u^{\prime},\omega^{\prime})$ we define $\mathcal{G}%
_{\alpha\alpha^{\prime}}^{ff}(\mathbf{j},i\omega)$. Instead of Eq. (\ref%
{Eq3.1.1}) we then have:
\begin{equation}
\mathcal{G}^{ff}(\mathbf{j},\alpha,u=-,\omega;\mathbf{j}^{\prime},\alpha^{%
\prime},u^{\prime},\omega^{\prime})=\mathcal{G}_{\alpha\alpha
^{\prime}}^{ff}(\mathbf{j},i\omega)\ \Delta\left( u+u^{\prime}\right) \
\Delta\left( \omega+\omega^{\prime}\right) \ \delta\left( \mathbf{j,j}%
^{\prime}\right) ,  \label{E5.13a}
\end{equation}
and most of the derivation employed in the previous section can be extended
to this case.

The contribution to\ $\mathcal{G}_{\alpha \alpha ^{\prime }}^{ff}(\mathbf{j}%
_{i},z)$ of the term in the series that has $n+1$ effective cumulants is
given again by (cf. Eq. (\ref{ApD45}) in Appendix \ref{ApD})
\begin{equation*}
\left\{ \left( \mathbf{M\cdot W}\right) ^{n}\cdot \mathbf{M}\right\}
_{\alpha \alpha ^{\prime }}=\left\{ \mathbf{M\cdot }\left( \mathbf{W\cdot M}%
\right) ^{n}\right\} _{\alpha \alpha ^{\prime }},
\end{equation*}%
where $\mathbf{M}$ is defined in Eq. (\ref{ApD43}):
\begin{equation}
\left\{ \mathbf{M}\right\} _{_{\alpha ,\alpha ^{\prime }}}=M_{\alpha \alpha
^{\prime }}^{eff}(\mathbf{j}_{i},z_{n},u),  \label{E3.25}
\end{equation}%
and $\mathbf{W}$ in Eqs. (\ref{ApD44},\ref{ApD21a})
\begin{equation}
\left\{ \mathbf{W}\right\} _{_{\alpha ^{\prime },\alpha }}=W_{\alpha
^{\prime },\alpha }\left( \sigma ,z_{n}\right) =\frac{1}{N_{s}}\sum_{\mathbf{%
k}}V^{\ast }(\alpha ^{\prime },\mathbf{k},\sigma )V(\alpha ,\mathbf{k}%
,\sigma )\ \mathcal{G}_{c,\sigma }^{0}\left( \mathbf{k},z_{n}\right) .
\label{E3.26}
\end{equation}%
As before we define a matrix
\begin{equation}
\mathbf{A=W\cdot M}  \label{E3.26a}
\end{equation}%
with components

\begin{equation}
A_{\alpha\alpha^{\prime}}\left( \mathbf{k},\sigma,z_{n}\right) \equiv\left(
\mathbf{W\cdot M}\right) _{\alpha\alpha^{\prime}},  \label{E3.27}
\end{equation}
and then we obtain a matrix with components $\mathcal{G}_{\alpha\alpha
^{\prime}}^{ff}(\mathbf{j}_{i},z_{n})$:
\begin{equation}
\mathbf{G}^{ff}=\mathbf{M\cdot}\left( \mathbf{I-A}\right) ^{-1}.
\label{E3.28}
\end{equation}
As before
\begin{equation}
\mathbf{M=}\left( \mathbf{I+G}^{ff}\cdot\mathbf{W}\right) ^{-1}\cdot\mathbf{G%
}^{ff}.  \label{E3.29}
\end{equation}
and if we substitute $\mathbf{G}^{ff}\left( \mathbf{j}_{i},z_{n}\right) $ by
$\mathbf{G}^{ff,at}(\mathbf{j}_{i},z_{n})=\mathbf{G}^{ff,at}\left(
z_{n}\right) $ we obtain an approximate effective cumulant for the impurity
problem.
\begin{equation}
\mathbf{M}^{ap}\left( z_{n}\right) \mathbf{=}\left( \mathbf{I+G}%
^{ff,at}\left( z_{n}\right) \cdot\mathbf{W}\right) ^{-1}\cdot \mathbf{G}%
^{ff,at}\left( z_{n}\right)  \label{E3.30}
\end{equation}
Introducing $\mathbf{M}^{ap}\left( z_{n}\right) $ into Eqs. (\ref{E3.26a},%
\ref{E3.28}) we obtain the approximate GF we shall use in our calculations.

\section{Calculation of the atomic Green's functions}

\label{S3}

When the conduction band has zero width and the hybridization is local (i.e.
independent of \textbf{k}), the eigenvalue problem of the Hamiltonian
introduced in Eqs. (\ref{Modelo5},\ref{Hibr}) has an exact solution \cite%
{FoglioF}, and the GFs can be calculated analytically. Taking $E_{\mathbf{k}%
,\sigma}=E_{0}^{a}$ and $V_{j\alpha,\mathbf{k},\sigma }=V_{j\alpha,\sigma}$
the problem becomes fully local, and one can use the Wannier representation
for the creation and annihilation operators $C_{j\mathbf{,}\sigma}^{\dagger}$
and $C_{j,\sigma}$ of the c-electrons. We then write $H_{r}=\sum_{j}H_{j}$,
where $H_{j}$ is the local Hamiltonian
\begin{equation}
H_{j}=\sum_{\sigma}\ \left\{ E_{0}^{a}\ C_{j\mathbf{,}\sigma}^{\dagger
}C_{j,\sigma}+\ \sum_{a}E_{ja}X_{j,aa}+\sum_{\alpha,\sigma}\left(
V_{j\alpha,\sigma}X_{j,\alpha}^{\dagger}C_{j\sigma}+V_{j\alpha,\sigma}^{\ast
}C_{j\sigma}^{\dagger}X_{j,\alpha}\right) \right\} ,  \label{Eq3.3}
\end{equation}
and the subindex $j$ can be dropped because we assume a uniform system.

We shall denote with $\left\vert n,r\right\rangle $ the eigenstates of the
Hamiltonian $H_{j}$ with eigenvalues $E_{n,r},$ where $n$ is the total
number of electrons in that state, and $r$ characterizes the different
states. These eigenstates satisfy
\begin{equation}
\mathcal{H}\ \left\vert n,r\right\rangle =\varepsilon_{n,r}\ \left\vert
n,r\right\rangle ,  \label{Eq3.4}
\end{equation}
\noindent where $\mathcal{H}$ corresponds to that in Eq. (\ref{Modelo7}) but
for a single site, and $\varepsilon_{n,r}=E_{n,r}-n\mu$ (cf. Eq.~(\ref%
{Eq1.13})). The states $\left\vert n,r\right\rangle $ are usually obtained
by diagonalization, and employing Eqs. (\ref{Fourier2},\ref{Eq3.4}) we find
\begin{align}
\exp\left[ -\beta\mathcal{H}\right] \hat{Y}(\gamma,\tau)\hat{Y}%
(\gamma^{\prime},0)\left\vert n,r\right\rangle &
=\sum_{n^{\prime}n^{\prime\prime}r^{\prime}r^{\prime\prime}}\exp\left[
\mathcal{H}\left( \tau-\beta\right) \right] \ \left\vert
n^{\prime\prime},r^{\prime\prime }\right\rangle \left\langle
n^{\prime\prime},r^{\prime\prime}\right\vert \ Y_{\gamma}\exp\left[ -%
\mathcal{H}\tau\right] \left\vert n^{\prime },r^{\prime}\right\rangle
\left\langle n^{\prime},r^{\prime}\right\vert \ Y_{\gamma^{\prime}}\
\left\vert n,r\right\rangle  \notag \\
& =\sum_{n^{\prime}n^{\prime\prime}r^{\prime}r^{\prime\prime}}\exp\left[
-\beta\varepsilon_{n^{\prime\prime},r^{\prime\prime}}+\left( \varepsilon
_{n^{\prime\prime},r^{\prime\prime}}-\varepsilon_{n^{\prime},r^{\prime}}%
\right) \tau\right] \left\langle n^{\prime\prime},r^{\prime\prime
}\right\vert \ Y_{\gamma}\left\vert n^{\prime},r^{\prime}\right\rangle
\left\langle n^{\prime},r^{\prime}\right\vert \ Y_{\gamma^{\prime}}\
\left\vert n,r\right\rangle \ \left\vert n^{\prime\prime},r^{\prime\prime
}\right\rangle  \label{Eq3.5}
\end{align}

Employing Eq.(\ref{EqC12}) of Appendix (\ref{ApC}) we calculate the Fourier
transform of $\left\langle \left(
X_{j,\alpha}(\tau)X_{j,\alpha^{\prime}}^{\dagger}(\tau^{\prime})\right)
_{+}\right\rangle _{\mathcal{H}}$:
\begin{equation}
\left\langle \left( X_{j,\alpha}(\omega_{s})\ X_{j,\alpha^{\prime}}^{\dagger
}(\omega_{s}^{\prime})\right) _{+}\right\rangle _{\mathcal{H}}=\ \Delta
\left( \omega_{s}+\omega_{s}^{\prime}\right) \int_{0}^{\beta}d\tau \
\left\langle \left( X_{j,\alpha}(\tau)X_{j,\alpha^{\prime}}^{\dagger
}(0)\right) _{+}\right\rangle _{\mathcal{H}}\exp\left[ i\tau\ \omega _{s}%
\right]  \label{Eq3.6}
\end{equation}
and we shall then abbreviate
\begin{equation}
\left\langle \left( X_{j,\alpha}(\omega_{s})\ X_{j,\alpha^{\prime}}^{\dagger
}(\omega_{s}^{\prime})\right) _{+}\right\rangle _{\mathcal{H}}=\Delta
(\omega_{s}+\omega_{s}^{\prime})\ \mathcal{G}_{\alpha\alpha^{%
\prime}}^{ff,at}(i\omega_{s}).  \label{Eq3.7}
\end{equation}

Employing the grand canonical potential $\Omega=-kT\ln\sum\exp(-\beta
\epsilon_{n,r})$ \cite{Martinez89} we find for $0\leq\tau\leq\beta$
\begin{equation}
\left\langle \left( X_{j,\alpha}(\tau)X_{j,\alpha^{\prime}}^{\dagger
}(0)\right) _{+}\right\rangle _{\mathcal{H}}=\exp\left( \beta\Omega\right) \
\left\{ \sum_{n,r}\left\langle n,r\right\vert \exp\left[ -\beta \mathcal{H}%
\right] \ X_{j,\alpha}(\tau)\ X_{j,\alpha^{\prime}}^{\dagger }(0)\
\left\vert n,r\right\rangle \right\} ,  \label{Eq3.8a}
\end{equation}
and from Eq. (\ref{Eq3.5})

\begin{equation}
\left\langle \left( X_{j,\alpha}(\tau)X_{j,\alpha^{\prime}}^{\dagger
}(0)\right) _{+}\right\rangle _{\mathcal{H}}=\exp\left( \beta\Omega\right) \
\left\{ \sum_{n,r,n^{\prime},r^{\prime}}\exp\left[ -\beta\varepsilon
_{n,r}+\left( \varepsilon_{n,r}-\varepsilon_{n^{\prime},r^{\prime}}\right)
\tau\right] \ \left\langle n,\sigma r\right\vert \ X_{j,\alpha}\left\vert
n^{\prime},r^{\prime}\right\rangle \left\langle n^{\prime},r^{\prime
}\right\vert \ X_{j,\alpha^{\prime}}^{\dagger}\ \left\vert n,r\right\rangle
\right\} .  \label{Eq3.9}
\end{equation}
Integrating Eq. (\ref{Eq3.9}) and using $\exp\left[ i\ \omega_{s}\beta\right]
=-1$ and the properties of the $X_{j,\alpha}$ we obtain from Eqs. (\ref%
{Eq3.6}-\ref{Eq3.7})
\begin{equation}
\mathcal{G}_{\alpha\alpha^{\prime}}^{ff,at}(i\omega_{s})=-e^{\beta\Omega}%
\sum_{n,r,r^{\prime}}\frac{\exp(-\beta\varepsilon_{n-1,r})+\exp(-\beta
\varepsilon_{n,r^{\prime}})}{i\omega_{s}+\varepsilon_{n-1,r}-\varepsilon
_{n,r^{\prime}}}\left\langle n-1,r\right\vert \ X_{j,\alpha}\left\vert
n,r^{\prime}\right\rangle \left\langle n,r^{\prime}\right\vert \ X_{j,\alpha
^{\prime}}^{\dagger}\ \left\vert n-1,r\right\rangle .  \label{Eq3.11}
\end{equation}

Equivalent expressions for $\mathcal{G}_{\sigma\sigma^{\prime}}^{cc,at}(i%
\omega_{s}),\mathcal{G}_{\alpha\sigma^{\prime}}^{fc,at}(i\omega _{s}),%
\mathcal{G}_{\sigma\alpha^{\prime}}^{cf,at}(i\omega_{s})$ are easily
obtained. One can also consider these functions as matrix elements of four
matrices $\mathbf{G}^{ff,at}(i\omega_{s})$, $\mathbf{G}^{cc,at}(i\omega_{s})$%
, $\mathbf{G}^{fc,at}(i\omega_{s})$ and $\mathbf{G}^{cf,at}(i\omega_{s})$,
and one can also define a larger matrix $\mathbf{G}^{at}(i\omega_{s})$ that
includes these four matrices, but it is not yet clear whether a formulation
that uses this larger matrix would have any advantage. We then define

\begin{equation}
\mathbf{G}^{at}(i\omega_{s})=
\begin{bmatrix}
\mathbf{G}^{ff,at}(i\omega_{s}) & \mathbf{G}^{fc,at}(i\omega_{s}) \\
\mathbf{G}^{cf,at}(i\omega_{s}) & \mathbf{G}^{cc,at}(i\omega_{s})%
\end{bmatrix}
,  \label{Eq3.11a}
\end{equation}
which would be a $6\times6$ matrix for finite $U$ and a $4\times4$ matrix
for infinite $U$.

\section{Detailed calculation of the approximate GF}

\label{S5}

\subsection{Introduction}

\label{S05_1}

The hybridization constant $V_{j,\mathbf{k},\sigma}$ in Eq. (\ref{Eq2}) is
given by Eq. (\ref{Eq3}),
\begin{equation*}
V_{j,\mathbf{k},\sigma}=\frac{1}{\sqrt{N_{s}}}V_{\sigma}(\mathbf{k})\exp{(i%
\mathbf{k}.\mathbf{R}_{j}),}
\end{equation*}
and when the Hubbard operators are introduced into Eq. (\ref{Eq2}) , the
hybridization Hamiltonian $H_{h}$ is transformed into Eq. (\ref{Hibr}):
\begin{equation*}
H_{h}=\sum_{jba,\vec{k}\sigma}\left( V_{jba,\mathbf{k}\sigma}X_{j,ba}^{%
\dagger}C_{\mathbf{k}\sigma}+V_{jba,\mathbf{k}\sigma}^{\ast}C_{\mathbf{k}%
\sigma}^{\dagger}X_{j,ba}\right) ,
\end{equation*}
with hybridization constant $V_{j\alpha,\mathbf{k}\sigma}$. The $\alpha
=(b,a)$\ describes the transition $\mid a>\rightarrow\mid b>$, where the
local state $\mid a>$ has one electron more than the state $\mid b>$. To
simplify the calculation one defines the parameters $v(j,\alpha,\mathbf{k}%
,\sigma ,u)$\ in Eq. (\ref{Ger3})
\begin{equation*}
\begin{array}{ccc}
v(j,\alpha,\mathbf{k},\sigma,+) & = & V_{j,ba,\mathbf{k},\sigma}, \\
v(j,\alpha,\mathbf{k},\sigma,-) & = & V_{j,ba,\mathbf{k},\sigma}^{\ast},%
\end{array}%
\end{equation*}
where $u=\pm$, and in the PAM we use $V(\alpha,\mathbf{k},\sigma,u)$,
defined in Eq. (\ref{Fourier13})
\begin{equation*}
v(j,\alpha,\mathbf{k},\sigma,u)=V(\alpha,\mathbf{k},\sigma,u)N_{s}^{-\frac {1%
}{2}}\exp(iu\mathbf{k}.\mathbf{R}_{j}).
\end{equation*}

After applying the rules for calculating the GF, it is convenient to return
to the explicit use of complex conjugates, and we introduce $V(\alpha ,%
\mathbf{k},\sigma )$ in Eq. (\ref{ApD12a})
\begin{equation}
V(\alpha ,\mathbf{k},\sigma )\equiv V(\alpha ,\mathbf{k},\sigma ,u=-),
\label{D12a}
\end{equation}
so that Eq. (\ref{ApD12b})
\begin{equation}
V^{\ast }(\alpha ,\mathbf{k},\sigma )\equiv V(\alpha ,\mathbf{k},\sigma ,u=+)
\label{D12b}
\end{equation}
follows from Eq. (\ref{Fourier13}).

There are four local states $\mid0>$, $\mid+>$, $\mid->$ and $\mid
d>=\mid+,->$ per site, and there are only four $X$ operators that destroy
one local electron at a given site. We use the index $I_{x}=1,2,3,4$ to
characterize these $X$ operators:

\begin{center}
\begin{equation}
\begin{tabular}{|l|l|l|l|l|}
\hline
$I_{x}$ & $1$ & $2$ & $3$ & $4$ \\ \hline
$\alpha=(b,a)$ & $(0,+)$ & $(0,-)$ & $(-,d)$ & $(+,d)$ \\ \hline
\end{tabular}
\label{E5.0}
\end{equation}
\end{center}

so that $I_{x}=1,3$ destroy one electron with spin up and $I_{x}=2,4$
destroy one electron with spin down. We use $\sigma=+$ and $\sigma=-$
instead of $\sigma=\uparrow$ and $\sigma=\downarrow$ to emphasize that the
spin belongs to a local electron.

The matrix $\mathbf{W}$, employed in the PAM calculation, is defined in Eq. (%
\ref{ApD24})
\begin{equation*}
\left\{ \mathbf{W}\right\} _{_{\alpha ^{\prime },\alpha }}=W_{\alpha
^{\prime },\alpha }\left( \mathbf{k},\sigma ,z_{n}\right) ,
\end{equation*}%
and its matrix elements are defined in Eq. (\ref{ApD21})
\begin{equation*}
W_{\alpha ^{\prime },\alpha }\left( \mathbf{k},\sigma ,z_{n}\right) =V^{\ast
}(\alpha ^{\prime },\mathbf{k},\sigma )V(\alpha ,\mathbf{k},\sigma )\
\mathcal{G}_{c,\sigma }^{0}\left( \mathbf{k},z_{n}\right) ,
\end{equation*}%
where \ $z_{n}=i\omega _{n}$ are the Matsubara frequencies. A related matrix
appears in the impurity case in Eq. (\ref{ApD44})
\begin{equation*}
\left\{ \mathbf{W}\right\} _{_{\alpha ^{\prime },\alpha }}=W_{\alpha
^{\prime },\alpha }\left( \sigma ,z_{n}\right) .
\end{equation*}%
with matrix elements defined in Eq. (\ref{ApD21a})
\begin{equation*}
W_{\alpha ^{\prime },\alpha }\left( \sigma ,z_{n}\right) =\frac{1}{N_{s}}%
\sum_{\mathbf{k}}V^{\ast }(\alpha ^{\prime },\mathbf{k},\sigma )V(\alpha ,%
\mathbf{k},\sigma )\ \mathcal{G}_{c,\sigma }^{0}\left( \mathbf{k}%
,z_{n}\right) .
\end{equation*}

The hybridization is spin independent in the Anderson model, so we have
\begin{equation*}
V(0\sigma,\mathbf{k},\bar{\sigma})=V(\bar{\sigma}d,\mathbf{k},\bar{\sigma }%
)=V(0\bar{\sigma},\mathbf{k},\sigma)=V(\sigma d,\mathbf{k},\sigma)=0.
\end{equation*}
We shall assume a purely local mixing, so that $V_{\sigma}(\mathbf{k})$ in
Eq. (\ref{Hibr}) \ is $\mathbf{k}$ independent, and when we introduce the
Hubbard operators we obtain
\begin{align}
V(0\sigma,\mathbf{k},\sigma) & =V,  \label{E5.1} \\
V(\bar{\sigma}d,\mathbf{k},\sigma) & =\sigma V,  \label{E5.2}
\end{align}
where we have also assumed that $V_{\sigma}(\mathbf{k})$ is independent of $%
\sigma=\pm1$.

As discussed in the introduction of Section \ref{S4},when the Hamiltonian is
spin independent or commutes with the $z$ component of the spin, the $%
4\times4$ matrices $\mathbf{G}^{ff}$, $\mathbf{M}$,~ $\mathbf{W}$ and $%
\mathbf{A=W.M}$ can be diagonalized into two $2\times2$ matrices, e.g.:

\begin{equation}
\mathbf{G}^{ff}=
\begin{pmatrix}
\mathbf{G}_{\uparrow}^{ff} & 0 \\
0 & \mathbf{G}_{\downarrow}^{ff}%
\end{pmatrix}
.  \label{E5.3}
\end{equation}
In this matrix the indices $I_{x}$ defined in Eq. (\ref{E5.0}) have been
rearranged, so that $I_{x}=1,3$ appear in $\mathbf{G}_{\uparrow}^{ff}$ and $%
I_{x}=2,4$ appear in $\mathbf{G}_{\downarrow}^{ff}$.

Employing Eqs. (\ref{E5.1},\ref{E5.2}) we find for the PAM (cf. Eq. (\ref%
{Eq3.14}))
\begin{align}
\mathbf{W}_{\uparrow}\left( \mathbf{k},z\right) & =\ \left\vert V\right\vert
^{2}\ \mathcal{G}_{c,\uparrow}^{0}\left( \mathbf{k},z\right) \
\begin{pmatrix}
1 & 1 \\
1 & 1%
\end{pmatrix}
,  \label{E5.4a} \\
\mathbf{W}_{\downarrow}\left( \mathbf{k},z\right) & =\ \left\vert
V\right\vert ^{2}\ \mathcal{G}_{c,\downarrow}^{0}\left( \mathbf{k},z\right)
\
\begin{pmatrix}
1 & -1 \\
-1 & 1%
\end{pmatrix}
.  \label{E5.4b}
\end{align}
For an impurity located at the origin we find \
\begin{align}
\mathbf{W}_{\uparrow}\left( z\right) & =\left\vert V\right\vert
^{2}\varphi_{\uparrow}(z)
\begin{pmatrix}
1 & 1 \\
1 & 1%
\end{pmatrix}
,  \label{E5.5a} \\
\mathbf{W}_{\downarrow}\left( z\right) & =\left\vert V\right\vert
^{2}\varphi_{\downarrow}(z)\
\begin{pmatrix}
1 & -1 \\
-1 & 1%
\end{pmatrix}
,  \label{E5.5b}
\end{align}
where
\begin{equation}
\varphi_{\sigma}(z)=\frac{1}{N_{s}}\sum_{\mathbf{k}}\ \mathcal{G}_{c,\sigma
}^{0}\left( \mathbf{k},z\right) .  \label{E5.6a}
\end{equation}

For a rectangular band with half-width $D$ in the interval $[A,B],$ with $%
B=A+2D$ we find (the minus sign of\emph{\ }$\mathcal{G}_{c,\sigma}^{0}\left(
\mathbf{k},z\right) $ is included in the logarithm)
\begin{equation}
\varphi_{\sigma}(z)=\frac{1}{2D}\ \ln\frac{z-B+\mu}{z+A+\mu},  \label{E5.6}
\end{equation}
where the $\mu$ appears in $\varphi_{\sigma}(z)$ because of the $\varepsilon
\left( \mathbf{k},\sigma\right) =E_{\mathbf{k,\sigma}}-\mu$ in $\mathcal{G}%
_{c,\sigma}^{0}\left( \mathbf{k},z\right) $ .

Both for the PAM (Eq. (\ref{Eq3.19})) and for the SIAM (Eq. (\ref{E3.28}))
we have the same relation for the submatrices in Eq. (\ref{E5.3}):
\begin{equation}
\mathbf{G}_{\sigma }^{ff}=\mathbf{M}_{\sigma }\mathbf{\cdot }\left( \mathbf{%
I-A}_{\sigma }\right) ^{-1},  \label{E5.7}
\end{equation}
and as before (cf. Section \ref{S4})
\begin{equation}
\mathbf{M}_{\sigma }\mathbf{=}\left( \mathbf{I+G}_{\sigma }^{ff}\cdot
\mathbf{W}_{\sigma }\right) ^{-1}\cdot \mathbf{G}_{\sigma }^{ff}.
\label{E5.8}
\end{equation}

\subsection{The approximate $\mathbf{G}_{\protect\sigma }^{ff,ap}$ GF for
the Periodic Anderson Model\label{S05_1a}}

The calculation of the exact effective cumulants $\mathbf{M}_{\sigma }$ is
as difficult as that of the exact $\mathbf{G}_{\sigma }^{ff}$, and the
atomic approach consists in using instead the $\mathbf{M}_{\sigma }$ of a
similar model that is exactly soluble. Taking a conduction band of zero
width at $\varepsilon _{o}=E_{0}-\mu $ and a local hybridization, the PAM
becomes a collection of independent atomic Anderson systems that can be
solved exactly, and we call this solution the atomic Anderson solution
(AAS). We then employ the AAS to calculate the corresponding exact Green's
function $\mathbf{G}_{\sigma }^{ff,at}(z)$, and we define an approximate
effective cumulant $\mathbf{M}_{\sigma }^{ap}$ by introducing this GF into
Eq.(\ref{E5.8}) using $\mathcal{G}_{c,\sigma }^{0,at}\left( i\omega \right)
=-1/\left( i\omega -\varepsilon _{0}\right) $ in the corresponding $\mathbf{W%
}^{at}$. This procedure overestimates the contribution of the $c$ electrons
\ \cite{Alascio79}\cite{Alascio80}, because we concentrate them at a single
energy level $E_{o}$, and to moderate this effect we replace $V^{2}$ by $%
\Delta ^{2}$ in the calculation of the $\mathbf{M}_{\sigma }^{ap}$, where $%
\Delta =\pi \left\vert V\right\vert ^{2}/2D$ is the Anderson parameter (but
we keep the full $\left\vert V\right\vert ^{2}$ when we substitute this $%
\mathbf{M}_{\sigma }^{ap}$ into Eq. (\ref{E5.7}) to calculate the
approximate $\mathbf{G}_{\sigma }^{ff,ap}$).


\subsubsection{Imaginary frequency and reciprocal space}

To calculate the approximate $\mathbf{G}_{\sigma }^{ff,ap}(\mathbf{k}%
,i\omega )$ for imaginary frequency and reciprocal space\ we then introduce
the $\mathbf{M}_{\sigma }^{ap}$ into Eq. (\ref{E5.7}), with $\mathbf{A}%
_{\sigma }^{ap}\mathbf{=W}_{\sigma }\mathbf{.M}_{\sigma }^{ap}$ and $\mathbf{%
W}_{\sigma }$ defined in Eqs. (\ref{E5.4a},\ref{E5.4b}) with the full $%
\left\vert V\right\vert ^{2}$. We can now write

\begin{equation}
\mathbf{M}_{\uparrow }^{ap}=
\begin{pmatrix}
m_{11} & m_{13} \\
m_{31} & _{m33}%
\end{pmatrix}
\hspace{20pt}\mathrm{;\hspace{20pt}}\mathbf{M}_{\downarrow }^{ap}=
\begin{pmatrix}
m_{22} & m_{24} \\
m_{42} & _{m44}%
\end{pmatrix}
,  \label{E5.9}
\end{equation}
and
\begin{equation}
\mathbf{A}_{\uparrow }^{ap}=
\begin{pmatrix}
a_{11} & a_{13} \\
a_{31} & a_{33}%
\end{pmatrix}
\hspace{20pt}\mathrm{;\hspace{20pt}}\mathbf{A}_{\downarrow }^{ap}=
\begin{pmatrix}
a_{22} & a_{24} \\
a_{42} & a_{33}%
\end{pmatrix}
,  \label{E5.10}
\end{equation}
and from Eqs.\ (\ref{E5.4a},\ref{E5.4b}) we find
\begin{equation}
\begin{tabular}{llll}
$a_{11}=\mathcal{G}_{c,\uparrow }^{0}\left( \mathbf{k},z\right) \left(
m_{11}+m_{31}\right) $ & \hspace{10pt}; & \hspace{10pt} & $a_{22}=\mathcal{G}%
_{c,\downarrow }^{0}\left( \mathbf{k},z\right) \left( m_{22}-m_{42}\right) $
\\
$a_{33}=\mathcal{G}_{c,\uparrow }^{0}\left( \mathbf{k},z\right) \left(
m_{33}+m_{13}\right) $ & \hspace{10pt}; & \hspace{10pt} & $a_{44}=\mathcal{G}%
_{c,\downarrow }^{0}\left( \mathbf{k},z\right) \left( m_{44}-m_{24}\right) $
\\
$a_{31}=a_{11}$ & \hspace{10pt}; & \hspace{10pt} & $a_{42}=-a_{22}$ \\
$a_{13}=a_{33}$ & \hspace{10pt}; & \hspace{10pt} & $a_{24}=-a_{44}$%
\end{tabular}
\ \ \ \ \ \ \ \ \ \ \   \label{E5.11a}
\end{equation}

To derive our approximate $\mathbf{G}_{\sigma }^{ff,ap}$ we substitute both $%
\mathbf{M}_{\sigma }^{ap}$ and $\mathbf{A}_{\sigma }^{ap}\mathbf{=W}_{\sigma
}\mathbf{.M}_{\sigma }^{ap}$\ into Eq. (\ref{E5.7}), and we obtain
\begin{equation}
\mathbf{G}_{\uparrow }^{ff,ap}(\mathbf{k},i\omega )=\frac{%
\begin{pmatrix}
m_{11} & m_{13} \\
m_{31} & _{m33}%
\end{pmatrix}
-\left\vert V\right\vert ^{2}\mathcal{G}_{c,\uparrow }^{0}\left( \mathbf{k}%
,z\right) \left( m_{11}m_{33}-m_{13}m_{31}\right)
\begin{pmatrix}
1 & -1 \\
-1 & 1%
\end{pmatrix}
}{1-\left\vert V\right\vert ^{2}\mathcal{G}_{c,\uparrow }^{0}\left( \mathbf{k%
},z\right) \left( m_{11}+m_{33}+m_{13}+m_{31}\right) },  \label{E5.12a}
\end{equation}
and
\begin{equation}
\mathbf{G}_{\downarrow }^{ff,ap}(\mathbf{k},i\omega )=\frac{%
\begin{pmatrix}
m_{22} & m_{24} \\
m_{42} & _{m44}%
\end{pmatrix}
-\left\vert V\right\vert ^{2}\mathcal{G}_{c,\downarrow }^{0}\left( \mathbf{k}%
,z\right) \left( m_{22}m_{44}-m_{24}m_{42}\right)
\begin{pmatrix}
1 & 1 \\
1 & 1%
\end{pmatrix}
}{1-\left\vert V\right\vert ^{2}\mathcal{G}_{c,\downarrow }^{0}\left(
\mathbf{k},z\right) \left( m_{22}+m_{44}-m_{24}-m_{42}\right) }.
\label{E5.13aa}
\end{equation}

Note that in this approach the $\mathbf{M}_{\sigma }^{ap}(i\omega )$ are
independent of $\mathbf{k},$ and that all the $\mathbf{k}$ dependence of our
approximate $\mathbf{G}_{\sigma }^{ff,ap}(\mathbf{k},i\omega )$ comes
through the $\mathcal{G}_{c,\sigma }^{0}\left( \mathbf{k},i\omega \right) $
in the Eqs. (\ref{E5.12a}-\ref{E5.13aa})

\subsubsection{Real space and imaginary frequency\label{S05_2a}}

In the previous section we derived the reciprocal space and imaginary
frequency GF for the PAM in the\ atomic approximation, but sometimes it is
necessary to use the GF in real space, with the $f$ electron being created
and destroyed at the same site. These GF are denoted by $G_{\sigma
}^{ff}(i\omega )$ and are given by
\begin{equation}
\mathbf{G}_{\sigma }^{ff}(i\omega )=\frac{1}{N}\sum_{\mathbf{k}}\mathbf{G}%
_{\sigma }^{ff}(\mathbf{k},i\omega )=\frac{1}{N}\sum_{\mathbf{k}}\mathbf{G}%
_{\sigma }^{ff}\left( \varepsilon \left( \mathbf{k}\right) ,i\omega \right) .
\label{eq2cap3}
\end{equation}
Considering a rectangular conduction band and transforming the sum into an
integral, we obtain
\begin{equation}
\mathbf{G}_{\sigma }^{ff}(i\omega )=\frac{1}{2D}\int_{-D-\mu }^{+D-\mu
}d\varepsilon \mathbf{G}_{\sigma }^{ff}(\varepsilon ,i\omega ),
\label{eq3cap3}
\end{equation}
where $\mu $ is the chemical potential and $D$ is the half-width\ of the
conduction band. Substituting Eqs. (\ref{E5.12a} and \ref{E5.13aa}) into Eq.
(\ref{eq3cap3}) we obtain that 
\begin{equation}
\mathbf{G}_{\uparrow }^{ff}(i\omega )=
\begin{pmatrix}
G_{11}^{ff} & G_{13}^{ff} \\
G_{31}^{ff} & G_{33}^{ff}%
\end{pmatrix}
=\frac{1}{2D}\int_{-D-\mu }^{+D-\mu }d\varepsilon \frac{\mathbf{M}_{\uparrow
}^{ap}-\left\vert V\right\vert ^{2}\mathcal{G}_{c,\uparrow }^{0}(\varepsilon
,i\omega )\left( m_{11}m_{33}-m_{13}m_{31}\right)
\begin{pmatrix}
1 & -1 \\
-1 & 1%
\end{pmatrix}
}{1-\left\vert V\right\vert ^{2}\mathcal{G}_{c,\uparrow }^{0}(\varepsilon
,i\omega )\left( m_{11}+m_{33}+m_{13}+m_{31}\right) },  \label{eq4cap3}
\end{equation}
and
\begin{equation}
\mathbf{G}_{\downarrow }^{ff}(i\omega )=
\begin{pmatrix}
G_{22}^{ff} & G_{24}^{ff} \\
G_{42}^{ff} & G_{44}^{ff}%
\end{pmatrix}
=\frac{1}{2D}\int_{-D-\mu }^{+D-\mu }d\varepsilon \frac{\mathbf{M}%
_{\downarrow }^{ap}-\left\vert V\right\vert ^{2}\mathcal{G}_{c,\downarrow
}^{0}(\varepsilon ,i\omega )\left( m_{22}m_{44}-m_{24}m_{42}\right)
\begin{pmatrix}
1 & 1 \\
1 & 1%
\end{pmatrix}
}{1-\left\vert V\right\vert ^{2}\mathcal{G}_{c,\downarrow }^{0}(\varepsilon
,i\omega )\left( m_{22}+m_{44}-m_{24}-m_{42}\right) }.  \label{eq5cap3}
\end{equation}
The variable $\varepsilon $ is present only in $\mathcal{G}_{c,\sigma }^{0}$%
\ and the integration is straightforward. We find
\begin{equation}
\mathbf{G}_{\uparrow }^{ff}(z)=\mathbf{M}_{\uparrow }^{ap}+\frac{V^{2}}{2D}%
ln\left( \frac{z-D+\mu +V^{2}M_{\uparrow }^{ff}}{z+D+\mu +V^{2}M_{\uparrow
}^{ff}}\right) \left[ \mathbf{M}_{\uparrow }^{ap}M_{\uparrow }^{ff}-
\begin{pmatrix}
1 & -1 \\
-1 & 1%
\end{pmatrix}
\Theta _{\uparrow }\right]  \label{eq6cap3a}
\end{equation}
\begin{equation}
\mathbf{G}_{\downarrow }^{ff}(z)=\mathbf{M}_{\downarrow }^{ap}+\frac{V^{2}}{%
2D}ln\left( \frac{z-D+\mu +V^{2}M_{\downarrow }^{ff}}{z+D+\mu
+V^{2}M_{\downarrow }^{ff}}\right) \left[ \mathbf{M}_{\downarrow
}^{ap}M_{\downarrow }^{ff}-
\begin{pmatrix}
1 & 1 \\
1 & 1%
\end{pmatrix}
\Theta _{\downarrow }\right]  \label{eq6cap3b}
\end{equation}
where
\begin{equation}
M_{\uparrow }^{ff}=m_{11}+m_{13}+m_{31}+m_{33},  \label{E5.13g1}
\end{equation}
\begin{equation}
M_{\downarrow }^{ff}=m_{22}+m_{44}-m_{24}-m_{42},  \label{E5.13g2}
\end{equation}
and
\begin{equation}
\Theta _{\uparrow }=m_{11}m_{33}-m_{13}m_{31,}  \label{E5.14g1}
\end{equation}
\begin{equation}
\Theta _{\downarrow }=m_{22}m_{33}-m_{24}m_{42}.  \label{E5.14g2}
\end{equation}

In Appendix \ref{ApE 0} we define and calculate the formal expressions of
the matrices $\mathbf{G}_{\sigma }^{fc,ap}(\mathbf{k},i\omega )$ (cf. Eq. (%
\ref{E5.16bbb},\ref{E5.16cb})) and $\mathbf{G}_{\sigma }^{cf,ap}(\mathbf{k}%
,i\omega )$ (cf. Eq. (\ref{E5.19bc},\ref{E5.19cc})) associated to the
crossed \ GF of the impurity, as well as the GF of the pure conduction
electron $\mathbf{G}_{\sigma }^{cc,ap}(\mathbf{k},i\omega )$ (cf. Eq. (\ref%
{E5.21bb},\ref{E5.21bd})). We can also describe the conduction electrons in
real space and imaginary time: the corresponding $\mathbf{G}_{\sigma
}^{fc,ap}(i\omega )$ are given in Eqs. (\ref{E5.32b},\ref{E5.33b}) and the $%
\mathbf{G}_{\sigma }^{cf,ap}(i\omega )$ in Eqs. (\ref{E5.34b},\ref{E5.35b}).
In a similar way we obtain $\mathbf{G}_{\sigma }^{cc,ap}(i\omega )$ (cf. Eq.
(\ref{E5.36b})), and we can use this relation to express all the other GF
(cf. Eqs. (\ref{E5.37aa}-\ref{E5.38bb},\ref{E5.40aa},\ref{E5.40bb})).

\subsection{The approximate $\mathbf{G}_{\protect\sigma }^{ff,ap}$ GF for
the Impurity Anderson Model}

\label{S05_2}

As in the case of the PAM, we substitute the $\mathbf{G}_{\sigma }^{ff}$ in
Eq. (\ref{E5.8}) by the exact solution $\mathbf{G}_{\sigma }^{ff,at}$ of the
problem with zero band width and local hybridization, and obtain the
corresponding effective cumulant $\mathbf{M}_{\sigma }^{ap}$. The conduction
band corresponding to this approximation then has $E_{\mathbf{k},\sigma
}=E_{0}^{a}$, so that for all values of $\mathbf{k}$ it is $\mathcal{G}%
_{c,\sigma }^{0}\left( \mathbf{k},z\right) =-1/\left( z-\varepsilon
_{o}\right) $ , where $\varepsilon _{o}=E_{0}^{a}-\mu $. From Eq. (\ref%
{E5.6a}) we then find $\varphi _{\sigma }(z)\rightarrow \varphi _{\sigma
}^{0}(z)=-1/\left( z-\varepsilon _{o}\right) $, and substituting into Eqs. (%
\ref{E5.5a},\ref{E5.5b}) we obtain the $\mathbf{W}_{\sigma }^{ap}$ that
should be used in Eq. (\ref{E5.8}) to calculate $\mathbf{M}_{\sigma }^{ap}$.
To define the approximate GF $\mathbf{G}_{\sigma }^{ff,ap}(i\omega )$\
introduced in the present work, we substitute this approximate $\mathbf{M}%
_{\sigma }^{ap}$ into the exact expression Eq. (\ref{E5.7}), but in this
equation we use the $\mathbf{W}_{\sigma }$ that corresponds to the
conduction band with full width. The Eqs. (\ref{E5.9},\ref{E5.10}) for $%
\mathbf{M}_{\sigma }^{ap}$ and $\mathbf{A}_{\sigma }^{ap}\mathbf{=W}_{\sigma
}\mathbf{.M}_{\sigma }^{ap}$\ are also valid for the SIAM, but using Eqs.\ (%
\ref{E5.5a},\ref{E5.5b}) we find different expression for the $a_{ij}$:
\begin{equation}
\begin{tabular}{llll}
$a_{11}=\varphi _{\uparrow }(z)\left( m_{11}+m_{31}\right) $ & \hspace{10pt};
& \hspace{10pt} & $a_{22}=\varphi _{\downarrow }(z)\left(
m_{22}-m_{42}\right) $ \\
$a_{33}=\varphi _{\uparrow }(z)\left( m_{33}+m_{13}\right) $ & \hspace{10pt};
& \hspace{10pt} & $a_{44}=\varphi _{\downarrow }(z)\left(
m_{44}-m_{24}\right) $ \\
$a_{31}=a_{11}$ & \hspace{10pt}; & \hspace{10pt} & $a_{42}=-a_{22}$ \\
$a_{13}=a_{33}$ & \hspace{10pt}; & \hspace{10pt} & $a_{24}=-a_{44}$%
\end{tabular}
\ \ \ \ \ \ \ \ \ \ \   \label{E5.11}
\end{equation}
If we now substitute these $\mathbf{M}_{\sigma }^{ap}$ and $\mathbf{A}%
_{\sigma }^{ap}$\ into Eq. (\ref{E5.7}), we obtain the approximate $\mathbf{G%
}_{\sigma }^{ff,ap}$ for the SIAM:
\begin{equation}
\mathbf{G}_{\uparrow }^{ff,ap}(i\omega )=\frac{%
\begin{pmatrix}
m_{11} & m_{13} \\
m_{31} & _{m33}%
\end{pmatrix}
-\left\vert V\right\vert ^{2}\varphi _{\uparrow }((i\omega ))\left(
m_{11}m_{33}-m_{13}m_{31}\right)
\begin{pmatrix}
1 & -1 \\
-1 & 1%
\end{pmatrix}
}{1-\left\vert V\right\vert ^{2}\varphi _{\uparrow }((i\omega ))\left(
m_{11}+m_{33}+m_{13}+m_{31}\right) },  \label{5.12}
\end{equation}
and
\begin{equation}
\mathbf{G}_{\downarrow }^{ff,ap}(i\omega )=\frac{%
\begin{pmatrix}
m_{22} & m_{24} \\
m_{42} & _{m44}%
\end{pmatrix}
-\left\vert V\right\vert ^{2}\varphi _{\downarrow }((i\omega ))\left(
m_{22}m_{44}-m_{24}m_{42}\right)
\begin{pmatrix}
1 & 1 \\
1 & 1%
\end{pmatrix}
}{1-\left\vert V\right\vert ^{2}\varphi _{\downarrow }((i\omega ))\left(
m_{22}+m_{44}-m_{24}-m_{42}\right) }.  \label{E5.13}
\end{equation}

In Appendix \ref{ApE 0} we define and calculate the formal expressions of
the matrices $\mathbf{G}_{\sigma }^{fc,ap}(\mathbf{j}_{i}=0,\mathbf{k}%
,i\omega )$ (cf. Eqs. (\ref{E5.16b},\ref{E5.16c})) and $\mathbf{G}_{\sigma
}^{cf,ap}(\mathbf{k},\mathbf{j}^{\prime }=0,i\omega )$ (cf. Eqs. (\ref%
{E5.19b},\ref{E5.19c})) associated to the crossed \ GF of the impurity, as
well as the GF of the pure conduction electron $\mathbf{G}_{\sigma }^{cc,ap}(%
\mathbf{k},\mathbf{k}^{\prime },i\omega )$ (cf. Eqs. (\ref{E5.21b},\ref%
{E5.21c})). We can also describe the conduction electrons in the Wannier
representation: the corresponding $\mathbf{G}_{\sigma }^{fc,ap}(i\omega )$
are given in Eqs. (\ref{E5.32},\ref{E5.33}) and the $\mathbf{G}_{\sigma
}^{cf,ap}(i\omega )$ in Eqs. (\ref{E5.34},\ref{E5.35}). In a similar way we
obtain $\mathbf{G}_{\sigma }^{cc,ap}(i\omega )$ (cf. Eq. (\ref{E5.36})), and
we can use this relation to express all the other GF (cf. Eqs. (\ref{E5.37a}-%
\ref{E5.38b},\ref{E5.40a},\ref{E5.40b})).

\appendix

\section{The sign of the contribution of a graph.}

\label{ApA}

For convenience we summarize here the Appendix C of reference \cite{FFM},
where one should have$^{\text{\ref{myfoot4}}}$ $v(\alpha ,\vec{k},\sigma
,\pm u)$\ instead of $V(\alpha ,\vec{k},\sigma ,\pm u)$ in the item\textbf{\
2(b)} of \textbf{Rule C.2}.

Here we discuss the sign that must be given to the contribution of a given
graph, and we are only interested in the case without external fields, i.e.
with $\xi=0$, but we shall keep some results for $\xi\neq0$ that are
convenient to understand $\xi=0$. The rules for drawing the graphs that
appear in the calculation of the averages $\left\langle \left(
Y(l_{1}),\cdots Y(l_{n})\right) _{+}\right\rangle ^{\xi}$ are given in Rules
3.3 and 3.4 in \cite{FFM}. In item 4 of those rules, the Fermi type lines
running to each vertex were paired in an arbitrary way for $\xi=0$, and
several open and closed loops were formed in this way, where all the open
loops must have two external vertices. A definite sense was arbitrarily
assigned to each of the loops, and we call this direction the
\textquotedblleft sense of the loop\textquotedblright. In the following
discussion we consider only Fermi-type operators, because the position of
the Bose-type operators does not affect the sign of the contribution, and we
shall mean Fermi-type operator when we say \textquotedblleft
operator\textquotedblright\ in the remaining of this Appendix. It is now
convenient to introduce two concepts that shall be useful in the present
computation.

\textbf{Definition C.1.}

A graph is in a ``perfect ordering'' when the following relations are
satisfied:

\begin{description}
\item[ 1. ] For all the open loops, $\tau$ increases in all the vertices of
the loop in the sense of the loop.

\item[ 2. ] For every closed loop, $\tau$ increases in the sense of the loop
for all the vertices but one (it is impossible to satisfy (1) for a closed
loop).

\item[ 3. ] All the $\tau$ in a given loop are either smaller or greater
than all the $\tau$ in all the other loops of the graph.$\bullet$
\end{description}

There are many ways to choose a perfect ordering of a graph, but the
particular choice is not important provided that we use always the same one
after it has been chosen.

\textbf{Definition C.2.}

Several Fermi-type operators of a graph contribution are in a ``perfect
order'' when:

\begin{description}
\item[ 1. ] The $Y$-operators are written from right to left following the
perfect ordering we have chosen for their graph.

\item[ 2. ] For the two operators of each internal edge (they have the same $%
\tau$) we write the $X$-operator to the left of the $C$-operator.$\bullet$
\end{description}

As a starting point we shall consider rules that are also valid for $%
\xi\neq0 $ because they are simpler to state although less systematic to
apply than those given\ by Hubbard \cite{Hubbard5} for $\xi=0$. We shall
consider explicitly the graphs for $\left\langle \left( \hat{Y}(1)\cdots\hat{%
Y}(r)\right) _{+}\right\rangle _{H}$, i.e. a GF with $r$ external operators $%
\hat {Y}(1)\cdots\hat{Y}(r)$ (but the rules are also valid for vacuum
graphs). The $n$-th order term of this GF contains the average

\begin{equation}  \label{Sinal1}
\left\langle \left( \hat Y(1)\cdots\hat Y(r)\left( H^{\prime
}(\tau_{1})\cdots H^{\prime}(\tau_{n})\right) \right) _{+}\right\rangle _{H}
\end{equation}

\noindent and the application of Theorem 3.1 from \cite{FFM} to this
equation gives all the $n$-th order graphs\footnote{%
In the absence of an external field (i.e. when $\xi=0$) the $%
H^{\prime}(\tau) $ in this equation is equal to $H_{h}(\tau)$ (cf. Eq. (\ref%
{Hibri1})).}.

\textbf{Rule C.1.}

To obtain the sign associated to a given graph, multiply the parities of the
following two permutations:

\begin{description}
\item[ 1. ] It takes the operators from the order used to write Eq.~(\ref%
{Sinal1}) into the perfect order.

\item[ 2. ] It takes the operators from the perfect order into the order in
which they appear in the final expression that gives the graph contribution.
\end{description}

As the operators $H_{h}$ are of the Bose type and can be moved freely inside
the ordered parenthesis in Eq.~(\ref{Sinal1}), in the first step it is
necessary to consider only the permutation that takes the external
Fermi-type operators to their perfect order. The \textbf{Rule C.1.} is just
the application of Theorem 3.1 in \cite{FFM} in two steps, and the only
reason to proceed in this way is that the perfect order of the $Y$-operators
in a graph provides a reference frame to organize the calculation.

For $\xi=0$ we shall give rules with the same labels employed by Hubbard
\cite{Hubbard5}, because of their similarity. In this case, there is only an
even number of lines running into each vertex, and for any CV this number is
two .This simplifies the treatment, and the first step is the same step 1)
employed in \textbf{Rule C.1}: this is just rule \textquotedblleft
d\textquotedblright\ of Hubbard.

To calculate the change of sign that corresponds to step 2) in \textbf{Rule
C.1} we proceed in three steps.

First we consider all the open loops that pass through each vertex, and note
that in the perfect order, the $X$-operator is to the left of the $C$%
-operator in all the internal edges. To be able to pair operators of the
same type at each vertex (otherwise the corresponding cumulant vanish) it is
necessary to change the order of these two operators (with a change of sign)
when the arrow in the edge points towards the CV. To correct for the sign
missing in Eq.~(\ref{Ger3}) one must also add a factor $\pm u$ to the $%
v(j,\alpha,\vec {k},\sigma,\pm u)$ in Rule 3.5.2.c in \cite{FFM}, and these
two factors correspond to Hubbard's rule \textquotedblleft
b\textquotedblright. In the present problem there are only two edges at each
CV, and when both are internal, the effect cancels out and the rule is not
necessary. To prove this result, note that according to Rule 3.6b$^{\prime}$
in \cite{FFM}, the cumulant

\begin{equation}  \label{Sinal2}
\left\langle \left( Y (c ; \vec{k}_{s}, \sigma_{2},-u_{2},\tau_{2} ) Y (c ;
\vec{k}_{s},\sigma_{1},+u_{1}, \tau_{1}) \right) _{+} \right\rangle _{c}
\end{equation}

\noindent at each CV, is already written with the $Y$-operators in the
perfect order, with the $-u_{2}$ corresponding to the outgoing arrow. The
contribution to Rule 3.5.2.c in \cite{FFM} of the two internal edges running
into the CV after correcting for the missing sign in Eq.~(\ref{Ger3}), is
then

\begin{equation}  \label{Sinal3}
(+u_{2})v(j_{2},\alpha_{2},k_{s},\sigma_{2},+u_{2})(-u_{1})v(j_{1},%
\alpha_{1},k_{s},\sigma_{1},-u_{1})
\end{equation}

As there is particle conservation, we have $u_{1}=u_{2}$, and when we
multiply into the minus sign due to the exchange of the X with the C
operators on the line with the arrow towards the CV, the overall sign is
always plus. Hubbard' s rule ``b'' is therefore not necessary for all the CV
with two internal lines.

For any CV with only one internal line (and therefore one external line
also), one must multiply the $v(j,\alpha,\vec k,\sigma,\pm u)$ into $\pm u$
and also into $-1$ when the internal edge points toward the CV . This is the
only effect that remains in the PAM of Hubbard's rule ``b''.

The discussion above fails for closed loops because $\tau$ can increase in
the sense of the loop in all vertices but one. After putting all the
operators in perfect order and then exchanging the $X$-operator with the $C$%
-operator for all the lines with arrows pointing to a CV, the first and last
operators in the resulting expression belong to the same vertex, and should
therefore be brought together . These two operators are separated by an even
number of Fermi operators, but bringing them together by an even permutation
would still leave them against the order of the loop, i.e.: the operator at
the left would correspond to the edge with the arrow pointing toward the
vertex. A permutation of odd parity is then necessary to put all the
operators of any closed loop in perfect order, and this is Hubbard's rule
``c''.

After the three steps discussed above, the $Y$-operators that were in the
order given by Eq.~(\ref{Sinal1}) are now paired at each vertex according to
the loops of the graph considered, each pair written in the sense of the
loop. We shall denote with $(\alpha_{s},\beta{_{s}})$ the two indices of the
$Y$-operators of each of those pairs, written already in the sense of the
loop,i.e.: $\beta_{s}\rightarrow\alpha_{s}$. All the pairs that correspond
to a given vertex are still separated by many pairs that belong to other
vertices of the graph, but it is only necessary an even permutation to put
together all the pairs of each vertex. The pair associated to each CV is
already in the same order of the cumulant of Rule $3.6$ $b^{\prime}$ in \cite%
{FFM}, and only remains to consider the cumulants associated to the FV. If
there are $p$ loops crossing an FV, we have already the corresponding
operators in the order $(\alpha_{1},\beta_{1}),\cdots(\alpha_{p},\beta_{p})$
while in the cumulant associated to that vertex by \textbf{Rule 3.6.}(2).a
in \cite{FFM} they are written in the order $Y(\gamma_{1})\cdots
Y(\gamma_{2p})$, where $\gamma _{1}\cdots\gamma_{2p}$, correspond to the same%
$(\alpha_{1},\beta{1}),\cdots(\alpha_{p},\beta{p})$ but in a different
order. It is then necessary to associate to each of these cumulants a $\pm$
given by the parity of the permutation that takes $(\alpha_{1},\beta_{{1}%
}),\cdots(\alpha_{p},\beta_{{p}})$ into $\gamma_{1}\cdots\gamma_{2p}$. This
is Hubbard's rule \textquotedblleft a\textquotedblright.

It is now convenient to put together the rules for the calculation of the
sign required by \textbf{Rule 3.7}.(2).d or \textbf{Rule 3.7a}.(2).e.

\textbf{Rule C.2}

To calculate the sign of a graph with $\xi= 0$

\begin{description}
\item[ 1. ] Define a perfect ordering for the graph according to Definition
C.1.

\item[ 2. ] The sign of the graph is the product of the following factors

\item[ (a) ] When there are $p$ loops crossing an FV, denote with $(\alpha
_{s},\beta_{s})$ the indices of the two $X$-operators of the $s$-th loop at
that vertex $(s=1,\cdots,p)$, written already in the sense of the loop
(i.e.: $\beta_{s}\rightarrow\alpha_{s}$). The $2p$ Fermi-type operators at
that FV appear in the cumulant of Rule 3.6a in the order $%
Y(\gamma_{1})\cdots Y(\gamma_{2p})$, where the $\gamma_{1}\cdots\gamma_{2p}$
are the same $(\alpha_{1},\beta_{1}),\cdots(\alpha_{p},\beta_{p})$ in a
different order. For each FV multiply into a $\pm1$ given by the parity of
the permutation that takes$(\alpha_{1},\beta_{1}),\cdots(\alpha_{p},%
\beta_{p})$ into $\gamma _{1},\cdots\gamma_{2p}$.

\item[ (b) ] For any CV with only one internal edge multiply the $V(\alpha,%
\vec{k},\sigma,\pm u)$ of Rule 3.6.(2).c into $(\pm u)$, and also into a
further $-1$ when the arrow of the internal edge points toward the CV.

\item[ (c) ] There is a factor $-1$ for every closed loop.

\item[ (d) ] If the graph is employed to calculate a GF with $r$ Fermi-type
operators written in the order $\hat{Y}(1)\cdots\hat{Y}(r)$, multiply into a
sign given by the parity of the permutation that takes $(Y(1)\cdots Y(r))$
into the same operators written in the perfect ordering chosen for the
graph. This item does not apply to vacuum graphs.
\end{description}


\section{Counting graphs and the symmetry factor.}

\label{ApB}

For convenience we reproduce here, with very minor changes, the Appendix D
of reference \cite{FFM}.

As discussed in appendix \ref{ApA} of reference \cite{FFM}, the n-th order
term of the perturbative expansion of the GF $\left\langle \left( \hat
{Y}%
(1)\cdots\hat{Y}(r)\right) _{+}\right\rangle _{H}$ contains the expression
in Eq.~(\ref{Sinal1}) of \cite{FFM}, and its contributions have the form
\begin{align}
& \frac{(-1)}{n!}^{n}\mathcal{Z}_{o}(\beta,\xi)\int_{0}^{\beta}d\tau_{1}%
\sum_{l_{1},l_{1}^{\prime}}V(l_{1},l_{1}^{\prime})\cdots\int_{0}^{\beta}d%
\tau_{n}  \notag \\
& \sum_{l_{n}\ ,\ l^{\prime}{}_{n}}V(l_{n},l^{\prime}{}_{n})\left\langle
\left( Y(l_{1})Y(l_{1}^{\prime})\cdots Y(l_{n})Y(l_{n}^{\prime})\right)
_{+}\right\rangle ^{\xi}  \label{Expansao2}
\end{align}
(cf. Eq. (3.11) in \cite{FFM}), but with the $r$ external operators $%
Y(1)\cdots Y(r)$ included in the averages. When the Theorem 3.1 in \cite{FFM}
is applied to these averages, the $n$-th order contribution can be
associated to a family of graphs, and many of them are disconnected and
composed of several connected graphs. We label each topologically distinct
connected graph with an index $\alpha$, and we use $n_{\alpha}$ to denote
the number of times that the $\alpha$ graph appears in the nth-order graph.
It is clear that there might be several identical contributions associated
to the same $n$-th order graph because all the $n!$ permutations of the
edges of a given graph give the same contribution. These identical
contributions should be counted as different contributions every time they
correspond to a different partition in cumulants. The correct number of
times that a topologically distinct graph of $n$-th order gives the same
contribution is then

\begin{equation}  \label{Simet1}
n! \prod^{\infty}_{\alpha= 1} \frac{1}{n_{\alpha}! g^{n_{\alpha }}_{\alpha}}
\end{equation}

\noindent where $g_{\alpha}$ is the symmetry factor of the connected graph $%
\alpha$ and is calculated using the Rule D.1 discussed below. To derive this
result one applies the same arguments employed in Ref.~\onlinecite{Wortis74}%
: the factor $2^{n}$ of that reference is not present in our expression
because the pair of vertices of any internal edge can not be exchanged (cf.
the definition of the coefficients of Eq.~(\ref{Hibri1}), discussed after
Eq.(\ref{Representacao6})).

To calculate the symmetry factor $g_{\alpha}$ it is enough to adapt the rule
given by Hubbard in Ref.~ \onlinecite{Hubbard5}, Appendix B . The
calculation seems rather obvious in simple cases, but it is convenient to
give the rule to deal with the more complicated ones.

\textbf{Definition D.1}

A vertex is said to be ``internal'' when all the lines running to it are
internal lines.$\bullet$

In the PAM, only Fermi lines can run into an internal vertex, because of the
form of the interaction (cf. Eq.~(\ref{Hibri1})).

\textbf{Rule D.1}

To calculate the symmetry factor g of a connected graph with $p_{f}$ and $%
p_{c}$ vertices FV and CV respectively:

\begin{enumerate}
\item Number the FV with $1,2,\cdots,p_{f}$ and the CV with $1,2,\dots,p_{c}
$ so that $1,2,\cdots,q_{f}$ correspond to all the internal FV and $%
1,2,\cdots,q_{c}$ to all the internal CV.

\item Form the $p_{f}\times p_{c}$ matrix $N$, with elements $N_{i,j}$,
where $N_{i,j}$ is the number of Fermi edges joining the FV $i$ to the CV $j$
.

\item Let $g_{1}$ be the order of the group of permutations $\mathcal{P}_{1}
$ of the $q_{f}\times q_{c}$ ordered pairs $(i,j)$, which has the property
that if any permutation of $\mathcal{P}_{1}$ is applied to the indices $%
i=1,2,\cdots,q_{f}$ and $j=1,2,\cdots,q_{c}$ of the matrix $N$, this matrix
is left unchanged.

\item The symmetry factor is then
\end{enumerate}

\begin{equation}
g=g_{1}\prod_{j=1}^{q_{f}}\prod_{j=1}^{q_{c}}(N_{i,j}!)  \label{Simet2}
\end{equation}

\section{The Fourier transform of Green's functions in imaginary time}

\label{ApC}

We are interested in the GFs defined in Eq. (\ref{Fourier1}) with only two
operators $\hat{Y}(\gamma,\tau)$:
\begin{equation}
\mathcal{G}(\gamma_{1},\tau_{1};\gamma_{2},\tau_{2})=\left\langle \left(
\hat{Y}(\gamma_{1},\tau_{1})\hat{Y}(\gamma_{2},\tau_{2})\right)
_{+}\right\rangle _{\mathcal{H}},  \label{EqC1}
\end{equation}
as well as in their Fourier transforms. Introducing
\begin{equation}
\Omega=-\frac{1}{\beta}\ln\left[ \exp\left( -\beta\mathcal{H}\right) \right]
,  \label{EqC2}
\end{equation}
we can write for ${\tau}_{1}>{\tau}_{2}$ (cf. Eq. (\ref{Fourier2}))
\begin{equation}
\mathcal{G}(\gamma_{1},\tau_{1};\gamma_{2},\tau_{2})=\exp\left( \beta
\Omega\right) Tr\left\{ \exp\left( -\beta\mathcal{H}\right) \exp{(\tau }_{1}{%
\mathcal{H})}Y_{\gamma_{1}}\exp{(-\tau}_{1}{\mathcal{H})}\exp{(\tau}_{2}{%
\mathcal{H})}Y_{\gamma_{2}}\exp{(-\tau}_{2}{\mathcal{H})}\right\}
\label{EqC3}
\end{equation}
and using the properties of the trace we obtain:
\begin{align}
\mathcal{G}(\gamma_{1},\tau_{1};\gamma_{2},\tau_{2}) & =\exp\left(
\beta\Omega\right) Tr\left\{ \exp\left( -\beta\mathcal{H}\right) \exp {(}%
\left( {\tau}_{1}-{\tau}_{2}\right) {\mathcal{H})}Y_{\gamma_{1}}\exp {(-}%
\left( {\tau}_{1}-{\tau}_{2}\right) {\mathcal{H})}Y_{\gamma_{2}}\right\} =
\notag \\
& =\left\langle \left( \hat{Y}(\gamma_{1},\tau_{1}-\tau_{2})\hat{Y}%
(\gamma_{2},0)\right) _{+}\right\rangle _{\mathcal{H}}=F\left( {\tau}_{1}-{%
\tau}_{2}\right) .  \label{EqC3a}
\end{align}

Similarly for \ ${\tau}_{1}<{\tau}_{2}$, we have (because both $Y_{\gamma}$
are of the Fermi type)
\begin{align}
\mathcal{G}(\gamma_{1},\tau_{1};\gamma_{2},\tau_{2}) & =\exp\left(
\beta\Omega\right) Tr\left\{ \exp\left( -\beta\mathcal{H}\right) (-1)\hat{Y}%
(\gamma_{2},\tau_{2})\hat{Y}(\gamma_{1},\tau_{1})\right\} =  \notag \\
& =-\exp\left( \beta\Omega\right) Tr\left\{ \exp\left( -\beta \mathcal{H}%
\right) \exp{(\tau}_{2}{\mathcal{H})}Y_{\gamma_{2}}\exp{(-\tau }_{2}{%
\mathcal{H})}\exp{(\tau}_{1}{\mathcal{H})}Y_{\gamma_{1}}\exp{(-\tau}_{1}{%
\mathcal{H})}\right\} =  \notag \\
& =-\exp\left( \beta\Omega\right) Tr\left\{ \exp{(}\left( {\tau}_{1}-{\tau}%
_{2}\right) {\mathcal{H})}Y_{\gamma_{1}}\exp{(-}\left( {\tau}_{1}-{\tau}%
_{2}\right) {\mathcal{H})}\exp\left( -\beta\mathcal{H}\right)
Y_{\gamma_{2}}\right\} =  \notag \\
& =-\exp\left( \beta\Omega\right) Tr\left\{ \exp\left( -\beta \mathcal{H}%
\right) \exp{(}\left( {\tau}_{1}-{\tau}_{2}+\beta\right) {\mathcal{H})}%
Y_{\gamma_{1}}\exp{(-}\left( {\tau}_{1}-{\tau}_{2}+\beta\right) {\mathcal{H})%
}Y_{\gamma_{2}}\right\} ,  \label{EqC4}
\end{align}
and as ${\tau}_{1}-{\tau}_{2}+\beta>0$ we have
\begin{equation}
\mathcal{G}(\gamma_{1},\tau_{1};\gamma_{2},\tau_{2})=-F\left( {\tau}_{1}-{%
\tau}_{2}+\beta\right) ,  \label{EqC5}
\end{equation}
and finally
\begin{equation}
\mathcal{G}(\gamma_{1},\tau_{1};\gamma_{2},\tau_{2})=\theta\left( {\tau}_{1}-%
{\tau}_{2}\right) F\left( {\tau}_{1}-{\tau}_{2}\right) -\theta\left( {\tau}%
_{2}-{\tau}_{1}\right) F\left( {\tau}_{1}-{\tau}_{2}+\beta\right) .
\label{EqC6}
\end{equation}

The time Fourier coefficients are given by
\begin{align}
\mathcal{G}(\gamma_{1},\omega_{1};\gamma_{2},\omega_{2})= & \frac{1}{\beta }%
\int_{0}^{\beta}d\tau_{1}\int_{0}^{\beta}d\tau_{2}\exp\left[ i\left(
\omega_{1}\tau_{1}+\omega_{2}\tau_{2}\right) \right] \left\langle \left(
\hat{Y}(\gamma_{1},\tau_{1})\hat{Y}(\gamma_{2},\tau_{2})\right)
_{+}\right\rangle _{\mathcal{H}}=  \notag \\
& =\frac{1}{\beta}\int_{0}^{\beta}d\tau_{1}\int_{0}^{\beta}d\tau_{2}\exp%
\left[ i\left( \omega_{1}\tau_{1}+\omega_{2}\tau_{2}\right) \right] \left(
\theta\left( {\tau}_{1}-{\tau}_{2}\right) F\left( {\tau}_{1}-{\tau }%
_{2}\right) -\theta\left( {\tau}_{2}-{\tau}_{1}\right) F\left( {\tau}_{1}-{%
\tau}_{2}+\beta\right) \right) .  \label{EqC7}
\end{align}
We change variables
\begin{align}
x & ={\tau}_{1}-{\tau}_{2}  \notag \\
y & ={\tau}_{1}  \label{EqC8}
\end{align}
and we find
\begin{align}
\mathcal{G}(\gamma_{1},\omega_{1};\gamma_{2},\omega_{2}) & =\frac{1}{\beta }%
\int_{0}^{\beta}dy\exp\left[ iy\left( \omega_{1}+\omega_{2}\right) \right]
\int_{-y}^{\beta-y}dx\exp\left[ ix\omega_{1}\right] \left( \theta\left( {x}%
\right) F\left( {x}\right) -\theta\left( {-x}\right) F\left( {x}%
+\beta\right) \right) =  \notag \\
& =\frac{1}{\beta}\int_{0}^{\beta}dy\exp\left[ iy\left(
\omega_{1}+\omega_{2}\right) \right] \left\{ \int_{0}^{\beta-y}dxF\left( {x}%
\right) \exp\left[ ix\omega_{1}\right] -\int_{-y}^{0}dxF\left( {x}%
+\beta\right) \exp\left[ ix\omega_{1}\right] \right\} .  \label{EqC9}
\end{align}
Changing variables ${x}+\beta=\xi$ in the second integral we obtain
\begin{equation}
\int_{-y}^{0}dxF\left( {x}+\beta\right) \exp\left[ ix\omega_{1}\right]
=\int_{\beta-y}^{\beta}dxF\left( {\xi}\right) \exp\left[ ix\omega _{1}\left(
\xi-\beta\right) \right]  \label{qC10}
\end{equation}
and employing $\exp\left[ ix\omega_{1}\beta\right] =-1$ we have$^{\text{\ref%
{myfoot4b}}}$%
\begin{align}
\mathcal{G}(\gamma_{1},\omega_{1};\gamma_{2},\omega_{2}) & =\frac{1}{\beta }%
\int_{0}^{\beta}dy\exp\left[ iy\left( \omega_{1}+\omega_{2}\right) \right]
\left\{ \int_{0}^{\beta-y}dxF\left( {x}\right) \exp\left[ ix\omega_{1}\right]
+\int_{\beta-y}^{\beta}dxF\left( {x}\right) \exp\left[ ix\omega_{1}\right]
\right\}  \notag \\
& =\Delta\left( \omega_{1}+\omega_{2}\right) \int_{0}^{\beta}dxF\left( {x}%
\right) \exp\left[ ix\omega_{1}\right]  \label{EqC11}
\end{align}
where $\omega_{1}$ and $\omega_{2}$ are given by Eq. (\ref{Fourier5}) for
Fermi-like operators. When the two operators $\hat{Y}(\gamma,\tau)$ are $X$%
-operators we can write
\begin{align}
& \frac{1}{\beta}\int_{0}^{\beta}d\tau_{1}\int_{0}^{\beta}d\tau_{2}\exp\left[
i\left( \omega_{1}\tau_{1}+\omega_{2}\tau_{2}\right) \right] \left\langle
\left(
X_{j_{1},\alpha_{1}}(\tau_{1})X_{j_{2},\alpha_{2}}^{\dagger}(\tau_{2})%
\right) _{+}\right\rangle _{\mathcal{H}}=  \notag \\
& =\ \Delta\left( \omega_{1}+\omega_{2}\right) \int_{0}^{\beta}d\tau
_{1}\left\langle \left(
X_{j_{1},\alpha_{1}}(\tau_{1})X_{j_{2},\alpha_{2}}^{\dagger}(0)\right)
_{+}\right\rangle _{\mathcal{H}}\exp\left[ i\tau _{1}\omega_{1}\right]
\label{EqC12}
\end{align}

\section{The exact GF as a function of effective cumulants.}

\label{ApD}

\subsection{Rules for reciprocal space and imaginary frequencies (Valid for
the PAM)}

\label{ApDS1}

In Section \ref{S02_4} we use the exact solution of the model with zero
band-width and local hybridization to approximate the effective cumulants
that appear in the exact GF. Here we shall use the prescriptions given in
Rule 3.7 of Section \ref{S2} to calculate the diagrams in imaginary
frequency and reciprocal space, to obtain the formal expression of the exact
GF
\begin{equation}
\left\langle \left( \hat{Y}(f;\mathbf{k},\alpha,u=-,\omega_{j})\hat {Y}(f;%
\mathbf{k}^{\prime},\alpha^{\prime},u^{\prime},\omega_{j}^{\prime })\right)
_{+}\right\rangle _{\mathcal{H}}=\mathcal{G}_{\alpha\alpha^{\prime }}^{ff}(%
\mathbf{k},i\omega_{j})\ \Delta\left( u+u^{\prime}\right) \Delta\left( u%
\mathbf{k}+u^{\prime}\mathbf{k}^{\prime}\right) \Delta\left(
\omega_{j}+\omega_{j}^{\prime}\right)  \label{ApD1}
\end{equation}
in terms of the corresponding effective cumulants, where $\omega_{j}$ and $%
\omega_{j}^{\prime}$ are the Matsubara frequencies (cf. Eq.~{\ref{Fourier5}}%
). As with the Feynman diagrams, one can rearrange all the diagrams that
contribute to the exact $\mathcal{G}_{\alpha\alpha^{\prime}}^{ff}(\mathbf{k}%
,i\omega_{j})$ by introducing effective cumulants $M_{\alpha
\alpha^{\prime}}^{eff}(\mathbf{k},\omega_{j})$, defined by the contributions
of all the diagrams of $\mathcal{G}_{\alpha\alpha^{\prime}}^{ff}(\mathbf{k}%
,i\omega_{j})$ that can not be separated by cutting a single edge (usually
called \textquotedblleft proper\textquotedblright\ or \textquotedblleft
irreducible\textquotedblright\ diagrams). The exact GF $\mathcal{G}%
_{\alpha\alpha^{\prime}}^{ff}(\mathbf{k},i\omega_{j})$ is then given by the
family of diagrams in reciprocal space corresponding to those given in
figure \ref{GffCHA}a for real space, but with the effective cumulants $%
M_{\alpha\alpha^{\prime}}^{eff}(\mathbf{k},\omega_{j})$ replacing the bare
cumulant $M_{\alpha\alpha^{\prime}}^{0}(\mathbf{k},\omega_{j})=-\delta
_{\alpha\alpha^{\prime}}\ D_{\alpha}/\left( i\omega_{j}+\varepsilon_{\alpha
}\right) $ at all the filled vertices (here $\varepsilon_{\alpha}=%
\varepsilon_{b}-\varepsilon_{a}$ when $\alpha=\left( b,a\right) $, cf.
appendix \ref{ApE}). To calculate the contribution of the diagram with $n+1$
effective cumulants we follow the steps in Rule 3.7:

\begin{description}
\item[1.] We label all the diagrams containing $n+1$ effective cumulants $%
M_{\alpha_{1}\alpha_{2}}^{eff}(\mathbf{k}_{1},\omega_{1},u_{1};\mathbf{k}%
_{2},\omega_{2},u_{2})$.
Because of Eq. (\ref{Fourier8}), this effective cumulant is proportional to $%
\Delta\left( \omega_{1}+\omega_{2}\right) $, and because of the $\delta$ in
Eq. (\ref{Fourier14}) it is also proportional to $\Delta\left( u_{1}\mathbf{k%
}_{1}+u_{2}\mathbf{k}_{2}\right) $. The particle conservation also requires
a $\Delta\left( u_{1}+u_{2}\right) $, and the labels we use are shown in
figure \ref{Fig2}.

\item We make the product of the following factors

\item[(a)] All the $\Delta\left( u+u^{\prime}\right) \Delta\left( u\mathbf{k}%
+u^{\prime}\mathbf{k}^{\prime}\right) \Delta\left( \omega
+\omega^{\prime}\right) $ that appear in Eq. (\ref{ApD1}) remain with the
effective cumulants, because they correspond to all the proper diagrams of $%
\mathcal{G}_{\alpha\alpha^{\prime}}^{ff}(\mathbf{k},i\omega)$. We then have
\begin{align}
& N_{s}\ \Delta\left( u^{\prime}\mathbf{k}^{\prime}-u_{1}\mathbf{k}%
_{1}\right) \Delta\left( u^{\prime}-u_{1}\right) \Delta\left(
\omega^{\prime}-\omega_{1}\right) \ M_{\alpha_{1}\alpha^{\prime}}^{eff}(%
\mathbf{k}_{1},-\omega_{1},-u_{1};\mathbf{k}^{\prime},\omega^{\prime
},u^{\prime})  \notag \\
\times & N_{s}\ \Delta\left( u_{1}^{\prime}\mathbf{k}_{1}^{\prime}-u_{2}%
\mathbf{k}_{2}\right) \Delta\left( u_{1}^{\prime}-u_{2}\right) \Delta\left(
\omega_{1}^{\prime}-\omega_{2}\right) \ M_{\alpha_{2}\alpha
_{1}^{\prime}}^{eff}(\mathbf{k}_{2},-\omega_{2},-u_{2};\mathbf{k}%
_{1}^{\prime },\omega_{1}^{\prime},u_{1}^{\prime})  \notag \\
\times & N_{s}\ \Delta\left( u_{2}^{\prime}\mathbf{k}_{2}^{\prime}-u_{3}%
\mathbf{k}_{3}\right) \Delta\left( u_{2}^{\prime}-u_{3}\right) \Delta\left(
\omega_{2}^{\prime}-\omega_{3}\right) \ M_{\alpha_{3}\alpha
_{2}^{\prime}}^{eff}(\mathbf{k}_{3},-\omega_{3},-u_{3};\mathbf{k}%
_{2}^{\prime },\omega_{2}^{\prime},u_{2}^{\prime})  \notag \\
& \vdots  \notag \\
\times & N_{s}\ \Delta\left( u_{n-1}^{\prime}\mathbf{k}_{n-1}^{\prime}-u_{n}%
\mathbf{k}_{n}\right) \Delta\left( u_{n-1}^{\prime}-u_{n}\right)
\Delta\left( \omega_{n-1}^{\prime}-\omega_{n}\right) \
M_{\alpha_{n}\alpha_{n-1}^{\prime}}^{eff}(\mathbf{k}_{n},-\omega_{n},-u_{n};%
\mathbf{k}_{n-1}^{\prime},\omega_{n-1}^{\prime},u_{n-1}^{\prime})  \notag \\
\times & N_{s}\ \Delta\left( u_{n}^{\prime}\mathbf{k}_{n}^{\prime }+u\
\mathbf{k}\right) \Delta\left( u_{n}^{\prime}+u\right) \Delta\left(
\omega_{n}^{\prime}+\omega\right) \ M_{\alpha\alpha_{n}^{\prime}}^{eff}(%
\mathbf{k},\omega,u;\mathbf{k}_{n}^{\prime},\omega_{n}^{\prime},u_{n}^{%
\prime}).  \label{ApD2}
\end{align}

\item[(b)] The contribution of the $n$ cumulants of conduction electrons
\begin{align}
& \quad\ \frac{1}{i\omega_{1}+u_{1}\ \varepsilon\left( \mathbf{k}%
_{1},\sigma_{1}\right) }\ \delta(\mathbf{k}_{1},\mathbf{k}%
_{1}^{\prime})\delta(u_{1},u_{1}^{\prime})\delta(\sigma_{1},\sigma_{1}^{%
\prime})\delta(\omega_{1},\omega_{1}^{\prime})  \notag \\
& \vdots  \notag \\
& \times\frac{1}{i\omega_{n}+u_{n}\ \varepsilon\left( \mathbf{k}%
_{n},\sigma_{n}\right) }\ \delta(\mathbf{k}_{n},\mathbf{k}%
_{n}^{\prime})\delta(u_{n},u_{n}^{\prime})\delta(\sigma_{n},\sigma_{n}^{%
\prime})\delta(\omega_{n},\omega_{n}^{\prime}).  \label{ApD3}
\end{align}

\item[(c)] The contribution of the\ $2n$ interaction edges
\begin{align}
& \quad V\left( \alpha_{1}^{\prime},\mathbf{k}_{1}^{\prime},\sigma
_{1}^{\prime},u_{1}^{\prime}\right) V\left( \alpha_{1},\mathbf{k}%
_{1},\sigma_{1},-u_{1}\right)  \notag \\
& \vdots  \notag \\
& \times V\left( \alpha_{n}^{\prime},\mathbf{k}_{n}^{\prime},\sigma
_{n}^{\prime},u_{n}^{\prime}\right) V\left( \alpha_{n},\mathbf{k}%
_{n},\sigma_{n},-u_{n}\right) .  \label{ApD4}
\end{align}
There is a factor $\left( -1\right) $ for each interaction parameter $%
V\left( \alpha_{s},\mathbf{k}_{s},\sigma_{s},u_{s}\right) $, and it cancels
out for the $\mathcal{G}_{\alpha\alpha^{\prime}}^{ff}(\mathbf{k},i\omega)$,
and the $\mathcal{G}_{\alpha\alpha^{\prime}}^{cc}(\mathbf{k},i\omega_{s})$,
but for $\mathcal{G}_{\alpha\alpha^{\prime}}^{cf}(\mathbf{k},i\omega_{s})$,
and $\mathcal{G}_{\alpha\alpha^{\prime}}^{fc}(\mathbf{k},i\omega_{s})$, one
of these factors $\left( -1\right) $ remains and a change of sign is
necessary$^{\text{\ref{myfoot3}}}$. This sign is not necessary in Eq. (\ref%
{ApD2}) because it cancels out like in the $\mathcal{G}_{\alpha
\alpha^{\prime}}^{ff}(\mathbf{k},i\omega)$.
\begin{figure}[ptb]
\begin{center}
\includegraphics[width=6.9647in]{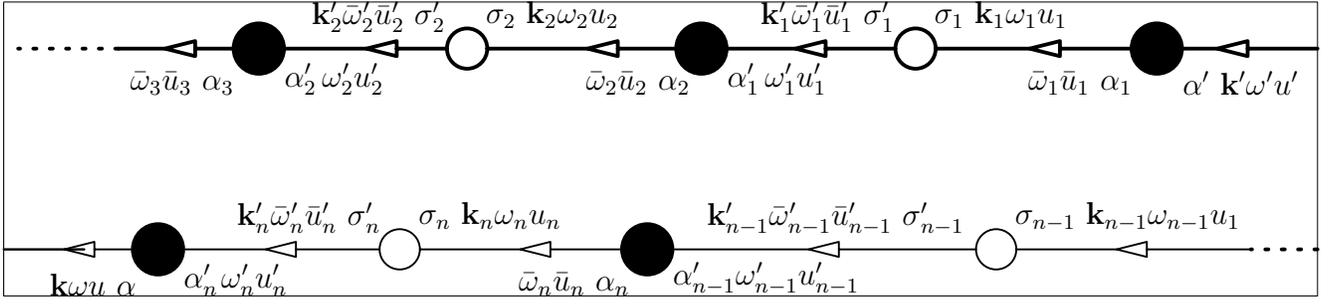}
\end{center}
\caption{Expansion of the exact GF employing effective cumulants. The figure
represents the collection of all the diagrams with $n+1$ effective
cumulants: we write $\mathbf{k}^{\prime},\protect\alpha^{\prime},u^{\prime},%
\protect\omega^{\prime }=\mathbf{k}_{0}^{\prime},\protect\alpha%
_{0}^{\prime},u_{0}^{\prime},\protect\omega _{0}^{\prime}$ and $\mathbf{k},%
\protect\alpha,u,\protect\omega=\mathbf{k}_{n+1},\protect\alpha _{n+1},\bar{u%
}_{n+1},\bar{\protect\omega}_{n+1}$}
\label{Fig2}
\end{figure}

\item[(d)] A factor $\pm1$ obtained employing the rules given in Appendix %
\ref{ApA}.

\item The graphs represented by figure \ref{Fig2} can be considered to be in
the perfect order of Rule C.1, and we can apply the first step of Rule C.1.
without any changes in sign. We assume that the contributions to the
effective cumulant $M_{\alpha\alpha^{\prime}}^{eff}(\mathbf{k},\omega_{s})$
have been already calculated with their correct sign, and we only have to
discuss the diagram with $n+1$ cumulants in figure \ref{Fig2} as a whole. As
discussed in Appendix \ref{ApA} is is not necessary to introduce any sign
change for all the CV joined by two internal lines, and only those joined by
only one internal line (and therefore joined also by an external line) have
to be considered. These CV appear only in the three GF of the type $\mathcal{%
G}_{\alpha\alpha^{\prime}}^{cc}(\mathbf{k},i\omega_{s})$, $\mathcal{G}%
_{\alpha\alpha^{\prime}}^{cf}(\mathbf{k},i\omega_{s})$, and $\mathcal{G}%
_{\alpha\alpha^{\prime}}^{fc}(\mathbf{k},i\omega_{s})$, and in disagreement
with previous results \cite{X-boson2}, no change of sign is required for
these three type of GF, so that in all cases we have only to multiply into

\item
\begin{equation}
+1.  \label{ApD5}
\end{equation}

\item[(e)] A factor $1/g$ calculated from appendix \ref{ApB}. We assume that
all the factors that appear in the contribution of the diagrams
corresponding to the effective cumulant $M_{\alpha\alpha^{\prime}}^{eff}(%
\mathbf{k},\omega_{s})$ have been already included in the $%
M_{\alpha\alpha^{\prime}}^{eff}(\mathbf{k},\omega_{s})$ itself, and we then
only need to use the $g$ corresponding to a chain, that is $g=1$. We can
then write
\begin{equation}
\frac{1}{g}=1.  \label{ApD6}
\end{equation}

\item[(f)] From the two FV external lines we have an extra factor
\begin{equation}
\frac{1}{N_{s}}.  \label{ApD7}
\end{equation}

\item[3.] We now have to sum the products with respect to:

\item[(a)] the momenta $\mathbf{k}_{s}$, the frequencies $\omega_{s}$ and
the indices $u_{s}$ of all the internal edges, and also divide each sum over
momenta into $\sqrt{N_{s}}$. This last contribution cancels exactly $n$
factors $N_{s}$,of all the $n+1$ factors $N_{s}$ that appear in Eq. (\ref%
{ApD2}). The extra factor $N_{s}$ is canceled by the spatial Fourier
transform of the external lines, and is taken care by the item 2.f of rule
3.7. When there is an external line running to a CV, there is a $1/\sqrt {%
N_{s}}$ associated to the internal line running to that CV, but there is no $%
1/\sqrt{N_{s}}$ associated to the external line, because the corresponding
operator has been already introduced in $k$ space.

\item Because of the delta functions in Eqs. (\ref{ApD2},\ref{ApD3}) we have
\begin{equation}
u^{\imath}=u_{s}=u_{s}^{\prime}=-u,  \label{ApD8}
\end{equation}
\begin{equation}
\mathbf{k}^{\prime}\mathbf{=k}_{s}=\mathbf{k}_{s}^{\prime}=\mathbf{k,}
\label{ApD9}
\end{equation}
and the Matsubara frequencies
\begin{equation}
\omega^{\prime}=\omega_{s}=\omega_{s}^{\prime}=-\omega  \label{ApD10}
\end{equation}
for all $s=0,1,\ldots,n$.

\item[(b)] We have the sum over all $\alpha_{s},\alpha_{s}^{\prime}$, for $%
s=1,2,\ldots,n$ and we shall use matrix notation to simplify the calculation.

\item[(c)] Because of the $\delta(\sigma_{s},\sigma_{s}^{\prime})$ in Eq. (%
\ref{ApD3}) and the spin conservation in the effective cumulants when $\left[
\sigma_{z};\mathcal{H}\right] =0$ (even if $\alpha_{s}^{\prime}\neq%
\alpha_{s+1}$) there are no sums left over the labels $\sigma_{s}.$

\item We shall now consider the contribution of the factors in Eq. (\ref%
{ApD2}); employing Eqs. (\ref{ApD8}-\ref{ApD10}) we can write for $s=n$%
\begin{equation}
M_{\alpha\alpha_{n}^{\prime}}^{eff}(\mathbf{k},\omega=-\omega_{n},u=-u_{n};%
\mathbf{k}_{n}^{\prime},\omega_{n}^{\prime}=\omega_{n},u_{n}^{%
\prime}=u_{n})=M_{\alpha\alpha_{n}^{\prime}}^{eff}(\mathbf{k},\omega ,u;%
\mathbf{k},\omega_{n}=-\omega,u_{n}=-u);  \label{ApD10a}
\end{equation}
for $s=1,2,\ldots,n-1$%
\begin{equation}
M_{\alpha_{s+1}\alpha_{s}^{\prime}}^{eff}(\mathbf{k}_{s+1},-\omega
_{s+1},-u_{s+1};\mathbf{k}_{s}^{\prime},\omega_{s}^{\prime},u_{s}^{\prime
})=M_{\alpha_{s+1}\alpha_{s}^{\prime}}^{eff}(\mathbf{k},\omega,u;\mathbf{k}%
,-\omega,-u);  \label{ApD10b}
\end{equation}
and finally
\begin{equation*}
M_{\alpha_{1}\alpha^{\prime}}^{eff}(\mathbf{k}_{1},-\omega_{1},-u_{1};%
\mathbf{k}^{\prime},\omega^{\prime},u^{\prime})=M_{\alpha_{1}\alpha^{%
\prime}}^{eff}(\mathbf{k},\omega,u;\mathbf{k},-\omega,-u).
\end{equation*}
If we now define
\begin{equation}
M_{\alpha\alpha^{\prime}}^{eff}(\mathbf{k},i\omega,u)\equiv M_{\alpha
\alpha^{\prime}}^{eff}(\mathbf{k},\omega,u;\mathbf{k},-\omega ,-u)
\label{ApD10c}
\end{equation}
we can write the contribution of the $n+1$ factors in Eq.\ (\ref{ApD2}) in
the following form (cf. 3 (a) for the cancelation of $n$ factors $N_{s}$):
\begin{equation}
M_{\alpha\alpha_{n}^{\prime}}^{eff}(\mathbf{k},i\omega,u)M_{\alpha_{n}%
\alpha_{n-1}^{\prime}}^{eff}(\mathbf{k},i\omega,u)\ldots
M_{\alpha_{2}\alpha_{1}^{\prime}}^{eff}(\mathbf{k},i\omega,u)M_{\alpha_{1}%
\alpha^{\prime}}^{eff}(\mathbf{k},i\omega,u)  \label{ApD10d}
\end{equation}
\end{description}

We still have to include in the contribution of the diagram with $n+1$
effective cumulants all the factors from Eqs. (\ref{ApD3},\ref{ApD4}).
Employing Eq. (\ref{Fourier13}) and Eq. (\ref{Ger3}) we have
\begin{equation}
V(\alpha,\mathbf{k},\sigma,-u)=V^{\ast}(\alpha,\mathbf{k},\sigma,u)
\label{ApD11}
\end{equation}
and it is then convenient to write
\begin{align}
V(\alpha,\mathbf{k},\sigma) & \equiv V(\alpha,\mathbf{k},\sigma ,-)
\label{ApD12a} \\
V^{\ast}(\alpha,\mathbf{k},\sigma) & \equiv V(\alpha,\mathbf{k},\sigma,+)
\label{ApD12b}
\end{align}

As before we use Eqs. (\ref{ApD8},\ref{ApD9}) and the conservation of $%
\sigma $ to simplify Eq. (\ref{ApD4}):
\begin{equation}
V\left( \alpha _{s}^{\prime },\mathbf{k}_{s}^{\prime },\sigma _{s}^{\prime
},u_{s}^{\prime }\right) V\left( \alpha _{s},\mathbf{k}_{s},\sigma
_{s},-u_{s}\right) =V\left( \alpha _{s}^{\prime },\mathbf{k},\sigma
,u_{s}\right) V\left( \alpha _{s},\mathbf{k},\sigma ,-u_{s}\right) ,
\label{ApD13}
\end{equation}
and we combine 2 (b),(c) introducing $\tilde{W}_{\alpha ,\alpha ^{\prime
}}\left( \mathbf{k},\sigma ,u;i\omega _{s}\right) $%
\begin{equation}
\tilde{W}_{\alpha ^{\prime },\alpha }\left( \mathbf{k},\sigma ,u^{\prime
};i\omega \right) =V\left( \alpha ^{\prime },\mathbf{k},\sigma ,u^{\prime
}\right) V\left( \alpha ,\mathbf{k},\sigma ,-u^{\prime }\right) \frac{1}{%
i\omega +u^{\prime }\ \varepsilon \left( \mathbf{k},\sigma \right) }.
\label{ApD14}
\end{equation}
Employing Eqs.(\ref{ApD12a},\ref{ApD12b}) we then obtain
\begin{equation}
\tilde{W}_{\alpha ^{\prime },\alpha }\left( \mathbf{k},\sigma ,u^{\prime
}=+;i\omega _{s}\right) =V^{\ast }(\alpha ^{\prime },\mathbf{k},\sigma
)V(\alpha ,\mathbf{k},\sigma )\frac{1}{i\omega _{s}+\ \varepsilon \left(
\mathbf{k},\sigma \right) }  \label{ApD15}
\end{equation}
and
\begin{equation}
\tilde{W}_{\alpha ^{\prime },\alpha }\left( \mathbf{k},\sigma ,u^{\prime
}=-;i\omega _{s}\right) =V(\alpha ^{\prime },\mathbf{k},\sigma )V^{\ast
}(\alpha ,\mathbf{k},\sigma )\frac{1}{i\omega _{s}-\ \varepsilon \left(
\mathbf{k},\sigma \right) }.  \label{ApD16}
\end{equation}

We return to the GF $\mathcal{G}_{\alpha \alpha ^{\prime }}^{ff}(\mathbf{k}%
,\omega )$\ defined in Eq. (\ref{ApD1}), which corresponds to $u^{\prime }=+$%
, so that the factors from Eqs. (\ref{ApD3},\ref{ApD4}) could be put then in
the form (cf. the labels in figure \ref{Fig2})
\begin{equation}
\tilde{W}_{\alpha _{n}^{\prime },\alpha _{n}}\left( \mathbf{k},\sigma
,+;i\omega _{n}\right) \tilde{W}_{\alpha _{n-1}^{\prime },\alpha
_{n-1}}\left( \mathbf{k},\sigma ,+;i\omega _{n-1}\right) \ldots \tilde{W}%
_{\alpha _{2}^{\prime },\alpha _{2}}\left( \mathbf{k},\sigma ,+;i\omega
_{2}\right) \tilde{W}_{\alpha _{1}^{\prime },\alpha _{1}}\left( \mathbf{k}%
,\sigma ,+;i\omega _{1}\right) ,  \label{ApD17}
\end{equation}
We now introduce the $c$-electron free GF $\mathcal{G}_{c,\sigma }^{0}\left(
\mathbf{k},z_{n}\right) $ (cf. Eq. (\ref{Eq3.14}))
\begin{equation}
\mathcal{G}_{c,\sigma }^{0}\left( \mathbf{k},z\right) =\frac{-1}{%
z-\varepsilon \left( \mathbf{k},\sigma \right) },  \label{ApD18}
\end{equation}
and from Eq. (\ref{ApD10}) we have $\omega _{s}=-\omega $, so that
\begin{equation}
\frac{1}{i\omega _{s}+\ \varepsilon \left( \mathbf{k},\sigma \right) }=\frac{%
-1}{-i\omega _{s}-\varepsilon \left( \mathbf{k},\sigma \right) }=\mathcal{G}%
_{c,\sigma }^{0}\left( \mathbf{k},-i\omega _{s}\right) =\mathcal{G}%
_{c,\sigma }^{0}\left( \mathbf{k},i\omega \right) .  \label{ApD19}
\end{equation}
We can then substitute in Eq. (\ref{ApD15})
\begin{equation}
\tilde{W}_{\alpha _{s}^{\prime },\alpha _{s}}\left( \mathbf{k},\sigma
,+;i\omega _{s}\right) =V^{\ast }(\alpha _{s}^{\prime },\mathbf{k},\sigma
)V(\alpha _{s},\mathbf{k},\sigma )\ \mathcal{G}_{c,\sigma }^{0}\left(
\mathbf{k},i\omega \right)  \label{ApD20}
\end{equation}
and then, to calculate the contribution in Eq. (\ref{ApD17}), it is more
convenient to use the quantity
\begin{equation}
W_{\alpha ^{\prime },\alpha }\left( \mathbf{k},\sigma ,z\right) =V^{\ast
}(\alpha ^{\prime },\mathbf{k},\sigma )V(\alpha ,\mathbf{k},\sigma )\
\mathcal{G}_{c,\sigma }^{0}\left( \mathbf{k},z\right)  \label{ApD21}
\end{equation}
where now the complex variable $z$ takes the place of the Matsubara
frequency $i\omega $. The contribution in Eq. (\ref{ApD17}) takes the form
\begin{equation}
W_{\alpha _{n}^{\prime },\alpha _{n}}\left( \mathbf{k},\sigma ;z\right)
W_{\alpha _{n-1}^{\prime },\alpha _{n-1}}\left( \mathbf{k},\sigma ;z\right)
\ldots W_{\alpha _{2}^{\prime },\alpha _{2}}\left( \mathbf{k},\sigma
;z\right) W_{\alpha _{1}^{\prime },\alpha _{1}}\left( \mathbf{k},\sigma
;z\right) .  \label{ApD22}
\end{equation}

To simplify the calculation we now introduce the two matrices (cf. Section %
\ref{S02_4})
\begin{equation}
\left\{ \mathbf{M}\right\} _{_{\alpha,\alpha^{\prime}}}=M_{\alpha
\alpha^{\prime}}^{eff}(\mathbf{k},z,u),  \label{ApD23}
\end{equation}
and
\begin{equation}
\left\{ \mathbf{W}\right\} _{_{\alpha^{\prime},\alpha}}=W_{\alpha^{\prime
},\alpha}\left( \mathbf{k},\sigma,z\right) .  \label{ApD24}
\end{equation}
The contribution of the diagram with $n+1$ cumulants is then (using the
Einstein convention of sum of repeated subindexes)
\begin{align}
& \left( +1\right) \times\frac{1}{g}\times\left\{ \mathbf{M}\right\}
_{\alpha\alpha_{n}^{\prime}}\left\{ \mathbf{W}\right\} _{_{\alpha
_{n}^{\prime},\alpha_{n}}}\left\{ \mathbf{M}\right\}
_{\alpha_{n}\alpha_{n-1}^{\prime}}\left\{ \mathbf{W}\right\}
_{\alpha_{n-1}^{\prime },\alpha_{n-1}}\ldots\left\{ \mathbf{W}\right\}
_{\alpha_{2}^{\prime},\alpha_{2}}\left\{ \mathbf{M}\right\}
_{\alpha_{2}\alpha_{1}^{\prime}}\left\{ \mathbf{W}\right\}
_{_{\alpha_{1}^{\prime},\alpha_{1}}}\left\{ \mathbf{M}\right\}
_{\alpha_{1}\alpha^{\prime}}=  \notag \\
& \left\{ \left( \mathbf{M\cdot W}\right) ^{n}\cdot\mathbf{M}\right\}
_{\alpha,\alpha^{\prime}}=\left\{ \mathbf{M}\cdot\left( \mathbf{W\cdot M}%
\right) ^{n}\right\} _{\alpha,\alpha^{\prime}}  \label{ApD25}
\end{align}

\subsection{Rules for real space and imaginary frequencies (Valid for the
impurity)}

\label{ApDS2}

We shall now obtain the formal expression of the exact GF $\left\langle
\left( \hat{Y}(f;\mathbf{j},\alpha,u=-,\omega_{s})\hat{Y}(f;\mathbf{j}%
^{\prime},\alpha^{\prime},u^{\prime},\omega_{s}^{\prime})\right)
_{+}\right\rangle _{\mathcal{H}}$ in terms of effective cumulants for the
system (SIAM) with a single impurity at site $\mathbf{j}_{i}$, and we shall
use the prescriptions given in Rule 3.7a of Section \ref{S2} to calculate
the diagrams in imaginary frequency and real space. The diagrams are
topologically the same employed for the PAM, \ and we write
\begin{equation}
\mathcal{G}_{\alpha\alpha^{\prime}}^{ff}(\mathbf{j},i\omega_{s},u;\mathbf{j}%
^{\prime},i\omega_{s}^{\prime},u^{\prime})\equiv\left\langle \left( \hat {Y}%
(f;\mathbf{j},\alpha,u,\omega_{s})\hat{Y}(f;\mathbf{j}^{\prime},\alpha^{%
\prime},u^{\prime},\omega_{s}^{\prime})\right) _{+}\right\rangle _{\mathcal{H%
}},  \label{ApD26a}
\end{equation}
but as there are local $f$ states only at the site $\mathbf{j}_{i}$, we must
then have $\mathbf{j}^{\prime}=\mathbf{j=j}_{i}$, and we write

\begin{equation}
\mathcal{G}_{\alpha\alpha^{\prime}}^{ff}(\mathbf{j},i\omega_{s},u=-;\mathbf{j%
}^{\prime},i\omega_{s}^{\prime},u^{\prime})=\mathcal{G}_{\alpha\alpha^{%
\prime}}^{ff}(\mathbf{j}_{i},i\omega_{s})\ \Delta\left( u+u^{\prime}\right)
\ \delta\left( \mathbf{j}_{i},\mathbf{j}\right) \ \delta\left( \mathbf{j}%
_{i},\mathbf{j}^{\prime}\right) \Delta\left( \omega_{s}+\omega
_{s}^{\prime}\right) .  \label{ApD26b}
\end{equation}

As in the previous Section \ref{ApDS1} one can rearrange all the diagrams
that contribute to the exact $\mathcal{G}_{\alpha\alpha^{\prime}}^{ff}(%
\mathbf{j},i\omega_{s},u;\mathbf{j}^{\prime},i\omega_{s}^{\prime},u^{\prime
})$ by introducing effective cumulants $M_{\alpha\alpha^{\prime}}^{ff}(%
\mathbf{j},i\omega_{s},u;\mathbf{j}^{\prime},i\omega_{s}^{\prime
},u^{\prime})$, defined by the contributions of all the diagrams of $%
\mathcal{G}_{\alpha\alpha^{\prime}}^{ff}(\mathbf{j},i\omega_{s},u;\mathbf{j}%
^{\prime},i\omega_{s}^{\prime},u^{\prime})$ that can not be separated by
cutting a single edge (usually called \textquotedblleft
proper\textquotedblright\ or \textquotedblleft
irreducible\textquotedblright\ diagrams). The exact GF $\mathcal{G}%
_{\alpha\alpha^{\prime}}^{ff}(\mathbf{j},i\omega_{s},u;\mathbf{j}%
^{\prime},i\omega_{s}^{\prime},u^{\prime })$ is then given by the family of
diagrams in figure \ref{GffCHA}a, but with effective cumulants $%
M_{\alpha_{1}\alpha_{2}}^{eff}(\mathbf{j}_{1},\omega _{1},u_{1};\mathbf{j}%
_{2},\omega_{2},u_{2})$ replacing the bare GF $\mathcal{G}%
_{f,\alpha\alpha^{\prime}}^{0}(\mathbf{j}_{s},\omega_{s})$ at all the filled
vertices (as usual $\varepsilon_{\alpha}=\varepsilon_{b}-\varepsilon_{a}$
when $\alpha=\left( b,a\right) $, cf. appendix \ref{ApE}). To calculate the
contribution of the diagram with $n+1$ effective cumulants we follow the
steps in Rule 3.7a of Section \ref{S2.1}:

\begin{enumerate}
\item We label all the diagrams that appear in the expansion of the $%
\mathcal{G}_{\alpha_{1}\alpha_{2}}^{ff}(\mathbf{j}_{1},\omega_{1},u_{1};%
\mathbf{j}_{2},\omega_{2},u_{2})$ corresponding to the SIAM, and in figure %
\ref{Fig3} we show the diagram that contains just $n+1$ effective cumulants $%
M_{\alpha_{1}\alpha_{2}}^{eff}(\mathbf{j}_{1},\omega_{1},u_{1};\mathbf{j}%
_{2},\omega_{2},u_{2})$. Because of Eq. (\ref{Fourier8}), this effective
cumulant is proportional to $\Delta\left( \omega_{1}+\omega_{2}\right) $,
the particle conservation requires a $\Delta\left( u_{1}+u_{2}\right) $ and
for the case of an impurity at site $\mathbf{j}_{i}$ the contribution is
also proportional to $\delta\left( \mathbf{j}_{1},\mathbf{j}_{i}\right) \
\delta\left( \mathbf{j}_{2},\mathbf{j}_{i}\right) $, because there are $f$
states only at that site. We shall then use the GF $G_{\alpha\alpha^{%
\prime}}^{ff}(\mathbf{j}_{i},i\omega_{s})$ defined in Eq. (\ref{ApD26b}).

\item We make the product of the following factors

\begin{description}
\item[(a)] All the $\Delta\left( u+u^{\prime}\right) \delta\left( \mathbf{j}%
_{i},\mathbf{j}\right) \ \delta\left( \mathbf{j}_{i},\mathbf{j}%
^{\prime}\right) \Delta\left( \omega+\omega^{\prime}\right) $ that appear in
Eq. (\ref{ApD26b}) remain with the effective cumulants, because they appear
in the contributions of all the proper diagrams of $\mathcal{G}%
_{\alpha\alpha^{\prime}}^{ff}(\mathbf{j}_{i},i\omega)$. We then have
\begin{align}
& \ \Delta\left( u^{\prime}-u_{1}\right) \ \delta\left( \mathbf{j}_{i},%
\mathbf{j}_{1}\right) \delta\left( \mathbf{j}_{i},\mathbf{j}^{\prime
}\right) \Delta\left( \omega^{\prime}-\omega_{1}\right) \ M_{\alpha
_{1}\alpha^{\prime}}^{eff}(\mathbf{j}_{1},-\omega_{1},-u_{1};\mathbf{j}%
^{\prime},\omega^{\prime},u^{\prime})  \notag \\
\times & \Delta\left( u_{1}^{\prime}-u_{2}\right) \ \delta\left( \mathbf{j}%
_{i},\mathbf{j}_{2}\right) \delta\left( \mathbf{j}_{i},\mathbf{j}%
_{1}^{\prime}\right) \Delta\left( \omega_{1}^{\prime}-\omega _{2}\right) \
M_{\alpha_{2}\alpha_{1}^{\prime}}^{eff}(\mathbf{j}_{2},-\omega_{2},-u_{2};%
\mathbf{j}_{1}^{\prime},\omega_{1}^{\prime},u_{1}^{\prime })  \notag \\
\times & \ \Delta\left( u_{2}^{\prime}-u_{3}\right) \ \delta\left( \mathbf{j}%
_{i},\mathbf{j}_{3}\right) \delta\left( \mathbf{j}_{i},\mathbf{j}%
_{2}^{\prime}\right) \Delta\left( \omega_{2}^{\prime}-\omega _{3}\right) \
M_{\alpha_{3}\alpha_{2}^{\prime}}^{eff}(\mathbf{j}_{3},-\omega_{3},-u_{3};%
\mathbf{j}_{2}^{\prime},\omega_{2}^{\prime},u_{2}^{\prime })  \notag \\
& \vdots  \notag \\
\times & \ \Delta\left( u_{n-1}^{\prime}-u_{n}\right) \ \delta\left( \mathbf{%
j}_{i},\mathbf{j}_{n}\right) \delta\left( \mathbf{j}_{i},\mathbf{j}%
_{n-1}^{\prime}\right) \Delta\left( \omega_{n-1}^{\prime}-\omega_{n}\right)
\ M_{\alpha_{_{n}}\alpha_{n-1}^{\prime}}^{eff}(\mathbf{j}_{n},-%
\omega_{n},-u_{n};\mathbf{j}_{n-1}^{\prime},\omega
_{n-1}^{\prime},u_{n-1}^{\prime})  \notag \\
\times & \ \Delta\left( u_{n}^{\prime}+u\right) \ \delta\left( \mathbf{j}%
_{i},\mathbf{j}\right) \delta\left( \mathbf{j}_{i},\mathbf{j}%
_{n}^{\prime}\right) \Delta\left( \omega_{n}^{\prime}+\omega\right) \
M_{\alpha\alpha_{n}^{\prime}}^{eff}(\mathbf{j},\omega,u;\mathbf{j}%
_{n}^{\prime},\omega_{n}^{\prime},u_{n}^{\prime}).  \label{ApD27}
\end{align}

\item[(b)] The contribution of the $n$ cumulants of conduction electrons
\begin{align}
& \quad\ \frac{1}{i\omega_{1}+u_{1}\ \varepsilon\left( \mathbf{k}%
_{1},\sigma_{1}\right) }\ \delta(\mathbf{k}_{1},\mathbf{k}%
_{1}^{\prime})\delta(u_{1},u_{1}^{\prime})\delta(\sigma_{1},\sigma_{1}^{%
\prime})\delta(\omega_{1},\omega_{1}^{\prime})  \notag \\
& \vdots  \notag \\
& \times\frac{1}{i\omega_{n}+u_{n}\ \varepsilon\left( \mathbf{k}%
_{n},\sigma_{n}\right) }\ \delta(\mathbf{k}_{n},\mathbf{k}%
_{n}^{\prime})\delta(u_{n},u_{n}^{\prime})\delta(\sigma_{n},\sigma_{n}^{%
\prime})\delta(\omega_{n},\omega_{n}^{\prime}).  \label{ApD28}
\end{align}

\item[(c)] The contribution of the\ $2n$ interaction edges
\begin{align}
& \quad v\left( \mathbf{j}_{1}^{\prime},\alpha_{1}^{\prime},\mathbf{k}%
_{1}^{\prime},\sigma_{1}^{\prime},u_{1}^{\prime}\right) v\left( \mathbf{j}%
_{1},\alpha_{1},\mathbf{k}_{1},\sigma_{1},-u_{1}\right)  \notag \\
& \vdots  \notag \\
& \times v\left( \mathbf{j}_{n}^{\prime},\alpha_{n}^{\prime},\mathbf{k}%
_{n}^{\prime},\sigma_{n}^{\prime},u_{n}^{\prime}\right) v\left( \mathbf{j}%
_{n},\alpha_{n},\mathbf{k}_{n},\sigma_{n},-u_{n}\right) .  \label{ApD29}
\end{align}
As in Eq. (\ref{ApD4}) there is a factor $\left( -1\right) $ for each
interaction parameter$^{\text{\ref{myfoot3}}}$ $v\left( \mathbf{j}%
_{s},\alpha_{s},\mathbf{k}_{s},\sigma_{s},u_{s}\right) $,and they all cancel
out in pairs for the $\mathcal{G}_{\alpha\alpha^{\prime}}^{ff}(\mathbf{k}%
,i\omega)$, and the $\mathcal{G}_{\alpha\alpha^{\prime}}^{cc}(\mathbf{k}%
,i\omega_{s})$, but for $\mathcal{G}_{\alpha\alpha^{\prime}}^{cf}(\mathbf{k}%
,i\omega_{s})$, and $\mathcal{G}_{\alpha\alpha^{\prime}}^{fc}(\mathbf{k}%
,i\omega_{s})$, one of these factors $\left( -1\right) $ remains and a
change of sign is necessary. This sign is not necessary in Eq. (\ref{ApD27})
because is cancels out like in the $\mathcal{G}_{\alpha
\alpha^{\prime}}^{ff}(\mathbf{k},i\omega)$.

\item[(d)] The two external lines correspond to the same site $\mathbf{j}%
_{i} $ because this is the only site with local $f$ states, so that we have $%
\delta(\mathbf{j,j}_{i})\delta(\mathbf{j}^{\prime}\mathbf{,j}_{i})$.

\item[(e)] A factor $\pm1$ obtained employing the rules given in Appendix %
\ref{ApA}, and the same arguments used in Section \ref{ApDS1} can be applied
here. In particular, we assume that the contributions to the effective
cumulants $M_{\alpha\alpha^{\prime}}^{eff}(\mathbf{j},\omega_{s})$ have been
already calculated with their correct sign. The factor is therefore
\begin{equation}
+1.  \label{ApD30}
\end{equation}

\item[(f)] A factor $1/g$ calculated from appendix \ref{ApB}. We assume that
all the factors that appear in the contribution of all the diagrams
corresponding to the effective cumulants have been already included in those
cumulants, and we then only need to calculate the $g$ corresponding to a
chain, that is $g=1$. We can then write
\begin{equation}
\frac{1}{g}=1.  \label{ApD31}
\end{equation}
\end{description}

\begin{figure}[ptb]
\begin{center}
\includegraphics[width=15.3264cm]{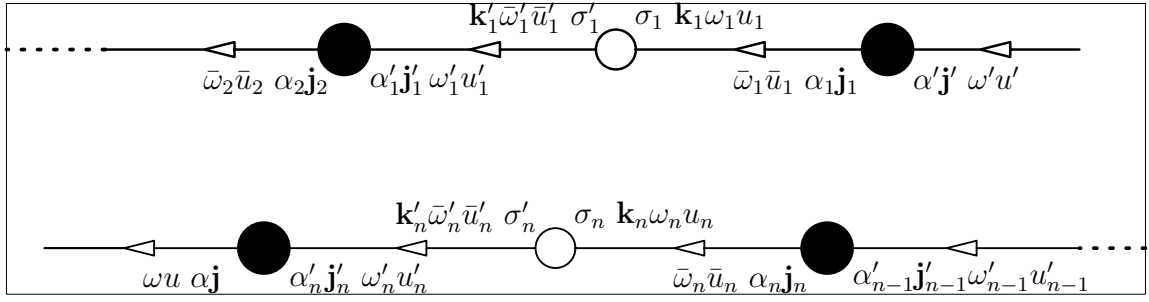}
\end{center}
\caption{GF diagrams in real space and imaginary frequency of the PAM, with $%
n+1$ effective cumulants: we write $\mathbf{j}^{\prime},\protect\alpha%
^{\prime },u^{\prime},\protect\omega^{\prime}=\mathbf{j}_{0}^{\prime},%
\protect\alpha_{0}^{\prime},u_{0}^{\prime},\protect\omega_{0}^{\prime}$ and $%
\mathbf{j},\protect\alpha,u,\protect\omega =\mathbf{j}_{n+1},\protect\alpha%
_{n+1},\bar{u}_{n+1},\bar{\protect\omega}_{n+1}$. The same diagrams describe
a single Anderson impurity (SIAM) at site $\mathbf{j}_{i}$ when there are
local $f$ states only at this site.}
\label{Fig3}
\end{figure}

\item Sum the resulting product with respect to
\end{enumerate}

\begin{description}
\item[ (a)] The site labels $\mathbf{j}_{s}$ of all the FV for $s=0,1,\ldots
,n$, but all these sums disappear because there is only one site $\mathbf{j}%
_{i}$ with local $f$ states:
\end{description}

\begin{equation}
\delta(\mathbf{j}_{i},\mathbf{j}_{s+1})\delta(\mathbf{j}_{i},\mathbf{j}%
_{s}^{\prime})  \label{ApD33}
\end{equation}
with $\mathbf{j}_{0}^{\prime}=\mathbf{j}^{\prime}$ and $\mathbf{j}_{n+1}=%
\mathbf{j}$.

\begin{description}
\item[ (b)] The momenta $\mathbf{k}_{s}$, the frequencies $\omega_{s}$ and
the indices $u_{s}$ of all the internal edges. Because of the delta
functions in Eqs. (\ref{ApD27},\ref{ApD28}) we have
\begin{equation}
u^{\imath}=u_{s}=u_{s}^{\prime}=-u,  \label{ApD32}
\end{equation}
\begin{equation}
\ \delta(\mathbf{k}_{s},\mathbf{k}_{s}^{\prime}),  \label{ApD32a}
\end{equation}
and
\begin{equation}
\omega^{\prime}=\omega_{s}=\omega_{s}^{\prime}=-\omega,  \label{ApD34}
\end{equation}
for $s=0,1,2,\ldots,n$.

\item Notice that for real space, the sum over momenta $\mathbf{k}_{s}$ does
not reduce to a single term, but there is a summation left at each CV, and
we shall discuss this summation at a later stage, because we have to
consider the dependence with $\mathbf{k}_{s}$ of the factors in Eqs. (\ref%
{ApD28},\ref{ApD29}).

\item[ (c)] We have the sum over all $\alpha_{s},\alpha_{s}^{\prime}$, for $%
s=1,2,\ldots,n$ and we shall use matrix notation to simplify the
calculation..

\item[ (d)] Because of the $\delta(\sigma_{s},\sigma_{s}^{\prime})$ in Eq. (%
\ref{ApD28}) and the spin conservation in the effective cumulants when $%
\left[ \sigma_{z};\mathcal{H}\right] =0$ (even if $\alpha_{s}^{\prime}\neq%
\alpha_{s+1}$) there are no sums left over the labels $\sigma_{s}.$
\end{description}

We shall now consider the contribution of the factors in Eq. (\ref{ApD27});
employing Eqs. (\ref{ApD32}-\ref{ApD34}) we can write

\begin{equation}
M_{\alpha\alpha_{n}^{\prime}}^{eff}(\mathbf{j},\omega=-\omega_{n},u=-u_{n};%
\mathbf{j}_{n}^{\prime},\omega_{n}^{\prime}=\omega_{n},u_{n}^{%
\prime}=u_{n})=M_{\alpha\alpha_{n}^{\prime}}^{eff}(\mathbf{j}_{i},\omega,u;%
\mathbf{j}_{i},\omega_{n}=-\omega,u_{n}=-u),  \label{ApD35}
\end{equation}
\begin{equation}
M_{\alpha_{1}\alpha^{\prime}}^{eff}(\mathbf{j}_{1},-\omega_{1},-u_{1};%
\mathbf{j}^{\prime},\omega^{\prime},u^{\prime})=M_{\alpha_{1}\alpha^{%
\prime}}^{eff}(\mathbf{j}_{1},\omega,u;\mathbf{j}_{1},-\omega,-u).
\label{ApD36}
\end{equation}
and
\begin{equation}
M_{\alpha_{s+1}\alpha_{s}^{\prime}}^{eff}(\mathbf{j}_{s+1},-\omega
_{s+1},-u_{s+1};\mathbf{j}_{s}^{\prime},\omega_{s}^{\prime},u_{s}^{\prime
})=M_{\alpha_{s+1}\alpha_{s}^{\prime}}^{eff}(\mathbf{j}_{s+1},\omega ,u;%
\mathbf{j}_{s+1},-\omega,-u),  \label{ApD37}
\end{equation}
for $s=1,2,\ldots,n-1$. If we now define
\begin{equation}
M_{\alpha\alpha^{\prime}}^{eff}(\mathbf{j},i\omega,u)\equiv M_{\alpha
\alpha^{\prime}}^{eff}(\mathbf{j},\omega,u;\mathbf{j},-\omega,-u)
\label{ApD38}
\end{equation}
(recall that in Eq. (\ref{ApD26b}) it is $u=-1$) we can write the
contribution of the $n+1$ factors in Eq.\ (\ref{ApD27}) in the following
form :
\begin{equation}
M_{\alpha\alpha_{n}^{\prime}}^{eff}(\mathbf{j}_{i},i\omega,u)M_{\alpha
_{n}\alpha_{n-1}^{\prime}}^{eff}(\mathbf{j}_{i},i\omega,u)\ldots M_{\alpha
_{2}\alpha_{1}^{\prime}}^{eff}(\mathbf{j}_{i},i\omega,u)M_{\alpha_{1}%
\alpha^{\prime}}^{eff}(\mathbf{j}_{i},i\omega,u).  \label{ApD39}
\end{equation}
All these factors are independent of all the $\mathbf{k}_{s}$ and $\mathbf{k}%
_{s}^{\prime}$, and can therefore be factored out from the summations over
these wave vectors.

We still have to include all the factors from Eqs. (\ref{ApD28},\ref{ApD29})
in the contribution of the diagram with $n+1$ effective cumulants. We
calculate $v\left( \mathbf{j}_{s},\alpha_{s},\mathbf{k}_{s},%
\sigma_{s},-u_{s}\right) v\left( \mathbf{j}_{s}^{\prime},\alpha_{s}^{\prime
},\mathbf{k}_{s}^{\prime},\sigma_{s}^{\prime},u_{s}^{\prime}\right) $
employing Eqs. (\ref{Fourier13},\ref{ApD32},\ref{ApD32a}) and the
conservation of $\sigma_{s}=\sigma$%
\begin{equation}
v\left( \mathbf{j}_{s}^{\prime},\alpha_{s}^{\prime},\mathbf{k}_{s}^{\prime
},\sigma_{s}^{\prime},u_{s}^{\prime}\right) v\left( \mathbf{j}%
_{s},\alpha_{s},\mathbf{k}_{s},\sigma_{s},-u_{s}\right) =\frac{1}{N_{s}}%
V(\alpha_{s}^{\prime},\mathbf{k}_{s},\sigma,u_{s}^{\prime})V(\alpha _{s},%
\mathbf{k}_{s},\sigma,-u_{s}).  \label{ApD40}
\end{equation}
Combining Eqs. (\ref{ApD28},\ref{ApD29},\ref{ApD40}) with Eqs.(\ref{ApD11}-%
\ref{ApD15},\ref{ApD18}) we write

\begin{align}
\tilde{W}_{\alpha ^{\prime },\alpha }\left( \mathbf{k},\sigma ,u^{\prime
}=+;i\omega _{s}\right) & =\frac{1}{N_{s}}V^{\ast }(\alpha ^{\prime },%
\mathbf{k},\sigma )V(\alpha ,\mathbf{k},\sigma )\frac{1}{i\omega _{s}+\
\varepsilon \left( \mathbf{k},\sigma \right) }  \notag \\
& =\frac{1}{N_{s}}V^{\ast }(\alpha ^{\prime },\mathbf{k},\sigma )V(\alpha ,%
\mathbf{k},\sigma )\mathcal{G}_{c,\sigma }^{0}\left( \mathbf{k},-i\omega
_{s}\right)  \label{ApD15a}
\end{align}
in place of Eq. (\ref{ApD15}):

We return to the GF $\mathcal{G}_{\alpha \alpha ^{\prime }}^{ff}(\mathbf{j}%
,i\omega _{s})$\ defined in Eq. (\ref{ApD26b}), which corresponds to $%
u^{\prime }=+$, so that the factors from Eqs. (\ref{ApD28},\ref{ApD29})
could be put then in the form (cf. the labels in figure \ref{Fig3})
\begin{equation}
\tilde{W}_{\alpha _{n}^{\prime },\alpha _{n}}\left( \mathbf{k}_{n},\sigma
,+;i\omega _{n}\right) \tilde{W}_{\alpha _{n-1}^{\prime },\alpha
_{n-1}}\left( \mathbf{k}_{n-1},\sigma ,+;i\omega _{n-1}\right) \ldots \tilde{%
W}_{\alpha _{2}^{\prime },\alpha _{2}}\left( \mathbf{k}_{2},\sigma
,+;i\omega _{2}\right) \tilde{W}_{\alpha _{1}^{\prime },\alpha _{1}}\left(
\mathbf{k}_{1},\sigma ,+;i\omega _{1}\right) ,  \label{ApD17a}
\end{equation}
and still have to be summed over all the $\mathbf{k}_{s}$. Employing Eqs.(%
\ref{ApD18}-\ref{ApD21}) we introduce
\begin{equation}
W_{\alpha ^{\prime },\alpha }\left( \sigma ,z\right) =\frac{1}{N_{s}}\sum_{%
\mathbf{k}}V^{\ast }(\alpha ^{\prime },\mathbf{k},\sigma )V(\alpha ,\mathbf{k%
},\sigma )\ \mathcal{G}_{c,\sigma }^{0}\left( \mathbf{k},z\right) ,
\label{ApD21a}
\end{equation}
where we have taken the sum over $\mathbf{k}$, used again $\omega
_{s}=-\omega $, and employed $z$ in place of the Matsubara frequency $%
i\omega $. The sum over all $\mathbf{k}_{s}$ of the contribution in Eq. (\ref%
{ApD17a}) then becomes
\begin{equation}
W_{\alpha _{n}^{\prime },\alpha _{n}}\left( \sigma ;z\right) W_{\alpha
_{n-1}^{\prime },\alpha _{n-1}}\left( \sigma ;z\right) \ldots W_{\alpha
_{2}^{\prime },\alpha _{2}}\left( \sigma ;z\right) W_{\alpha _{1}^{\prime
},\alpha _{1}}\left( \sigma ;z\right) .  \label{ApD42}
\end{equation}

To simplify the calculation we now introduce again two matrices (cf. Section %
\ref{S02_4})
\begin{equation}
\left\{ \mathbf{M}\right\} _{_{\alpha,\alpha^{\prime}}}=M_{\alpha
\alpha^{\prime}}^{eff}(\mathbf{j}_{i},z,u),  \label{ApD43}
\end{equation}
and
\begin{equation}
\left\{ \mathbf{W}\right\} _{_{\alpha^{\prime},\alpha}}=W_{\alpha^{\prime
},\alpha}\left( \sigma,z\right) .  \label{ApD44}
\end{equation}
The contribution of the diagram with $n+1$ cumulants takes then the same
form of Eq. (\ref{ApD25}), but with Eqs. (\ref{ApD43},\ref{ApD44}) in place
of Eqs. (\ref{ApD23},\ref{ApD24}):
\begin{align}
& \left\{ \mathbf{M}\right\} _{\alpha\alpha_{n}^{\prime}}\left\{ \mathbf{W}%
\right\} _{_{\alpha_{n}^{\prime},\alpha_{n}}}\left\{ \mathbf{M}\right\}
_{\alpha_{n}\alpha_{n-1}^{\prime}}\left\{ \mathbf{W}\right\}
_{\alpha_{n-1}^{\prime},\alpha_{n-1}}\ldots\left\{ \mathbf{W}\right\}
_{\alpha_{2}^{\prime},\alpha_{2}}\left\{ \mathbf{M}\right\}
_{\alpha_{2}\alpha_{1}^{\prime}}\left\{ \mathbf{W}\right\} _{_{\alpha
_{1}^{\prime},\alpha_{1}}}\left\{ \mathbf{M}\right\}
_{\alpha_{1}\alpha^{\prime}}=  \notag \\
& \left\{ \left( \mathbf{M\cdot W}\right) ^{n}\cdot\mathbf{M}\right\}
_{\alpha,\alpha^{\prime}}=\left\{ \mathbf{M}\cdot\left( \mathbf{W\cdot M}%
\right) ^{n}\right\} _{\alpha,\alpha^{\prime}}  \label{ApD45}
\end{align}

\section{Calculation of approximate GF}

\label{ApE 0}

Here we define and give the formal expression in the Atomic Approximation of
the GF for the PAM and the SIAM, that were left out in Section \ref{S05_2}.

\subsection{The other approximate GF for the PAM\label{ApE1}}

\subsubsection{In reciprocal space and imaginary frequencies}

The approximate GF $\mathcal{G}_{\alpha \sigma ^{\prime }}^{fc}(\mathbf{k}%
,i\omega )$ in reciprocal space and imaginary time is defined by
\begin{equation}
\left\langle \left( Y\left( f;\mathbf{k,}\alpha ,u=-,\omega \right) \
Y\left( c;\mathbf{k}^{\prime }\mathbf{,}\sigma ^{\prime },u^{\prime },\omega
^{\prime }\right) \right) _{+}\right\rangle =\mathcal{G}_{\alpha \sigma
^{\prime }}^{fc}(\mathbf{k},i\omega )\ \Delta \left( u+u^{\prime }\right) \
\Delta \left( \omega +\omega ^{\prime }\right) \ \delta \left( \mathbf{k,k}%
^{\prime }\right) ,  \label{E5.14b}
\end{equation}
and employing the \textbf{Rule 3.7} in Section \ref{S2.0} we obtain
\begin{equation}
\mathcal{G}_{\alpha \sigma ^{\prime }}^{fc}(\mathbf{k},i\omega
)=-\sum\limits_{\alpha ^{\prime }}\mathcal{G}_{\alpha \alpha ^{\prime
}}^{ff}(\mathbf{k},i\omega )\ V(\alpha ^{\prime },\mathbf{k},\sigma ^{\prime
},u=+)\ \mathcal{G}_{c,\sigma ^{\prime }}^{0}\left( \mathbf{k},i\omega
\right) ,  \label{E5.15b}
\end{equation}
where
\begin{equation}
V(\alpha ^{\prime },\mathbf{k},\sigma ^{\prime },u=+)=V^{\ast }(\alpha
^{\prime },\mathbf{k},\sigma ^{\prime }).  \label{E5.16bbc}
\end{equation}
We now introduce a column vector $\mathbf{G}_{\sigma ^{\prime }}^{fc}(%
\mathbf{k},i\omega )$ so that
\begin{equation}
\mathbf{G}_{\sigma }^{fc,ap}(\mathbf{k},i\omega )=
\begin{pmatrix}
\mathcal{G}_{0\sigma ,\sigma }^{fc}(\mathbf{k},i\omega ) \\
\mathcal{G}_{-\sigma d,\sigma }^{fc}(\mathbf{k},i\omega )%
\end{pmatrix}
,  \label{E5.16ab}
\end{equation}
where we have changed the dummy variable $\sigma ^{\prime }$ into $\sigma $,
and we must remember that\ $\mathcal{G}_{0\sigma ,\bar{\sigma}}^{fc}=%
\mathcal{G}_{\bar{\sigma}d,\bar{\sigma}}^{fc}=0.$

Substituting Eqs. (\ref{E5.16bbc},\ref{E5.1},\ref{E5.2}) into Eq. (\ref%
{E5.15b}) we obtain
\begin{equation}
\mathbf{G}_{\uparrow }^{fc,ap}(\mathbf{k},i\omega )=-V^{\ast }\frac{\mathcal{%
G}_{c,\uparrow }^{0}\left( \mathbf{k},i\omega \right)
\begin{pmatrix}
m_{11}+m_{13} \\
m_{31}+m_{33}%
\end{pmatrix}
}{1-\left\vert V\right\vert ^{2}\mathcal{G}_{c,\uparrow }^{0}\left( \mathbf{k%
},i\omega \right) \left( m_{11}+m_{33}+m_{13}+m_{31}\right) }
\label{E5.16bbb}
\end{equation}
\begin{equation}
\mathbf{G}_{\downarrow }^{fc,ap}(\mathbf{k},i\omega )=-V^{\ast }\frac{%
\mathcal{G}_{c,\downarrow }^{0}\left( \mathbf{k},i\omega \right)
\begin{pmatrix}
m_{22}-m_{24} \\
m_{42}-m_{44}%
\end{pmatrix}
}{1-\left\vert V\right\vert ^{2}\mathcal{G}_{c,\downarrow }^{0}\left(
\mathbf{k},i\omega \right) \left( m_{22}+m_{44}-m_{24}-m_{42}\right) }
\label{E5.16cb}
\end{equation}

We define the approximate $\mathcal{G}_{\alpha \sigma ^{\prime }}^{cf}(%
\mathbf{k},i\omega )$ with
\begin{equation}
\left\langle \left( Y\left( c;\mathbf{k,}\sigma ,u=-,\omega \right) \
Y\left( f;\mathbf{k}^{\prime }\mathbf{,}\alpha ^{\prime },u^{\prime },\omega
^{\prime }\right) \right) _{+}\right\rangle =\mathcal{G}_{\sigma \alpha
^{\prime }}^{cf}(\mathbf{k},i\omega )\ \Delta \left( u+u^{\prime }\right) \
\Delta \left( \omega +\omega ^{\prime }\right) \ \delta \left( \mathbf{k,k}%
^{\prime }\right) ,  \label{E5.17b}
\end{equation}
and from the \textbf{Rule 3.7} in Section \ref{S2.0} we obtain
\begin{equation}
\mathcal{G}_{\sigma \alpha ^{\prime }}^{cf}(\mathbf{k},i\omega
)=-\sum\limits_{\alpha _{1}}\mathcal{G}_{c,\sigma }^{0}\left( \mathbf{k}%
,i\omega \right) V(\alpha _{1},\mathbf{k},\sigma ,u=-)\mathcal{G}_{\alpha
_{1},\alpha ^{\prime }}^{ff}(\mathbf{k},i\omega ),  \label{E5.18b}
\end{equation}
where
\begin{equation}
V(\alpha _{1},\mathbf{k},\sigma ,u=-)=V(\alpha _{1},\mathbf{k},\sigma ).
\label{E5.19bb}
\end{equation}
We now introduce a row vector $\mathbf{G}_{\sigma }^{cf,ap}(\mathbf{k}%
,i\omega )$ so that $\left\{ \mathbf{G}_{\sigma }^{cf,ap}(\mathbf{k},i\omega
)\right\} _{\alpha ^{\prime }}=\mathcal{G}_{\sigma \alpha ^{\prime }}^{cf}(%
\mathbf{k},i\omega )$, and then
\begin{equation}
\mathbf{G}_{\sigma }^{cf,ap}(\mathbf{k},i\omega )=
\begin{pmatrix}
\mathcal{G}_{\sigma ,0\sigma }^{cf}(\mathbf{k},i\omega ) & ,\mathcal{G}%
_{\sigma ,-\sigma d}^{cf}(\mathbf{k},i\omega )%
\end{pmatrix}
.  \label{E5.19ac}
\end{equation}
Substituting Eqs. (\ref{E5.19bb},\ref{E5.1},\ref{E5.2}) into Eq. (\ref%
{E5.18b}) we obtain
\begin{equation}
\mathbf{G}_{\uparrow }^{cf,ap}(\mathbf{k},i\omega )=-V\frac{\mathcal{G}%
_{c,\uparrow }^{0}\left( \mathbf{k},i\omega \right)
\begin{pmatrix}
m_{11}+m_{31} & ,m_{13}+m_{33}%
\end{pmatrix}
}{1-\left\vert V\right\vert ^{2}\mathcal{G}_{c,\uparrow }^{0}\left( \mathbf{k%
},i\omega \right) \left( m_{11}+m_{33}+m_{13}+m_{31}\right) }
\label{E5.19bc}
\end{equation}
\begin{equation}
\mathbf{G}_{\downarrow }^{cf,ap}(\mathbf{k},i\omega )=-V\frac{\mathcal{G}%
_{c,\downarrow }^{0}\left( \mathbf{k},i\omega \right)
\begin{pmatrix}
m_{22}-m_{42} & ,m_{24}-m_{44}%
\end{pmatrix}
}{1-\left\vert V\right\vert ^{2}\mathcal{G}_{c,\downarrow }^{0}\left(
\mathbf{k},i\omega \right) \left( m_{22}+m_{44}-m_{24}-m_{42}\right) }
\label{E5.19cc}
\end{equation}

Finally, we define the approximate $\mathcal{G}_{\sigma }^{cc}(\mathbf{k}%
,i\omega )$ with
\begin{equation}
\left\langle \left( Y\left( c;\mathbf{k,}\sigma ,u=-,\omega \right) \
Y\left( c;\mathbf{k}^{\prime }\mathbf{,}\sigma ^{\prime },u^{\prime },\omega
^{\prime }\right) \right) _{+}\right\rangle =\mathcal{G}_{\sigma }^{cc}(%
\mathbf{k},i\omega )\ \Delta \left( u+u^{\prime }\right) \ \Delta \left(
\omega +\omega ^{\prime }\right) \ \delta \left( \mathbf{k}^{\prime }\mathbf{%
,k}\right) \text{ }\delta \left( \sigma ,\sigma ^{\prime }\right) ,
\label{E20.c}
\end{equation}
and using \textbf{Rule 3.7} in Section \ref{S2.0}) we obtain
\begin{align}
\mathcal{G}_{\sigma }^{cc}(\mathbf{k},\mathbf{k}^{\prime },i\omega )& =%
\mathcal{G}_{c,\sigma }^{0}\left( \mathbf{k},i\omega \right) \times  \notag
\\
& \left\{ 1+\sum_{\alpha _{1},\alpha _{1}^{\prime }}V(\alpha _{1},\mathbf{k}%
,\sigma ,u=-)\mathcal{G}_{\alpha _{1}\alpha _{1}^{\prime }}^{ff}(\mathbf{k}%
,i\omega )V(\alpha _{1}^{\prime },\mathbf{k}^{\prime },\sigma ,u=+)\mathcal{G%
}_{c,\sigma }^{0}\left( \mathbf{k}^{\prime },i\omega \right) \right\} \delta
\left( \mathbf{k,k}^{\prime }\right) .  \label{E5.21cc}
\end{align}
We also introduce the scalar $\mathbf{G}_{\sigma }^{cc,ap}(\mathbf{k}%
,i\omega )$ so that
\begin{equation}
\mathbf{G}_{\sigma }^{cc,ap}(\mathbf{k},i\omega )=\mathcal{G}_{\sigma }^{cc}(%
\mathbf{k},\mathbf{k},i\omega ).  \label{5.21ba}
\end{equation}
Substituting Eqs. (\ref{E5.16bbc},\ref{E5.19bb},\ref{E5.1},\ref{E5.2}) into
Eq. (\ref{E5.21cc}) we obtain
\begin{eqnarray}
\mathbf{G}_{\uparrow }^{cc,ap}(\mathbf{k},i\omega ) &=&\mathcal{G}%
_{c,\uparrow }^{0}\left( \mathbf{k},i\omega \right) +\frac{\left\vert
V\right\vert ^{2}\mathcal{G}_{c,\uparrow }^{0}\left( \mathbf{k},i\omega
\right) \left( m_{11}+m_{33}+m_{13}+m_{31}\right) }{1-\left\vert
V\right\vert ^{2}\mathcal{G}_{c,\uparrow }^{0}\left( \mathbf{k},i\omega
\right) \left( m_{11}+m_{33}+m_{13}+m_{31}\right) }\mathcal{G}_{c,\uparrow
}^{0}\left( \mathbf{k},i\omega \right)  \notag \\
&=&\frac{\mathcal{G}_{c,\uparrow }^{0}\left( \mathbf{k},i\omega \right) }{%
1-\left\vert V\right\vert ^{2}\mathcal{G}_{c,\uparrow }^{0}\left( \mathbf{k}%
,i\omega \right) \left( m_{11}+m_{33}+m_{13}+m_{31}\right) },
\label{E5.21bb}
\end{eqnarray}
and

\begin{eqnarray}
\mathbf{G}_{\downarrow }^{cc,ap}(\mathbf{k},i\omega ) &=&\mathcal{G}%
_{c,\downarrow }^{0}\left( \mathbf{k},i\omega \right) +\frac{\left\vert
V\right\vert ^{2}\mathcal{G}_{c\downarrow }^{0}\left( \mathbf{k},i\omega
\right) \left( m_{22}+m_{44}-m_{24}-m_{42}\right) }{1-\left\vert
V\right\vert ^{2}\mathcal{G}_{c,\downarrow }^{0}\left( \mathbf{k},i\omega
\right) \left( m_{22}+m_{44}-m_{24}-m_{42}\right) }\mathcal{G}_{c,\downarrow
}^{0}\left( \mathbf{k},i\omega \right)  \notag \\
&=&\frac{\mathcal{G}_{c,\downarrow }^{0}\left( \mathbf{k},i\omega \right) }{%
1-\left\vert V\right\vert ^{2}\mathcal{G}_{c,\downarrow }^{0}\left( \mathbf{k%
},i\omega \right) \left( m_{22}+m_{44}-m_{24}-m_{42}\right) },
\label{E5.21bd}
\end{eqnarray}

\subsubsection{In real space and imaginary frequencies}

We follow the same procedure used in section \ref{S05_2a} to derive the GF
in real space when the $f$ electron is created and destroyed at the same
site. Considering again a rectangular conduction band we find $\mathbf{G}%
_{\sigma }^{fc}(i\omega )$ by integrating Eqs. (\ref{E5.16bbb}-\ref{E5.16cb}%
):

\begin{equation*}
\mathbf{G}_{\uparrow }^{fc,ap}(i\omega )=
\begin{pmatrix}
\mathcal{G}_{0\uparrow ,\uparrow }^{fc}(i\omega ) \\
\mathcal{G}_{\downarrow d,\uparrow }^{fc}(i\omega )%
\end{pmatrix}
\equiv -\frac{V^{\ast }}{N_{s}}\sum_{\mathbf{k}}\frac{\mathcal{G}%
_{c,\uparrow }^{0}\left( \mathbf{k},i\omega \right) }{1-\left\vert
V\right\vert ^{2}\mathcal{G}_{c,\uparrow }^{0}\left( \mathbf{k},i\omega
\right) M_{\uparrow }^{ff}}
\begin{pmatrix}
m_{11}+m_{13} \\
m_{31}+m_{33}%
\end{pmatrix}
,
\end{equation*}
so that
\begin{equation}
\mathbf{G}_{\uparrow }^{fc,ap}(i\omega )=-\frac{V^{\ast }}{2D}\ln \left(
\frac{A_{\sigma }(i\omega )+D-\mu }{A_{\sigma }(i\omega )-D-\mu }\right)
\begin{pmatrix}
m_{11}+m_{13} \\
m_{31}+m_{33}%
\end{pmatrix}
,  \label{E5.32b}
\end{equation}
where
\begin{equation}
A_{\sigma }(i\omega )=-i\omega -\left\vert V\right\vert ^{2}M_{\sigma }^{ff},
\label{E5.32c}
\end{equation}
and in the same way we obtain
\begin{equation}
\mathbf{G}_{\downarrow }^{fc,ap}(\mathbf{k},i\omega )=-\frac{V^{\ast }}{2D}%
\ln \left( \frac{A_{\sigma }(i\omega )+D-\mu }{A_{\sigma }(i\omega )-D-\mu }%
\right)
\begin{pmatrix}
m_{22}-m_{24} \\
m_{42}-m_{44}%
\end{pmatrix}
.  \label{E5.33b}
\end{equation}
In a similar way we obtain the $\mathbf{G}_{\sigma }^{cf,ap}(i\omega )=
\begin{pmatrix}
\mathcal{G}_{\sigma ,0\sigma }^{fc}(i\omega ) & ,\mathcal{G}_{\sigma ,\bar{%
\sigma}d}^{fc}(i\omega )%
\end{pmatrix}
$ by integrating Eqs. (\ref{E5.19bc},\ref{E5.19cc}):
\begin{equation}
\mathbf{G}_{\uparrow }^{cf,ap}(i\omega )=-\frac{V}{2D}\ln \left( \frac{%
A_{\uparrow }(i\omega )+D-\mu }{A_{\uparrow }(i\omega )-D-\mu }\right)
\begin{pmatrix}
m_{11}+m_{31} & ,m_{13}+m_{33}%
\end{pmatrix}
,  \label{E5.34b}
\end{equation}
and
\begin{equation}
\mathbf{G}_{\downarrow }^{cf,ap}(i\omega )=-\frac{V}{2D}\ln \left( \frac{%
A_{\downarrow }(i\omega )+D-\mu }{A_{\downarrow }(i\omega )-D-\mu }\right)
\begin{pmatrix}
m_{22}-m_{42} & ,m_{24}-m_{44}%
\end{pmatrix}
.  \label{E5.35b}
\end{equation}

To obtain $\mathbf{G}_{\sigma }^{cc}(i\omega )$ we integrate Eqs. (\ref%
{E5.21bb},\ref{E5.21bd}):

\begin{equation}
\mathbf{G}_{\sigma }^{cc}(i\omega )=\frac{1}{2D}\ln \left( \frac{A_{\sigma
}(i\omega )+D-\mu }{A_{\sigma }(i\omega )-D-\mu }\right) .  \label{E5.36b}
\end{equation}

Employing this relation, we can then write
\begin{equation}
\mathbf{G}_{\uparrow }^{fc,ap}(i\omega )=-V^{\ast }\ \mathbf{G}_{\uparrow
}^{cc,ap}(i\omega )\
\begin{pmatrix}
m_{11}+m_{13} \\
m_{31}+m_{33}%
\end{pmatrix}
,  \label{E5.37aa}
\end{equation}
\begin{equation}
\mathbf{G}_{\downarrow }^{fc,ap}(i\omega )=-V^{\ast }\ \mathbf{G}%
_{\downarrow }^{cc,ap}(i\omega )\
\begin{pmatrix}
m_{22}-m_{24} \\
m_{42}-m_{44}%
\end{pmatrix}
,  \label{E5.37bb}
\end{equation}
\begin{equation}
\mathbf{G}_{\uparrow }^{cf,ap}(i\omega )=-V\ \mathbf{G}_{\uparrow
}^{cc,ap}(i\omega )\
\begin{pmatrix}
m_{11}+m_{31} & ,m_{13}+m_{33}%
\end{pmatrix}
,  \label{E5.38aa}
\end{equation}
\begin{equation}
\mathbf{G}_{\downarrow }^{cf,ap}(i\omega )=-V\ \mathbf{G}_{\downarrow
}^{cc,ap}(i\omega )
\begin{pmatrix}
m_{22}-m_{42} & ,m_{24}-m_{44}%
\end{pmatrix}
,  \label{E5.38bb}
\end{equation}
\begin{equation}
\mathbf{G}_{\uparrow }^{ff}(i\omega )=\mathbf{M}_{\uparrow }^{ap}+\left\vert
V\right\vert ^{2}\mathbf{G}_{\uparrow }^{cc}(i\omega )\left[ \mathbf{M}%
_{\uparrow }^{ap}M_{\uparrow }^{ff}-
\begin{pmatrix}
1 & -1 \\
-1 & 1%
\end{pmatrix}
\Theta _{\uparrow }\right]  \label{E5.40aa}
\end{equation}
\begin{equation}
\mathbf{G}_{\downarrow }^{ff}(i\omega )=\mathbf{M}_{\downarrow
}^{ap}+\left\vert V\right\vert ^{2}\mathbf{G}_{\downarrow }^{cc}(i\omega )%
\left[ \mathbf{M}_{\downarrow }^{ap}M_{\downarrow }^{ff}-
\begin{pmatrix}
1 & 1 \\
1 & 1%
\end{pmatrix}
\Theta _{\downarrow }\right]  \label{E5.40bb}
\end{equation}

\subsubsection{Green's functions with the usual Fermi operators $f$ and $%
f^{\dagger }$.}

It is interesting to calculate the Gf of the usual Fermi operators, related
to the Hubbard operators through
\begin{equation}
f=X_{0\sigma }+\sigma \ X_{-\sigma d}  \label{E5.41}
\end{equation}
where for typographical convenienc we use $-\sigma $ in place of $\bar{\sigma%
}$. It is straightforward to obtain
\begin{equation}
\left\langle \left\langle f_{\sigma };f_{\sigma }^{\dagger }\right\rangle
\right\rangle _{z}\left( \mathbf{k}\right) =\frac{M_{\sigma }^{ff}}{%
1-\left\vert V\right\vert ^{2}\ \mathcal{G}_{c,\sigma }^{0}\left( \mathbf{k}%
,z\right) \ M_{\sigma }^{ff}}  \label{E5.30c}
\end{equation}
where we used Eqs. (\ref{E5.13g1},\ref{E5.13g2}).

In a symilar way we find
\begin{equation}
\left\langle \left\langle f_{\sigma };c_{\sigma }^{\dagger }(\mathbf{k}%
)\right\rangle \right\rangle _{z}=-V^{\ast }\frac{\mathcal{G}_{c,\sigma
}^{0}\left( \mathbf{k},z\right) M_{\sigma }^{ff}}{1-\left\vert V\right\vert
^{2}\ \mathcal{G}_{c,\sigma }^{0}\left( \mathbf{k},z\right) \ M_{\sigma
}^{ff}},  \label{E5.33c}
\end{equation}
\begin{equation}
\left\langle \left\langle c_{\sigma }(\mathbf{k});f_{\sigma }^{\dagger
}\right\rangle \right\rangle _{z}=-V\frac{\mathcal{G}_{c,\sigma }^{0}\left(
\mathbf{k},z\right) M_{\sigma }^{ff}}{1-\left\vert V\right\vert ^{2}\
\mathcal{G}_{c,\sigma }^{0}\left( \mathbf{k},z\right) \ M_{\sigma }^{ff}}.
\label{E5.34c}
\end{equation}
and
\begin{equation}
\left\langle \left\langle c_{\sigma }(\mathbf{k});c_{\sigma }^{\dagger }(%
\mathbf{k}^{\prime })\right\rangle \right\rangle _{z}=\delta \left( \mathbf{%
k,k}^{\prime }\right) \frac{\mathcal{G}_{c,\sigma }^{0}\left( \mathbf{k}%
^{\prime },z\right) }{1-\left\vert V\right\vert ^{2}\ \mathcal{G}_{c,\sigma
}^{0}\left( \mathbf{k},z\right) \ M_{\sigma }^{ff}}.  \label{E5.35c}
\end{equation}

As in the previous section we calculate the GF in imaginary frequency and
real space\ when the $f$ electron is created and destroyed at the same site
\begin{equation*}
\left\langle \left\langle f_{\sigma };f_{\sigma }^{\dagger }\right\rangle
\right\rangle _{z}=M_{\sigma }^{ff}\left[ 1+\left\vert V\right\vert
^{2}M_{\sigma }^{ff}\mathbf{G}_{\sigma }^{cc}(i\omega )\right] .
\end{equation*}
We also obtain

\begin{equation}
\left\langle \left\langle f_{\sigma };c_{\sigma }^{\dagger }\right\rangle
\right\rangle _{z}=-V^{\ast }\ \mathbf{G}_{\sigma }^{cc}(i\omega )\
M_{\sigma }^{ff},
\end{equation}
and
\begin{equation}
\left\langle \left\langle c_{\sigma };f_{\sigma }^{\dagger }\right\rangle
\right\rangle _{z}=-V\ \mathbf{G}_{\sigma }^{cc}(i\omega )\ M_{\sigma }^{ff}.
\end{equation}

The $\left\langle \left\langle c_{\sigma };c_{\sigma }^{\dagger
}\right\rangle \right\rangle _{z}$ is given by $\mathbf{G}_{\sigma }^{cc}(z)$
in Eq. (\ref{E5.36b})
\begin{equation}
\left\langle \left\langle c_{\sigma };c_{\sigma }^{\dagger }\right\rangle
\right\rangle _{z}=\frac{1}{2D}\ln \left( \frac{A_{\sigma }(z)+D-\mu }{%
A_{\sigma }(z)-D-\mu }\right) .
\end{equation}

\subsection{The other approximate GF for the SIAM\label{ApE2}}

\subsubsection{With $f$ electrons in real space, and imaginary frequencies}

The approximate GF $\mathcal{G}_{\alpha \sigma ^{\prime }}^{fc}(\mathbf{j}%
_{i},\mathbf{k}^{\prime },i\omega )$ with the impurity at $\mathbf{j}_{i}$
is defined by
\begin{equation}
\left\langle \left( Y\left( f;\mathbf{j,}\alpha ,u=-,\omega \right) \
Y\left( c;\mathbf{k}^{\prime }\mathbf{,}\sigma ^{\prime },u^{\prime },\omega
^{\prime }\right) \right) _{+}\right\rangle =\mathcal{G}_{\alpha \sigma
^{\prime }}^{fc}(\mathbf{j},\mathbf{k}^{\prime },i\omega )\ \Delta \left(
u+u^{\prime }\right) \ \Delta \left( \omega +\omega ^{\prime }\right) \
\delta \left( \mathbf{j,j}_{i}\right) ,  \label{E5.14}
\end{equation}
and employing the \textbf{Rule 3.7a} in Section \ref{S2.1} we obtain
\begin{equation}
\mathcal{G}_{\alpha \sigma ^{\prime }}^{fc}(\mathbf{j}_{i}=0,\mathbf{k}%
,i\omega )=-\sum\limits_{\alpha ^{\prime }}\mathcal{G}_{\alpha \alpha
^{\prime }}^{ff}(\mathbf{j}_{i}=0,i\omega )\ v(\mathbf{j}=0,\alpha ^{\prime
},\mathbf{k},\sigma ^{\prime },u=+)\ \mathcal{G}_{c,\sigma ^{\prime
}}^{0}\left( \mathbf{k},i\omega \right) ,  \label{E5.15}
\end{equation}
where
\begin{equation}
v(\mathbf{j}=0,\alpha ^{\prime },\mathbf{k},\sigma ^{\prime },u=+)=N_{s}^{-%
\frac{1}{2}}V^{\ast }(\alpha ^{\prime },\mathbf{k},\sigma ^{\prime }).
\label{E5.16}
\end{equation}
We now introduce a column vector $\mathbf{G}_{\sigma ^{\prime }}^{fc}(%
\mathbf{j}_{i}=0,\mathbf{k},i\omega )$ so that
\begin{equation}
\mathbf{G}_{\sigma }^{fc,ap}(\mathbf{j}_{i}=0,\mathbf{k},i\omega )=
\begin{pmatrix}
\mathcal{G}_{0\sigma ,\sigma }^{fc}(\mathbf{j}_{i}=0,\mathbf{k},i\omega ) \\
\mathcal{G}_{-\sigma d,\sigma }^{fc}(\mathbf{j}_{i}=0,\mathbf{k},i\omega )%
\end{pmatrix}
,  \label{E5.16a}
\end{equation}
where we have changed the dummy variable $\sigma ^{\prime }$ into $\sigma $,
and we must remember that\ $\mathcal{G}_{0\sigma ,\bar{\sigma}}^{fc}=%
\mathcal{G}_{\bar{\sigma}d,\bar{\sigma}}^{fc}=0.$

Substituting Eqs. (\ref{E5.16},\ref{E5.1},\ref{E5.2}) into Eq. (\ref{E5.15})
we obtain
\begin{equation}
\mathbf{G}_{\uparrow }^{fc,ap}(\mathbf{k},i\omega )=-\frac{V^{\ast }}{\sqrt{%
N_{s}}}\mathcal{G}_{c,\uparrow }^{0}\left( \mathbf{k},i\omega \right) \frac{%
\begin{pmatrix}
m_{11}+m_{13} \\
m_{31}+m_{33}%
\end{pmatrix}
}{1-\left\vert V\right\vert ^{2}\varphi _{\uparrow }(i\omega )\left(
m_{11}+m_{33}+m_{13}+m_{31}\right) }  \label{E5.16b}
\end{equation}
\begin{equation}
\mathbf{G}_{\downarrow }^{fc,ap}(\mathbf{k},i\omega )=-\frac{V^{\ast }}{%
\sqrt{N_{s}}}\mathcal{G}_{c,\downarrow }^{0}\left( \mathbf{k},i\omega
\right) \frac{%
\begin{pmatrix}
m_{22}-m_{24} \\
m_{42}-m_{44}%
\end{pmatrix}
}{1-\left\vert V\right\vert ^{2}\varphi _{\downarrow }(i\omega )\left(
m_{22}+m_{44}-m_{24}-m_{42}\right) }  \label{E5.16c}
\end{equation}

We define the approximate $\mathcal{G}_{\alpha \sigma ^{\prime }}^{cf}(%
\mathbf{k},\mathbf{j}_{i}^{\prime },i\omega )$ with
\begin{equation}
\left\langle \left( Y\left( c;\mathbf{k,}\sigma ,u=-,\omega \right) \
Y\left( f;\mathbf{j}^{\prime }\mathbf{,}\alpha ^{\prime },u^{\prime },\omega
^{\prime }\right) \right) _{+}\right\rangle =\mathcal{G}_{\sigma \alpha
^{\prime }}^{cf}(\mathbf{k},\mathbf{j}^{\prime },i\omega )\ \Delta \left(
u+u^{\prime }\right) \ \Delta \left( \omega +\omega ^{\prime }\right) \
\delta \left( \mathbf{j}^{\prime }\mathbf{,j}_{i}\right) ,  \label{E5.17}
\end{equation}
and from the \textbf{Rule 3.7a} in Section \ref{S2.1} we obtain
\begin{equation}
\mathcal{G}_{\sigma \alpha ^{\prime }}^{cf}(\mathbf{k},\mathbf{j}%
_{i}=0,i\omega )=-\sum\limits_{\alpha _{1}}\mathcal{G}_{c,\sigma }^{0}\left(
\mathbf{k},i\omega \right) v(\mathbf{j}=0,\alpha _{1},\mathbf{k},\sigma ,u=-)%
\mathcal{G}_{\alpha _{1},\alpha ^{\prime }}^{ff}(\mathbf{j}_{i}=0,i\omega ),
\label{E5.18}
\end{equation}
where
\begin{equation}
v(\mathbf{j}=0,\alpha _{1},\mathbf{k},\sigma ,u=-)=N_{s}^{-\frac{1}{2}%
}V(\alpha _{1},\mathbf{k},\sigma ).  \label{E5.19}
\end{equation}
We now introduce a row vector $\mathbf{G}_{\sigma }^{cf,ap}(\mathbf{k},%
\mathbf{j}_{i}=0,i\omega )$ so that $\left\{ \mathbf{G}_{\sigma }^{cf,ap}(%
\mathbf{k},\mathbf{j}_{i}=0,i\omega )\right\} _{\alpha ^{\prime }}=\mathcal{G%
}_{\sigma \alpha ^{\prime }}^{cf}(\mathbf{k},\mathbf{j}_{i}=0,i\omega )$,
and then
\begin{equation}
\mathbf{G}_{\sigma }^{cf,ap}(\mathbf{k},\mathbf{j}_{i}=0,i\omega )=
\begin{pmatrix}
\mathcal{G}_{\sigma ,0\sigma }^{cf}(\mathbf{k},\mathbf{j}_{i}=0,i\omega ) & ,%
\mathcal{G}_{\sigma ,-\sigma d}^{cf}(\mathbf{k},\mathbf{j}_{i}=0,i\omega )%
\end{pmatrix}
.  \label{E5.19a}
\end{equation}
Substituting Eqs. (\ref{E5.16},\ref{E5.1},\ref{E5.2}) into Eq. (\ref{E5.18})
we obtain
\begin{equation}
\mathbf{G}_{\uparrow }^{cf,ap}(\mathbf{k},i\omega )=-\frac{V}{\sqrt{N_{s}}}%
\mathcal{G}_{c,\uparrow }^{0}\left( \mathbf{k},i\omega \right) \frac{%
\begin{pmatrix}
m_{11}+m_{31} & ,m_{13}+m_{33}%
\end{pmatrix}
}{1-\left\vert V\right\vert ^{2}\varphi _{\uparrow }(i\omega )\left(
m_{11}+m_{33}+m_{13}+m_{31}\right) }  \label{E5.19b}
\end{equation}
\begin{equation}
\mathbf{G}_{\downarrow }^{cf,ap}(\mathbf{k},i\omega )=-\frac{V}{\sqrt{N_{s}}}%
\mathcal{G}_{c,\downarrow }^{0}\left( \mathbf{k},i\omega \right) \frac{%
\begin{pmatrix}
m_{22}-m_{42} & ,m_{24}-m_{44}%
\end{pmatrix}
}{1-\left\vert V\right\vert ^{2}\varphi _{\downarrow }(i\omega )\left(
m_{22}+m_{44}-m_{24}-m_{42}\right) }  \label{E5.19c}
\end{equation}

Finally, we define the approximate $\mathcal{G}_{\sigma }^{cc}(\mathbf{k},%
\mathbf{k}^{\prime },i\omega )$ with
\begin{equation}
\left\langle \left( Y\left( c;\mathbf{k,}\sigma ,u=-,\omega \right) \
Y\left( c;\mathbf{k}^{\prime }\mathbf{,}\sigma ^{\prime },u^{\prime },\omega
^{\prime }\right) \right) _{+}\right\rangle =\mathcal{G}_{\sigma }^{cc}(%
\mathbf{k},\mathbf{k}^{\prime },i\omega )\ \Delta \left( u+u^{\prime
}\right) \ \Delta \left( \omega +\omega ^{\prime }\right) \ \delta \left(
\mathbf{j}^{\prime }\mathbf{,j}_{i}\right) \text{ }\delta \left( \sigma
,\sigma ^{\prime }\right) ,  \label{E5.20}
\end{equation}
and using \textbf{Rule 3.7a} in Section \ref{S2.1}) we obtain
\begin{align}
\mathcal{G}_{\sigma }^{cc}(\mathbf{k},\mathbf{k}^{\prime },i\omega )& =%
\mathcal{G}_{c,\sigma }^{0}\left( \mathbf{k},i\omega \right) \times  \notag
\\
& \left\{ \delta \left( \mathbf{k,k}^{\prime }\right) +\sum_{\alpha
_{1},\alpha _{1}^{\prime }}v(\mathbf{j}=0,\alpha _{1},\mathbf{k},\sigma ,u=-)%
\mathcal{G}_{\alpha _{1}\alpha _{1}^{\prime }}^{ff}(\mathbf{j},i\omega )v(%
\mathbf{j}=0,\alpha _{1}^{\prime },\mathbf{k}^{\prime },\sigma ,u=+)\mathcal{%
G}_{c,\sigma }^{0}\left( \mathbf{k}^{\prime },i\omega \right) \right\} .
\label{E5.21}
\end{align}
We also introduce the scalar $\mathbf{G}_{\sigma }^{cc,ap}(\mathbf{k},%
\mathbf{k}^{\prime },i\omega )$ so that
\begin{equation}
\mathbf{G}_{\sigma }^{cc,ap}(\mathbf{k},\mathbf{k}^{\prime },i\omega )=%
\mathcal{G}_{\sigma }^{cc}(\mathbf{k},\mathbf{k}^{\prime },i\omega ).
\label{E5.21a}
\end{equation}
Substituting Eqs. (\ref{E5.16},\ref{E5.1},\ref{E5.2}) into Eq. (\ref{E5.21})
we obtain
\begin{equation}
\mathbf{G}_{\uparrow }^{cc,ap}(\mathbf{k},\mathbf{k}^{\prime },i\omega )=%
\mathcal{G}_{c,\uparrow }^{0}\left( \mathbf{k},i\omega \right) \delta \left(
\mathbf{k,k}^{\prime }\right) +\frac{\left\vert V\right\vert ^{2}}{N_{s}}%
\mathcal{G}_{c,\uparrow }^{0}\left( \mathbf{k},i\omega \right) \frac{\left(
m_{11}+m_{33}+m_{13}+m_{31}\right) }{1-\left\vert V\right\vert ^{2}\varphi
_{\uparrow }(i\omega )\left( m_{11}+m_{33}+m_{13}+m_{31}\right) }\mathcal{G}%
_{c,\uparrow }^{0}\left( \mathbf{k}^{\prime },i\omega \right)  \label{E5.21b}
\end{equation}

\begin{equation}
\mathbf{G}_{\downarrow }^{cc,ap}(\mathbf{k},\mathbf{k}^{\prime },i\omega )=%
\mathcal{G}_{c,\downarrow }^{0}\left( \mathbf{k},i\omega \right) \delta
\left( \mathbf{k,k}^{\prime }\right) +\frac{\left\vert V\right\vert ^{2}}{%
N_{s}}\mathcal{G}_{c,\downarrow }^{0}\left( \mathbf{k},i\omega \right) \frac{%
\left( mm_{22}+m_{44}-m_{24}-m_{42}\right) }{1-\left\vert V\right\vert
^{2}\varphi _{\downarrow }(i\omega )\left(
m_{22}+m_{44}-m_{24}-m_{42}\right) }\mathcal{G}_{c,\downarrow }^{0}\left(
\mathbf{k}^{\prime },i\omega \right)  \label{E5.21c}
\end{equation}

\subsubsection{Green's functions with the conduction electron in the Wannier
representation.}

In the impurity case, it is more convenient to use the GF with the
conduction electrons in the Wannier representation, localized at the
impurity site. To that purpose, we employ Eq. (\ref{Fourier9})
\begin{equation}
C_{j\sigma }^{\dagger }=\frac{1}{\sqrt{N_{s}}}\sum_{\mathbf{k}}\exp (-i%
\mathbf{k}\cdot \mathbf{R}_{j})C_{\mathbf{k}\sigma }^{\dagger }\qquad ,
\tag{($\rightarrow $23)}
\end{equation}
and as before we use Eqs. (\ref{E5.13g1},\ref{E5.13g2}): $M_{\uparrow
}^{ff}=\left( m_{11}+m_{33}+m_{13}+m_{31}\right) $, and $M_{\downarrow
}^{ff}=\left( m_{22}+m_{44}-m_{24}-m_{42}\right) .$

We now apply Eq. (\ref{Fourier9}) to Eq. (\ref{E5.16b}) with $\mathbf{R}%
_{i}=0$, and find
\begin{equation}
\mathbf{G}_{\uparrow }^{fc,ap}(i\omega )=%
\begin{pmatrix}
\mathcal{G}_{0\uparrow ,\uparrow }^{fc}(i\omega ) \\
\mathcal{G}_{\downarrow d,\uparrow }^{fc}(i\omega )%
\end{pmatrix}%
\equiv -\frac{V^{\ast }}{N_{s}}\sum_{\mathbf{k}}\mathcal{G}_{c,\uparrow
}^{0}\left( \mathbf{k},i\omega \right) \frac{%
\begin{pmatrix}
m_{11}+m_{13} \\
m_{31}+m_{33}%
\end{pmatrix}%
}{1-\left\vert V\right\vert ^{2}\varphi _{\uparrow }(i\omega )M_{\uparrow
}^{ff}}.  \label{E5.31}
\end{equation}%
Employing Eq. (\ref{E5.6a}) we obtain

\begin{equation}
\mathbf{G}_{\uparrow }^{fc,ap}(i\omega )=-V^{\ast }\frac{\varphi _{\uparrow
}(i\omega )}{1-\left\vert V\right\vert ^{2}\varphi _{\uparrow }(i\omega
)M_{\uparrow }^{ff}}%
\begin{pmatrix}
m_{11}+m_{13} \\
m_{31}+m_{33}%
\end{pmatrix}%
.
\end{equation}

and in a symilar way we find
\begin{equation}
\mathbf{G}_{\downarrow }^{fc,ap}(i\omega )=-V^{\ast }\frac{\varphi
_{\downarrow }(i\omega )}{1-\left\vert V\right\vert ^{2}\varphi _{\downarrow
}(i\omega )M_{\downarrow }^{ff}}%
\begin{pmatrix}
m_{22}-m_{24} \\
m_{42}-m_{44}%
\end{pmatrix}%
.  \label{E5.33}
\end{equation}%
To obtain the $\mathbf{G}_{\sigma }^{cf,ap}(i\omega )=%
\begin{pmatrix}
\mathcal{G}_{\sigma ,0\sigma }^{fc}(i\omega ) & ,\mathcal{G}_{\sigma ,\bar{%
\sigma}d}^{fc}(i\omega )%
\end{pmatrix}%
$ we employ $C_{j\sigma }=\frac{1}{\sqrt{N_{s}}}\sum_{\mathbf{k}}\exp (+i%
\mathbf{k}\cdot \mathbf{R}_{j})C_{\mathbf{k}\sigma }$, and in a similar way
we find for $\mathbf{R}_{i}=0$%
\begin{equation}
\mathbf{G}_{\uparrow }^{cf,ap}(i\omega )=-V\frac{\varphi _{\uparrow
}(i\omega )}{1-\left\vert V\right\vert ^{2}\varphi _{\uparrow }(i\omega
)M_{\uparrow }^{ff}}%
\begin{pmatrix}
m_{11}+m_{31} & ,m_{13}+m_{33}%
\end{pmatrix}%
,  \label{E5.34}
\end{equation}%
and
\begin{equation}
\mathbf{G}_{\downarrow }^{cf,ap}(i\omega )=-V\frac{\varphi _{\downarrow
}(i\omega )}{1-\left\vert V\right\vert ^{2}\varphi _{\downarrow }(i\omega
)M_{\downarrow }^{ff}}%
\begin{pmatrix}
m_{22}-m_{42} & ,m_{24}-m_{44}%
\end{pmatrix}%
.  \label{E5.35}
\end{equation}

The remaining $\mathbf{G}_{\sigma}^{cc,ap}(i\omega)$ are now easily obtained
\begin{equation}
\mathbf{G}_{\sigma}^{cc,ap}(i\omega)=\frac{\varphi_{\sigma}(z)}{1-\left\vert
V\right\vert ^{2}\varphi_{\sigma}(z)M_{\sigma}^{ff}}.  \label{E5.36}
\end{equation}
Employing this equation we can now write
\begin{equation}
\mathbf{G}_{\uparrow}^{fc,ap}(i\omega)=-V^{\ast}\ \mathbf{G}_{\uparrow
}^{cc,ap}(i\omega)\
\begin{pmatrix}
m_{11}+m_{13} \\
m_{31}+m_{33}%
\end{pmatrix}
,  \label{E5.37a}
\end{equation}
\begin{equation}
\mathbf{G}_{\downarrow}^{fc,ap}(i\omega)=-V^{\ast}\ \mathbf{G}_{\downarrow
}^{cc,ap}(i\omega)\
\begin{pmatrix}
m_{22}-m_{24} \\
m_{42}-m_{44}%
\end{pmatrix}
,  \label{E5.37b}
\end{equation}
\begin{equation}
\mathbf{G}_{\uparrow}^{cf,ap}(i\omega)=-V\ \mathbf{G}_{\uparrow}^{cc,ap}(i%
\omega)\
\begin{pmatrix}
m_{11}+m_{31} & ,m_{13}+m_{33}%
\end{pmatrix}
,  \label{E5.38a}
\end{equation}
\begin{equation}
\mathbf{G}_{\downarrow}^{cf,ap}(i\omega)=-V\ \mathbf{G}_{\uparrow}^{cc,ap}(i%
\omega)
\begin{pmatrix}
m_{22}-m_{42} & ,m_{24}-m_{44}%
\end{pmatrix}
,  \label{E5.38b}
\end{equation}
and also

\begin{equation}
\mathbf{G}_{\uparrow }^{ff,ap}(i\omega )=\frac{%
\begin{pmatrix}
m_{11} & m_{13} \\
m_{31} & _{m33}%
\end{pmatrix}
}{1-\left\vert V\right\vert ^{2}\varphi _{\uparrow }(i\omega )M_{\uparrow
}^{ff}}-\left\vert V\right\vert ^{2}\ \mathbf{G}_{\uparrow }^{cc}(i\omega )\
\left( m_{11}m_{33}-m_{13}m_{31}\right)
\begin{pmatrix}
1 & -1 \\
-1 & 1%
\end{pmatrix}
,  \label{E5.39a}
\end{equation}

\begin{equation}
\mathbf{G}_{\downarrow }^{ff,ap}(i\omega )=\frac{%
\begin{pmatrix}
m_{22} & m_{24} \\
m_{42} & _{m44}%
\end{pmatrix}
}{1-\left\vert V\right\vert ^{2}\varphi _{\downarrow }(i\omega
)M_{\downarrow }^{ff}}-\left\vert V\right\vert ^{2}\ \mathbf{G}_{\downarrow
}^{cc}(i\omega )\ \left( m_{22}m_{44}-m_{24}m_{42}\right)
\begin{pmatrix}
1 & 1 \\
1 & 1%
\end{pmatrix}
.  \label{E5.39b}
\end{equation}
We can now rewrite these two equation in the form (cf. Eq. (\ref{E5.9})
\begin{equation}
\mathbf{G}_{\uparrow }^{ff,ap}(i\omega )=\mathbf{M}_{\uparrow
}^{ap}+\left\vert V\right\vert ^{2}\ \mathbf{G}_{\uparrow }^{cc}(i\omega )\
\left\{ M_{\uparrow }^{ff}\ \mathbf{M}_{\uparrow }^{ap}-\left(
m_{11}m_{33}-m_{13}m_{31}\right)
\begin{pmatrix}
1 & -1 \\
-1 & 1%
\end{pmatrix}
\right\} ,  \label{E5.40a}
\end{equation}
\begin{equation}
\mathbf{G}_{\downarrow }^{ff,ap}(i\omega )=\mathbf{M}_{\downarrow
}^{ap}+\left\vert V\right\vert ^{2}\ \mathbf{G}_{\downarrow }^{cc}(i\omega
)\ \left\{ M_{\downarrow }^{ff}\ \mathbf{M}_{\downarrow }^{ap}-\left(
m_{22}m_{44}-m_{24}m_{42}\right)
\begin{pmatrix}
1 & 1 \\
1 & 1%
\end{pmatrix}
\right\} .  \label{E5.40b}
\end{equation}

\subsubsection{Green's functions with the usual Fermi operators $f$ and $%
f^{\dagger }$.}

As with the PAM, we calculate the Gf of the usual Fermi operators, (cf. Eq. (%
\ref{E5.41})). It is straightforward to obtain
\begin{equation}
\left\langle \left\langle f_{\sigma };f_{\sigma }^{\dagger }\right\rangle
\right\rangle _{z}=\frac{M_{\sigma }^{ff}}{1-\left\vert V\right\vert ^{2}\
\varphi _{\sigma }(z)\ M_{\sigma }^{ff}}  \label{E5.30}
\end{equation}
where we used Eqs. (\ref{E5.13g1},\ref{E5.13g2}). In a symilar way we find
\begin{equation}
\left\langle \left\langle f_{\sigma };c_{\sigma }^{\dagger }(\mathbf{k}%
)\right\rangle \right\rangle _{z}=-\frac{V^{\ast }}{\sqrt{N_{s}}}\mathcal{G}%
_{c,\sigma }^{0}\left( \mathbf{k},z\right) \frac{M_{\sigma }^{ff}}{%
1-\left\vert V\right\vert ^{2}\ \varphi _{\sigma }(z)\ M_{\sigma }^{ff}},
\label{E5.33a}
\end{equation}
\begin{equation}
\left\langle \left\langle c_{\sigma }(\mathbf{k});f_{\sigma }^{\dagger
}\right\rangle \right\rangle _{z}=-\frac{V}{\sqrt{N_{s}}}\mathcal{G}%
_{c,\sigma }^{0}\left( \mathbf{k},z\right) \frac{M_{\sigma }^{ff}}{%
1-\left\vert V\right\vert ^{2}\ \varphi _{\sigma }(z)\ M_{\sigma }^{ff}}.
\label{E5.34a}
\end{equation}
and
\begin{equation}
\left\langle \left\langle c_{\sigma }(\mathbf{k});c_{\sigma }^{\dagger }(%
\mathbf{k}^{\prime })\right\rangle \right\rangle _{z}=\mathcal{G}_{c,\sigma
}^{0}\left( \mathbf{k},i\omega \right) \delta \left( \mathbf{k,k}^{\prime
}\right) +\frac{\left\vert V\right\vert ^{2}}{N_{s}}\mathcal{G}_{c,\sigma
}^{0}\left( \mathbf{k},z\right) \frac{M_{\sigma }^{ff}}{1-\left\vert
V\right\vert ^{2}\ \varphi _{\sigma }(z)\ M_{\sigma }^{ff}}\mathcal{G}%
_{c,\sigma }^{0}\left( \mathbf{k}^{\prime },z\right) .  \label{E5.35a}
\end{equation}

As in the previous section we calculate the GF with the conduction electron
in the Wannier representation

\begin{equation}
\left\langle \left\langle f_{\sigma};c_{\mathbf{j}=0\sigma}^{\dagger
}\right\rangle \right\rangle _{z}=-V^{\ast}\ \mathbf{G}_{\sigma}^{cc}(i%
\omega)\ M_{\sigma}^{ff},
\end{equation}
and
\begin{equation}
\left\langle \left\langle c_{\mathbf{j}=0\sigma};f_{\sigma}^{\dagger
}\right\rangle \right\rangle _{z}=-V\ \mathbf{G}_{\sigma}^{cc}(i\omega )\
M_{\sigma}^{ff}.
\end{equation}

The $\left\langle \left\langle c_{\mathbf{j}=0\sigma};c_{\mathbf{j}=0\sigma
}^{\dagger}\right\rangle \right\rangle _{z}$ is given by $\mathbf{G}_{\sigma
}^{cc}(i\omega)$ in Eq. (\ref{E5.36})
\begin{equation}
\left\langle \left\langle c_{\mathbf{j}=0\sigma};c_{\mathbf{j}=0\sigma
}^{\dagger}\right\rangle \right\rangle _{z}=\frac{\varphi_{\sigma}(z)}{%
1-\left\vert V\right\vert ^{2}\varphi_{\sigma}(z)M_{\sigma}^{ff}}.
\end{equation}

\subsection{Summary of the approximate GF for the PAM}

\subsubsection{GF in reciprocal space and imaginary frequency}

\begin{equation}
\mathbf{G}_{\uparrow }^{ff,ap}(\mathbf{k},i\omega )=\frac{%
\begin{pmatrix}
m_{11} & m_{13} \\
m_{31} & _{m33}%
\end{pmatrix}
-\left\vert V\right\vert ^{2}\mathcal{G}_{c,\uparrow }^{0}\left( \mathbf{k}%
,z\right) \left( m_{11}m_{33}-m_{13}m_{31}\right)
\begin{pmatrix}
1 & -1 \\
-1 & 1%
\end{pmatrix}
}{1-\left\vert V\right\vert ^{2}\mathcal{G}_{c,\uparrow }^{0}\left( \mathbf{k%
},z\right) \left( m_{11}+m_{33}+m_{13}+m_{31}\right) },  \label{E5.22a}
\end{equation}
and
\begin{equation}
\mathbf{G}_{\downarrow }^{ff,ap}(\mathbf{k},i\omega )=\frac{%
\begin{pmatrix}
m_{22} & m_{24} \\
m_{42} & _{m44}%
\end{pmatrix}
-\left\vert V\right\vert ^{2}\mathcal{G}_{c,\downarrow }^{0}\left( \mathbf{k}%
,z\right) \left( m_{22}m_{44}-m_{24}m_{42}\right)
\begin{pmatrix}
1 & 1 \\
1 & 1%
\end{pmatrix}
}{1-\left\vert V\right\vert ^{2}\mathcal{G}_{c,\downarrow }^{0}\left(
\mathbf{k},z\right) \left( m_{22}+m_{44}-m_{24}-m_{42}\right) }.
\label{E5.23a}
\end{equation}

For
\begin{equation*}
\mathbf{G}_{\sigma }^{fc,ap}(\mathbf{k},i\omega )=
\begin{pmatrix}
\mathcal{G}_{0\sigma ,\sigma }^{fc}(\mathbf{k},i\omega ) \\
\mathcal{G}_{-\sigma d,\sigma }^{fc}(\mathbf{k},i\omega )%
\end{pmatrix}
,
\end{equation*}
\begin{equation}
\mathbf{G}_{\uparrow }^{fc,ap}(\mathbf{k},i\omega )=-V^{\ast }\frac{\mathcal{%
G}_{c,\uparrow }^{0}\left( \mathbf{k},i\omega \right)
\begin{pmatrix}
m_{11}+m_{13} \\
m_{31}+m_{33}%
\end{pmatrix}
}{1-\left\vert V\right\vert ^{2}\mathcal{G}_{c,\uparrow }^{0}\left( \mathbf{k%
},i\omega \right) \left( m_{11}+m_{33}+m_{13}+m_{31}\right) }  \label{E5.24a}
\end{equation}

\begin{equation}
\mathbf{G}_{\downarrow }^{fc,ap}(\mathbf{k},i\omega )=-V^{\ast }\frac{%
\mathcal{G}_{c,\downarrow }^{0}\left( \mathbf{k},i\omega \right)
\begin{pmatrix}
m_{22}-m_{24} \\
m_{42}-m_{44}%
\end{pmatrix}
}{1-\left\vert V\right\vert ^{2}\mathcal{G}_{c,\downarrow }^{0}\left(
\mathbf{k},i\omega \right) \left( m_{22}+m_{44}-m_{24}-m_{42}\right) }
\label{E5.25a}
\end{equation}
For $\mathbf{G}_{\sigma }^{cf,ap}(\mathbf{k},i\omega )=
\begin{pmatrix}
\mathcal{G}_{\sigma ,0\sigma }^{fc}(\mathbf{k},i\omega ) & ,\mathcal{G}%
_{\sigma ,\bar{\sigma}d}^{fc}(\mathbf{k},i\omega )%
\end{pmatrix}
$:

\begin{equation}
\mathbf{G}_{\uparrow }^{cf,ap}(\mathbf{k},i\omega )=-V\frac{\mathcal{G}%
_{c,\uparrow }^{0}\left( \mathbf{k},i\omega \right)
\begin{pmatrix}
m_{11}+m_{31} & ,m_{13}+m_{33}%
\end{pmatrix}
}{1-\left\vert V\right\vert ^{2}\mathcal{G}_{c,\uparrow }^{0}\left( \mathbf{k%
},i\omega \right) \left( m_{11}+m_{33}+m_{13}+m_{31}\right) }  \label{E5.26a}
\end{equation}
\begin{equation}
\mathbf{G}_{\downarrow }^{cf,ap}(\mathbf{k},i\omega )=-V\frac{\mathcal{G}%
_{c,\downarrow }^{0}\left( \mathbf{k},i\omega \right)
\begin{pmatrix}
m_{22}-m_{42} & ,m_{24}-m_{44}%
\end{pmatrix}
}{1-\left\vert V\right\vert ^{2}\mathcal{G}_{c,\downarrow }^{0}\left(
\mathbf{k},i\omega \right) \left( m_{22}+m_{44}-m_{24}-m_{42}\right) }
\label{E5.27a}
\end{equation}

\begin{equation}
\mathbf{G}_{\uparrow }^{cc,ap}(\mathbf{k},i\omega )=\frac{\mathcal{G}%
_{c,\uparrow }^{0}\left( \mathbf{k},i\omega \right) }{1-\left\vert
V\right\vert ^{2}\mathcal{G}_{c,\uparrow }^{0}\left( \mathbf{k},i\omega
\right) \left( m_{11}+m_{33}+m_{13}+m_{31}\right) }  \label{5.28a}
\end{equation}

\begin{equation}
\mathbf{G}_{\downarrow }^{cc,ap}(\mathbf{k},i\omega )=\frac{\mathcal{G}%
_{c,\downarrow }^{0}\left( \mathbf{k},i\omega \right) }{1-\left\vert
V\right\vert ^{2}\mathcal{G}_{c,\downarrow }^{0}\left( \mathbf{k},i\omega
\right) \left( m_{22}+m_{44}-m_{24}-m_{42}\right) }  \label{E5.29a}
\end{equation}

\subsubsection{Green's functions in real space and imaginary frequency.}

\begin{equation}
\mathbf{G}_{\uparrow }^{ff}(i\omega )=\mathbf{M}_{\uparrow }^{ap}+\frac{%
\left\vert V\right\vert ^{2}}{2D}ln\left( \frac{A_{\uparrow }(i\omega
)+D-\mu }{A_{\uparrow }(i\omega )-D-\mu }\right) \left[ \mathbf{M}_{\uparrow
}^{ap}M_{\uparrow }^{ff}-
\begin{pmatrix}
1 & -1 \\
-1 & 1%
\end{pmatrix}
\Theta _{\uparrow }\right]  \label{E5.30ab}
\end{equation}
\begin{equation}
\mathbf{G}_{\downarrow }^{ff}(i\omega )=\mathbf{M}_{\downarrow }^{ap}+\frac{%
\left\vert V\right\vert ^{2}}{2D}ln\left( \frac{A_{\downarrow }(i\omega
)+D-\mu }{A_{\downarrow }(i\omega )-D-\mu }\right) \left[ \mathbf{M}%
_{\downarrow }^{ap}M_{\downarrow }^{ff}-
\begin{pmatrix}
1 & 1 \\
1 & 1%
\end{pmatrix}
\Theta _{\downarrow }\right]  \label{E5.31ab}
\end{equation}
For
\begin{equation*}
\mathbf{G}_{\sigma }^{fc,ap}(i\omega )=
\begin{pmatrix}
\mathcal{G}_{0\sigma ,\sigma }^{fc}(i\omega ) \\
\mathcal{G}_{-\sigma d,\sigma }^{fc}(i\omega )%
\end{pmatrix}
,
\end{equation*}
\begin{equation}
\mathbf{G}_{\uparrow }^{fc,ap}(i\omega )=-\frac{V^{\ast }}{2D}\ln \left(
\frac{A_{\uparrow }(i\omega )+D-\mu }{A_{\uparrow }(i\omega )-D-\mu }\right)
\begin{pmatrix}
m_{11}+m_{13} \\
m_{31}+m_{33}%
\end{pmatrix}
,  \label{E5.32ab}
\end{equation}
\begin{equation}
\mathbf{G}_{\downarrow }^{fc,ap}(i\omega )=-\frac{V^{\ast }}{2D}\ln \left(
\frac{A_{\downarrow }(i\omega )+D-\mu }{A_{\downarrow }(i\omega )-D-\mu }%
\right)
\begin{pmatrix}
m_{22}-m_{24} \\
m_{42}-m_{44}%
\end{pmatrix}
,  \label{E5.33ab}
\end{equation}
For $\mathbf{G}_{\sigma }^{cf,ap}(i\omega )=
\begin{pmatrix}
\mathcal{G}_{\sigma ,0\sigma }^{fc}(i\omega ) & ,\mathcal{G}_{\sigma ,\bar{%
\sigma}d}^{fc}(i\omega )%
\end{pmatrix}
$:
\begin{equation}
\mathbf{G}_{\uparrow }^{cf,ap}(i\omega )=-\frac{V}{2D}\ln \left( \frac{%
A_{\uparrow }(i\omega )+D-\mu }{A_{\uparrow }(i\omega )-D-\mu }\right)
\begin{pmatrix}
m_{11}+m_{31} & ,m_{13}+m_{33}%
\end{pmatrix}
,  \label{5.34ab}
\end{equation}
\begin{equation}
\mathbf{G}_{\downarrow }^{cf,ap}(i\omega )=-\frac{V}{2D}\ln \left( \frac{%
A_{\downarrow }(i\omega )+D-\mu }{A_{\downarrow }(i\omega )-D-\mu }\right)
\begin{pmatrix}
m_{22}-m_{42} & ,m_{24}-m_{44}%
\end{pmatrix}
.  \label{E5.35ab}
\end{equation}
\begin{equation}
\mathbf{G}_{\sigma }^{cc}(i\omega )=\frac{1}{2D}\ln \left( \frac{A_{\sigma
}(i\omega )+D-\mu }{A_{\sigma }(i\omega )-D-\mu }\right) .  \label{E5.36ab}
\end{equation}
where
\begin{equation*}
A_{\sigma }(i\omega )=-i\omega -\left\vert V\right\vert ^{2}M_{\sigma }^{ff},
\end{equation*}
\begin{equation*}
M_{\uparrow }^{ff}=m_{11}+m_{13}+m_{31}+m_{33},
\end{equation*}
\begin{equation*}
M_{\downarrow }^{ff}=m_{22}+m_{44}-m_{24}-m_{42},
\end{equation*}
\begin{equation*}
\Theta _{\uparrow }=m_{11}m_{33}-m_{13}m_{31,}
\end{equation*}
\begin{equation*}
\Theta _{\downarrow }=m_{22}m_{33}-m_{24}m_{42}.
\end{equation*}

\subsubsection{Green's functions with the usual Fermi operators $f$ and $%
f^{\dagger }$}

\begin{equation}
\left\langle \left\langle f_{\sigma };f_{\sigma }^{\dagger }\right\rangle
\right\rangle _{z}\left( \mathbf{k}\right) =\frac{M_{\sigma }^{ff}}{%
1-\left\vert V\right\vert ^{2}\ \mathcal{G}_{c,\sigma }^{0}\left( \mathbf{k}%
,z\right) \ M_{\sigma }^{ff}}  \label{E5.43c}
\end{equation}
\begin{equation}
\left\langle \left\langle f_{\sigma };c_{\sigma }^{\dagger }(\mathbf{k}%
)\right\rangle \right\rangle _{z}=-V^{\ast }\frac{\mathcal{G}_{c,\sigma
}^{0}\left( \mathbf{k},z\right) M_{\sigma }^{ff}}{1-\left\vert V\right\vert
^{2}\ \mathcal{G}_{c,\sigma }^{0}\left( \mathbf{k},z\right) \ M_{\sigma
}^{ff}},  \label{E5.44cc}
\end{equation}
\begin{equation}
\left\langle \left\langle c_{\sigma }(\mathbf{k});f_{\sigma }^{\dagger
}\right\rangle \right\rangle _{z}=-V\frac{\mathcal{G}_{c,\sigma }^{0}\left(
\mathbf{k},z\right) M_{\sigma }^{ff}}{1-\left\vert V\right\vert ^{2}\
\mathcal{G}_{c,\sigma }^{0}\left( \mathbf{k},z\right) \ M_{\sigma }^{ff}}.
\label{E5.45cc}
\end{equation}
\begin{equation}
\left\langle \left\langle c_{\sigma }(\mathbf{k});c_{\sigma }^{\dagger }(%
\mathbf{k}^{\prime })\right\rangle \right\rangle _{z}=\delta \left( \mathbf{%
k,k}^{\prime }\right) \frac{\mathcal{G}_{c,\sigma }^{0}\left( \mathbf{k}%
^{\prime },z\right) }{1-\left\vert V\right\vert ^{2}\ \mathcal{G}_{c,\sigma
}^{0}\left( \mathbf{k},z\right) \ M_{\sigma }^{ff}}.  \label{E5.46cc}
\end{equation}

In imaginary frequency and real space\ when the $f$ electron is created and
destroyed at the same site
\begin{equation}
\left\langle \left\langle f_{\sigma };f_{\sigma }^{\dagger }\right\rangle
\right\rangle _{z}=M_{\sigma }^{ff}\left[ 1+\left\vert V\right\vert
^{2}M_{\sigma }^{ff}\mathbf{G}_{\sigma }^{cc}(i\omega )\right] .
\label{E5.43ca}
\end{equation}
\begin{equation}
\left\langle \left\langle f_{\sigma };c_{\sigma }^{\dagger }(\mathbf{k}%
)\right\rangle \right\rangle _{z}=-V^{\ast }\ \mathbf{G}_{\sigma
}^{cc}(i\omega )\ M_{\sigma }^{ff}.  \label{E5.44ca}
\end{equation}
\begin{equation}
\left\langle \left\langle c_{\sigma };f_{\sigma }^{\dagger }\right\rangle
\right\rangle _{z}=-V\ \mathbf{G}_{\sigma }^{cc}(i\omega )\ M_{\sigma }^{ff}.
\label{E5.45c}
\end{equation}

\begin{equation}
\left\langle \left\langle c_{\sigma };c_{\sigma }^{\dagger }\right\rangle
\right\rangle _{z}=\frac{1}{2D}\ln \left( \frac{A_{\sigma }(z)+D-\mu }{%
A_{\sigma }(z)-D-\mu }\right) .  \label{E5.46ca}
\end{equation}


\subsection{Summary of the approximate GF for the SIAM}

\subsubsection{GF with conduction electrons in $\mathbf{k}$ space}

\begin{equation}
\mathbf{G}_{\uparrow}^{ff,ap}=\frac{%
\begin{pmatrix}
m_{11} & m_{13} \\
m_{31} & _{m33}%
\end{pmatrix}
-\left\vert V\right\vert ^{2}\varphi_{\uparrow}(i\omega)\left(
m_{11}m_{33}-m_{13}m_{31}\right)
\begin{pmatrix}
1 & -1 \\
-1 & 1%
\end{pmatrix}
}{1-\left\vert V\right\vert ^{2}\varphi_{\uparrow}(i\omega)\left(
m_{11}+m_{33}+m_{13}+m_{31}\right) }  \label{E5.22}
\end{equation}

\begin{equation}
\mathbf{G}_{\downarrow}^{ff,ap}=\frac{%
\begin{pmatrix}
m_{22} & m_{24} \\
m_{42} & _{m44}%
\end{pmatrix}
-\left\vert V\right\vert ^{2}\varphi_{\downarrow}(i\omega)\left(
m_{22}m_{44}-m_{24}m_{42}\right)
\begin{pmatrix}
1 & 1 \\
1 & 1%
\end{pmatrix}
}{1-\left\vert V\right\vert ^{2}\varphi_{\downarrow}(i\omega)\left(
m_{22}+m_{44}-m_{24}-m_{42}\right) }.  \label{E5.23}
\end{equation}

\begin{equation}
\mathbf{G}_{\uparrow}^{fc}(\mathbf{k},i\omega)=-\frac{V^{\ast}}{\sqrt{N_{s}}}%
\mathcal{G}_{c,\uparrow}^{0}\left( \mathbf{k},i\omega\right) \frac{%
\begin{pmatrix}
m_{11}+m_{13} \\
m_{31}+m_{33}%
\end{pmatrix}
}{1-\left\vert V\right\vert ^{2}\varphi_{\uparrow}(i\omega)\left(
m_{11}+m_{33}+m_{13}+m_{31}\right) }  \label{E5.24}
\end{equation}

\begin{equation}
\mathbf{G}_{\downarrow }^{fc}(\mathbf{k},i\omega )=-\frac{V^{\ast }}{\sqrt{%
N_{s}}}\mathcal{G}_{c,\downarrow }^{0}\left( \mathbf{k},i\omega \right)
\frac{%
\begin{pmatrix}
m_{22}-m_{24} \\
m_{42}-m_{44}%
\end{pmatrix}
}{1-\left\vert V\right\vert ^{2}\varphi _{\downarrow }(i\omega )\left(
m_{22}+m_{44}-m_{24}-m_{42}\right) }  \label{E5.25}
\end{equation}

\begin{equation}
\mathbf{G}_{\uparrow }^{cf}(\mathbf{k},i\omega )=-\frac{V}{\sqrt{N_{s}}}%
\mathcal{G}_{c,\uparrow }^{0}\left( \mathbf{k},i\omega \right) \frac{%
\begin{pmatrix}
m_{11}+m_{31} & ,m_{13}+m_{33}%
\end{pmatrix}
}{1-\left\vert V\right\vert ^{2}\varphi _{\uparrow }(i\omega )\left(
m_{11}+m_{33}+m_{13}+m_{31}\right) }  \label{E5.26}
\end{equation}
\begin{equation}
\mathbf{G}_{\downarrow }^{cf}(\mathbf{k},i\omega )=-\frac{V}{\sqrt{N_{s}}}%
\mathcal{G}_{c,\downarrow }^{0}\left( \mathbf{k},i\omega \right) \frac{%
\begin{pmatrix}
m_{22}-m_{42} & ,m_{24}-m_{44}%
\end{pmatrix}
}{1-\left\vert V\right\vert ^{2}\varphi _{\downarrow }(i\omega )\left(
m_{22}+m_{44}-m_{24}-m_{42}\right) }  \label{E5.27}
\end{equation}

\begin{equation}
\mathbf{G}_{\uparrow}^{cc}(\mathbf{k},\mathbf{k}^{\prime},i\omega )=\mathcal{%
G}_{c,\uparrow}^{0}\left( \mathbf{k},i\omega\right) \delta\left( \mathbf{k,k}%
^{\prime}\right) +\frac{\left\vert V\right\vert ^{2}}{N_{s}}\mathcal{G}%
_{c,\uparrow}^{0}\left( \mathbf{k},i\omega\right) \frac{\left(
m_{11}+m_{33}+m_{13}+m_{31}\right) }{1-\left\vert V\right\vert
^{2}\varphi_{\uparrow}(i\omega)\left( m_{11}+m_{33}+m_{13}+m_{31}\right) }%
\mathcal{G}_{c,\uparrow}^{0}\left( \mathbf{k}^{\prime},i\omega\right)
\label{5.28}
\end{equation}

\begin{equation}
\mathbf{G}_{\downarrow }^{cc}(\mathbf{k},\mathbf{k}^{\prime },i\omega )=%
\mathcal{G}_{c,\downarrow }^{0}\left( \mathbf{k},i\omega \right) \delta
\left( \mathbf{k,k}^{\prime }\right) +\frac{\left\vert V\right\vert ^{2}}{%
N_{s}}\mathcal{G}_{c,\downarrow }^{0}\left( \mathbf{k},i\omega \right) \frac{%
\left( mm_{22}+m_{44}-m_{24}-m_{42}\right) }{1-\varphi _{\downarrow
}(i\omega )\left( m_{22}+m_{44}-m_{24}-m_{42}\right) }\mathcal{G}%
_{c,\downarrow }^{0}\left( \mathbf{k}^{\prime },i\omega \right)
\label{E5.29}
\end{equation}

\subsubsection{Green's functions with conduction electron in the Wannier
representation.}

\begin{equation}
\mathbf{G}_{\uparrow }^{fc,ap}(i\omega )=-V^{\ast }\frac{\varphi _{\uparrow
}(i\omega )}{1-\left\vert V\right\vert ^{2}\varphi _{\uparrow }(i\omega
)M_{\uparrow }^{ff}}
\begin{pmatrix}
m_{11}+m_{13} \\
m_{31}+m_{33}%
\end{pmatrix}
,  \label{E5.30a}
\end{equation}
\begin{equation}
\mathbf{G}_{\downarrow }^{fc}(i\omega )=-V^{\ast }\frac{\varphi _{\downarrow
}(i\omega )}{1-\left\vert V\right\vert ^{2}\varphi _{\downarrow }(i\omega
)M_{\downarrow }^{ff}}
\begin{pmatrix}
m_{22}-m_{24} \\
m_{42}-m_{44}%
\end{pmatrix}
.  \label{E5.30b}
\end{equation}
\begin{equation}
\mathbf{G}_{\uparrow }^{cf}(i\omega )=-V\frac{\varphi _{\uparrow }(i\omega )%
}{1-\left\vert V\right\vert ^{2}\varphi _{\uparrow }(i\omega )M_{\uparrow
}^{ff}}
\begin{pmatrix}
m_{11}+m_{31} & ,m_{13}+m_{33}%
\end{pmatrix}
,
\end{equation}
\begin{equation}
\mathbf{G}_{\downarrow }^{cf}(i\omega )=-V\frac{\varphi _{\downarrow
}(i\omega )}{1-\left\vert V\right\vert ^{2}\varphi _{\downarrow }(i\omega
)M_{\downarrow }^{ff}}
\begin{pmatrix}
m_{22}-m_{42} & ,m_{24}-m_{44}%
\end{pmatrix}
.
\end{equation}
\begin{equation}
\mathbf{G}_{\sigma }^{cc}(i\omega )=\frac{\varphi _{\sigma }(z)}{%
1-\left\vert V\right\vert ^{2}\varphi _{\sigma }(z)M_{\sigma }^{ff}}.
\end{equation}

\subsubsection{Green's functions with the usual Fermi operators $f$ and $%
f^{\dagger}$.}

\begin{equation}
\left\langle \left\langle f_{\sigma};f_{\sigma}^{\dagger}\right\rangle
\right\rangle _{z}=\frac{M_{\sigma}^{ff}}{1-\left\vert V\right\vert ^{2}\
\varphi_{\sigma}(z)\ M_{\sigma}^{ff}}  \label{E5.43}
\end{equation}
\begin{equation}
\left\langle \left\langle f_{\sigma};c_{\mathbf{j}=0\sigma}^{\dagger
}\right\rangle \right\rangle _{z}=-V^{\ast}\ \mathbf{G}_{\sigma}^{cc}(i%
\omega)\ M_{\sigma}^{ff},  \label{E5.44}
\end{equation}
\begin{equation}
\left\langle \left\langle c_{\mathbf{j}=0\sigma};f_{\sigma}^{\dagger
}\right\rangle \right\rangle _{z}=-V\ \mathbf{G}_{\sigma}^{cc}(i\omega )\
M_{\sigma}^{ff}.  \label{E5.45}
\end{equation}

\begin{equation}
\left\langle \left\langle c_{\mathbf{j}=0\sigma};c_{\mathbf{j}=0\sigma
}^{\dagger}\right\rangle \right\rangle _{z}=\frac{\varphi_{\sigma}(z)}{%
1-\left\vert V\right\vert ^{2}\varphi_{\sigma}(z)M_{\sigma}^{ff}}.
\label{E5.46}
\end{equation}


\section{The free $f$-electron GF}

\label{ApE}

We first collect a few equations from the previous appendices \ (cf. Eqs. (%
\ref{EqC1},\ref{EqC3a}))

\begin{equation}
\mathcal{G}(\gamma_{1},\tau_{1};\gamma_{2},\tau_{2})=\left\langle \left(
\hat{Y}(\gamma_{1},\tau_{1})\hat{Y}(\gamma_{2},\tau_{2})\right)
_{+}\right\rangle _{\mathcal{H}}==\left\langle \left( \hat{Y}%
(\gamma_{1},\tau_{1}-\tau_{2})\hat{Y}(\gamma_{2},0)\right) _{+}\right\rangle
_{\mathcal{H}}=F\left( {\tau}_{1}-{\tau}_{2}\right) ,  \label{EqE2}
\end{equation}
(cf. Eqs. (\ref{EqC3}))

\begin{equation}
\mathcal{G}(\gamma_{1},\tau_{1};\gamma_{2},\tau_{2})=\exp\left( \beta
\Omega\right) Tr\left\{ \exp\left( -\beta\mathcal{H}\right) \exp{(\tau }_{1}{%
\mathcal{H})}Y_{\gamma_{1}}\exp{(-\tau}_{1}{\mathcal{H})}\exp{(\tau}_{2}{%
\mathcal{H})}Y_{\gamma_{2}}\exp{(-\tau}_{2}{\mathcal{H})}\right\} ,
\label{EqE3}
\end{equation}
(cf. Eq. (\ref{EqC11}))

\begin{equation}
\mathcal{G}(\gamma_{1},\omega_{1};\gamma_{2},\omega_{2})=\Delta\left(
\omega_{1}+\omega_{2}\right) \int_{0}^{\beta}dxF\left( {x}\right) \exp\left[
ix\omega_{1}\right] ,  \label{EqE4}
\end{equation}
and (cf. Eq. \ref{ApD1})

\begin{equation}
\left\langle \left( \hat{Y}(f;\mathbf{k},\alpha,u=-,\omega)\hat {Y}(f;%
\mathbf{k}^{\prime},\alpha^{\prime},u^{\prime},\omega^{\prime})\right)
_{+}\right\rangle _{\mathcal{H}}=\mathcal{G}_{\alpha\alpha^{\prime}}^{ff}(%
\mathbf{k},i\omega)\ \Delta\left( u+u^{\prime}\right) \Delta\left( u\mathbf{k%
}+u^{\prime}\mathbf{k}^{\prime}\right) \Delta\left( \omega
+\omega^{\prime}\right) .  \label{EqE5}
\end{equation}

These equations correspond to the exact GF, and for the free GF (i.e. with
no hybridization) we shall now prove that

\begin{equation}
M_{\alpha\alpha^{\prime}}^{0}(\mathbf{k},\omega)\equiv\mathcal{G}%
_{\alpha\alpha^{\prime}}^{ff,0}(\mathbf{k},\omega)=-\delta_{\alpha
\alpha^{\prime}}\ D_{\alpha}/\left( i\omega-\varepsilon_{\alpha}\right) ,
\label{EqE6}
\end{equation}
where
\begin{equation}
\varepsilon_{\alpha}=\varepsilon_{\left( b,a\right) }=\mathcal{\varepsilon }%
_{b}-\mathcal{\varepsilon}_{a}.  \label{EqE7}
\end{equation}
Substituting the general $\hat{Y}(\gamma,\tau)$ operators by $X$ operators
we have

\begin{align}
F_{\gamma,\gamma^{\prime}}^{0}\left( {\tau}\right) & =\left\langle \left(
\hat{Y}(\gamma,\tau)\hat{Y}(\gamma^{\prime},0)\right) _{+}\right\rangle _{%
\mathcal{H}_{0}}==\left\langle \left( X_{j,\alpha}(\tau)X_{j,\alpha
^{\prime}}^{\dagger}(0)\right) _{+}\right\rangle _{\mathcal{H}_{0}}=  \notag
\\
& =\exp\left( \beta\Omega\right) Tr\left\{ \exp\left( -\beta \mathcal{H}%
_{0}\right) \exp{(\tau}\mathcal{H}_{0}{)}X_{j,\left( b,a\right) }\exp{(-\tau}%
\mathcal{H}_{0}{)}X_{j,\left( b^{\prime},a^{\prime}\right)
}^{\dagger}\right\} .  \label{EqE8}
\end{align}
Now we use a basis $\left\{ \left\vert c\right\rangle \right\} $ of
eigenstates of $\mathcal{H}_{0}$ to calculate the trace:
\begin{align}
F_{\alpha,\alpha^{\prime}}^{0}\left( {\tau}\right) & =\exp\left(
\beta\Omega\right) \sum_{c}\left\langle c\right\vert \exp\left( -\beta%
\mathcal{H}_{0}\right) \exp{(\tau}\mathcal{H}_{0}{)}X_{j,\left( b,a\right)
}\exp{(-\tau}\mathcal{H}_{0}{)}X_{j,\left( a^{\prime},b^{\prime }\right)
}\left\vert c\right\rangle =  \notag \\
& =\exp\left( \beta\Omega\right)
\sum_{c}\delta_{b^{\prime},c}\delta_{a^{\prime},a}\delta_{c,b}\exp{(}\left( {%
\tau-\beta}\right) \mathcal{\varepsilon}_{b}{)}\exp{(-\tau}\mathcal{%
\varepsilon}_{a^{\prime}}{)=}  \notag \\
& =\delta_{a^{\prime},a}\delta_{b^{\prime},b}\exp\left( \beta\Omega\right)
\exp{(-\beta}\mathcal{\varepsilon}_{b}{)}\exp{(\tau}\left( \mathcal{%
\varepsilon}_{b}-\mathcal{\varepsilon}_{a}\right) {)\equiv}%
\delta_{\alpha\alpha^{\prime}}F_{\alpha}^{0}\left( {\tau}\right) .
\label{EqE9}
\end{align}

We calculate the Fourier transform:
\begin{align}
\mathcal{G}_{\alpha\alpha^{\prime}}^{ff,0}(\mathbf{k},i\omega) &
=\delta_{\alpha\alpha^{\prime}}\int_{0}^{\beta}dxF_{\alpha}^{0}\left( {x}%
\right) \exp\left[ ix\omega\right] =\delta_{\alpha\alpha^{\prime}}\exp\left(
\beta\Omega\right) \exp{(-\beta}\mathcal{\varepsilon}_{b}{)}%
\int_{0}^{\beta}dx\exp{(x}\left( \mathcal{\varepsilon}_{b}-\mathcal{%
\varepsilon}_{a}\right) {)}\exp\left[ ix\omega\right] =  \notag \\
& =\delta_{\alpha\alpha^{\prime}}\exp\left( \beta\Omega\right) \exp {(-\beta}%
\mathcal{\varepsilon}_{b}{)}\frac{1}{i\omega+\varepsilon_{\alpha}}\left\{
\exp{(\beta}\left( i\omega+\mathcal{\varepsilon}_{\alpha}\right) {)-1}%
\right\} ,  \label{EqE10}
\end{align}
and employing $\exp{(i\ \beta}\omega{)=-1}$ we find\footnote{%
Rewriting Eq. (\ref{EqE5}) for the unperturbed case, but with $%
u^{\prime}\rightarrow -u^{\prime}$ and $\omega^{\prime}\rightarrow-\omega^{%
\prime}$ (after exchanging primed and unprimed variables), we have
\begin{equation*}
\left\langle \left( Y(f;\mathbf{k}^{\prime},\alpha^{\prime},-u^{\prime
}=-,-\omega^{\prime})\ Y(f;;\mathbf{k},\alpha,u,\omega)\right)
_{+}\right\rangle _{c}=\mathcal{G}_{\alpha\alpha^{\prime}}^{ff,0}(\mathbf{k}%
^{\prime},-\omega^{\prime})\ \Delta\left( u-u^{\prime}\right) \Delta\left( u%
\mathbf{k}-u^{\prime}\mathbf{k}^{\prime}\right) \Delta\left( \omega
-\omega^{\prime}\right) ,
\end{equation*}
and employing Eq. (\ref{EqE11}) we find:
\begin{equation*}
\mathcal{G}_{\alpha\alpha^{\prime}}^{ff,0}(\mathbf{k}^{\prime},-\omega
^{\prime})=-\ \delta_{\alpha\alpha^{\prime}}\frac{D_{\alpha}}{-i\omega
^{\prime}+\left( \varepsilon_{b}-\varepsilon_{a}\right) }=\ \delta
_{\alpha\alpha^{\prime}}\frac{D_{\alpha}}{i\omega^{\prime}+\left(
\varepsilon_{a}-\varepsilon_{b}\right) }.
\end{equation*}
For the conduction electrons we should then have (putting $D_{\alpha}=1$ and
$\left( \varepsilon_{a}-\varepsilon_{b}\right) =\varepsilon\left( \mathbf{k}%
_{1},\sigma_{1}\right) -0=\varepsilon\left( \mathbf{k}_{1},\sigma_{1}\right)
$)
\begin{align*}
\left\langle \left( C(\mathbf{k}_{1}^{\prime},\sigma_{1}^{\prime},-u_{1}^{%
\prime}=-,-\omega_{1}^{\prime})\ C(\mathbf{k}_{1},\sigma_{1},u_{1},%
\omega_{1})\right) _{+}\right\rangle _{c} & =\mathcal{G}_{\sigma }^{cc,0}(%
\mathbf{k}_{1}^{\prime},-\omega_{1}^{\prime})\ \Delta\left(
u_{1}-u_{1}^{\prime}\right) \Delta\left( u_{1}\mathbf{k}_{1}-u_{1}^{\prime }%
\mathbf{k}_{1}^{\prime}\right) \Delta\left( \omega_{1}-\omega_{1}^{\prime
}\right) \\
& =\frac{1}{i\omega_{1}+\varepsilon\left( \mathbf{k}_{1},\sigma_{1}\right) }%
\delta(\mathbf{k}_{1},\mathbf{k}_{1}^{\prime})\delta(u_{1},u_{1}^{\prime
})\delta(\sigma_{1},\sigma_{1}^{\prime})\delta(\omega_{1},\omega_{1}^{\prime
}),
\end{align*}
which is just Eqs. (\ref{Fourier16},\ref{Fourier16a}) for $u_{1}=+$. Note
that one then has
\begin{equation*}
\mathcal{G}_{\alpha\alpha^{\prime}}^{ff,0}(\mathbf{k}^{\prime},\omega^{%
\prime })=\ \delta_{\alpha\alpha^{\prime}}\frac{-D_{\alpha}}{%
i\omega^{\prime}-\left( \varepsilon_{a}-\varepsilon_{b}\right) },
\end{equation*}
and (cf. Eq. (\ref{Eq3.14}))
\begin{equation*}
\mathcal{G}_{\sigma}^{cc,0}(\mathbf{k}_{1}^{\prime},\omega_{1}^{\prime})=%
\frac{-1}{i\omega_{1}^{\prime}-\varepsilon\left( \mathbf{k}_{1}^{\prime
},\sigma_{1}^{\prime}\right) }\equiv\mathcal{G}_{c,\sigma}^{0}\left( \mathbf{%
k}_{1}^{\prime},\omega_{1}^{\prime}\right) ,
\end{equation*}
which are used in another context.\label{myfoot6}}
\begin{equation}
\mathcal{G}_{\alpha\alpha^{\prime}}^{ff,0}(\mathbf{k},\omega)=-\ \delta
_{\alpha\alpha^{\prime}}\frac{\exp\left( \beta\Omega\right) \left[ \exp{%
(-\beta}\mathcal{\varepsilon}_{b}{)+}\exp{(-\beta}\mathcal{\varepsilon }_{a}{%
)}\right] }{i\omega+\varepsilon_{\alpha}}==-\ \delta_{\alpha \alpha^{\prime}}%
\frac{D_{\alpha}}{i\omega+\varepsilon_{\alpha}}=-\ \delta
_{\alpha\alpha^{\prime}}\frac{D_{\alpha}}{i\omega+\left( \varepsilon
_{b}-\varepsilon_{a}\right) },  \label{EqE11}
\end{equation}
where
\begin{equation}
D_{\alpha}=\left\langle X_{aa}+X_{bb}\right\rangle =\exp\left( \beta
\Omega\right) \left[ \exp{(-\beta}\mathcal{\varepsilon}_{b}{)+}\exp{(-\beta }%
\mathcal{\varepsilon}_{a}{)}\right] .  \label{EqE12}
\end{equation}

\section{The approximate GF for U$\rightarrow\infty$ and a rectangular band
in the impurity case.}

We assume that the dispersion relation of the conduction electrons
corresponds to a rectangular band, with
\begin{equation}
-D\leq E_{\mathbf{k},\sigma}\leq D.  \label{EqF1}
\end{equation}
In the present case we have only a single $\alpha=\left( 0,\sigma\right) $,
and we assume that $V(\alpha^{\prime},\mathbf{k},\sigma)V^{\ast}(\alpha,%
\mathbf{k},\sigma)=\left\vert V\right\vert ^{2}$, so that from Eqs. (\ref%
{ApD18},\ref{ApD21a})
\begin{equation}
W\left( z\right) =\frac{1}{N_{s}}\sum_{\mathbf{k}}\left\vert V\right\vert
^{2}\ \frac{-1}{z-\varepsilon\left( \mathbf{k},z\right) }=-\frac{\left\vert
V\right\vert ^{2}}{2D}\int\limits_{-D-\mu}^{D-\mu}\frac{1}{z-x}dx.
\label{EqF2}
\end{equation}
To avoid singularities we employ $z=\omega+is$ with $s>0$, so that
\begin{equation}
W(\omega+is)=-\frac{\left\vert V\right\vert ^{2}}{2D}\ln\left[ \frac {%
\omega+is-\left( D-\mu\right) }{\omega+is-\left( -D-\mu\right) }\right] .
\label{EqF3}
\end{equation}
and (cf. \ref{Eq3.15})
\begin{equation}
A(\omega+is)=-\frac{\left\vert V\right\vert ^{2}}{2D}\ln\left[ \frac {%
\omega+is-\left( D-\mu\right) }{\omega+is-\left( -D-\mu\right) }\right] \
M^{eff}(\omega+is)\mathbf{\ .}  \label{EqF3a}
\end{equation}
For the band with zero width we take $E_{\mathbf{k},\sigma}=E_{0}^{a}$ (cf.
Section \ref{S3}), so that from Eq. (\ref{EqF2})
\begin{equation}
W^{at}(z)=\frac{-\left\vert V\right\vert ^{2}}{z-E_{0}^{a}+\mu}.
\label{EqF4}
\end{equation}
We then have from Eq. (\ref{EqF3a}) (cf. also Eq. (\ref{E3.30})):

\begin{equation}
\mathbf{M}^{ap}\left( z\right) \mathbf{=}\left( \mathbf{I+G}^{ff,at}\left(
z\right) \cdot\mathbf{W}^{at}\right) ^{-1}\cdot\mathbf{G}^{ff,at}\left(
z\right) =\frac{\left( z-E_{0}^{a}+\mu\right) \mathbf{G}^{ff,at}\left(
z\right) }{z-E_{0}^{a}+\mu-\left\vert V\right\vert ^{2}\mathbf{G}%
^{ff,at}\left( z\right) },  \label{EqF5}
\end{equation}
and substituting Eq. (\ref{EqF5}) in Eq. (\ref{EqF3a}) (cf. Eq. (\ref{E3.26a}%
)) we find
\begin{equation}
A^{at}=-\frac{\left\vert V\right\vert ^{2}}{2D}\ln\left[ \frac{\omega
+is-\left( D-\mu\right) }{\omega+is-\left( -D-\mu\right) }\right] \frac{%
\left( z-E_{0}^{a}+\mu\right) \mathbf{G}^{ff,at}\left( z\right) }{%
z-E_{0}^{a}+\mu-\left\vert V\right\vert ^{2}\mathbf{G}^{ff,at}\left(
z\right) }.  \label{EqF6}
\end{equation}
Employing Eq. (\ref{E3.28}) we can now write the approximate GF as
\begin{align}
\mathbf{G}^{ff} & =\mathbf{M\cdot}\left( \mathbf{I-A}\right) ^{-1}=  \notag
\\
& =\frac{\left( z-E_{0}^{a}+\mu\right) \mathbf{G}^{ff,at}\left( z\right) }{%
\left( z-E_{0}^{a}+\mu\right) -\left\vert V\right\vert ^{2}\mathbf{G}%
^{ff,at}\left( z\right) +\frac{\left\vert V\right\vert ^{2}}{2D}\ln\left[
\frac{\omega+is-\left( D-\mu\right) }{\omega+is-\left( -D-\mu\right) }\right]
\left( z-E_{0}^{a}+\mu\right) \mathbf{G}^{ff,at}\left( z\right) }.
\label{EqF7}
\end{align}

\subsubsection*{\textbf{Employing the GF of the PAM with a band of zeroth
width to calculate the impurity} $\mathbf{M}^{ap}\left( z\right) $ .}

For the PAM we have

\begin{equation}
\mathbf{M}^{ap}\left( z\right) \mathbf{=}\left( \mathbf{I+G}^{ff,at}\left(
z\right) \cdot\mathbf{W}\right) ^{-1}\cdot\mathbf{G}^{ff,at}\left( z\right) .
\label{3.23a}
\end{equation}
where
\begin{equation}
\left\{ \mathbf{W}\right\} _{\alpha^{\prime}\alpha}\equiv W_{\alpha^{\prime
}\alpha}\left( \mathbf{k},\sigma,z\right) =V_{\alpha^{\prime}}\left( \mathbf{%
k},\sigma\right) \ V_{\alpha}^{\ast}\left( \mathbf{k},\sigma\right) \
G_{c,\sigma}^{0}\left( \mathbf{k},z\right) .  \label{EqF8}
\end{equation}
Employing the conduction electron free GF for a zeroth width band
\begin{equation}
\mathcal{G}_{c,\sigma}^{0}\left( \mathbf{k},z\right) =\frac{-1}{%
z-\varepsilon\left( \mathbf{k},z\right) }=\frac{-1}{z-E_{0}^{a}+\mu},
\label{EqF9}
\end{equation}
and assuming that $V_{\alpha^{\prime}}\left( \mathbf{k},\sigma\right) \
V_{\alpha}^{\ast}\left( \mathbf{k},\sigma\right) =\left\vert V\right\vert
^{2}$ we obtain
\begin{equation}
W=\frac{-\left\vert V\right\vert ^{2}}{z-E_{0}^{a}+\mu},  \label{EqF10}
\end{equation}
which is exactly what we obtained in Eq. (\ref{EqF4}).

\subsection{References}

\bibliographystyle{apsrev}
\bibliography{GFfiniteU100712}

\begin{table}[ptb]
\begin{tabular}[t]{||l||c||c|c|c|c||c|c|c|c|c|c||c|c|c|c||c||c||}
\hline\hline
& $\quad\left\vert 1\right\rangle $ & $\quad\left\vert 2\right\rangle $ & $%
\quad\left\vert 3\right\rangle $ & $\quad\left\vert 4\right\rangle $ & $%
\quad\left\vert 5\right\rangle $ & $\left\vert 6\right\rangle $ & $%
\left\vert 7\right\rangle $ & $\left\vert 8\right\rangle $ & $\left\vert
9\right\rangle $ & $\left\vert 10\right\rangle $ & $\left\vert
11\right\rangle $ & $\left\vert 12\right\rangle $ & $\left\vert
13\right\rangle $ & $\left\vert 14\right\rangle $ & $\left\vert
15\right\rangle $ & $\left\vert 16\right\rangle $ &  \\ \hline\hline
$\left\langle 1\right\vert $ &  & $Y_{6}$ & $Y_{5}$ & $Y_{2}$ & $Y_{1}$ &  &
&  &  &  &  &  &  &  &  &  & \multicolumn{1}{||l||}{$\left\vert
1\right\rangle =\left\vert 0,0\right\rangle $} \\ \hline\hline
$\left\langle 2\right\vert $ &  &  &  &  &  & $Y_{5}$ & $Y_{2}$ & . & $Y_{1}$
& . & . &  &  &  &  &  & \multicolumn{1}{||l||}{$\left\vert 2\right\rangle
=\left\vert 0,\downarrow\right\rangle $} \\ \hline
$\left\langle 3\right\vert $ &  &  &  &  &  & $-Y_{6}$ & . & $Y_{2}$ & . & $%
Y_{1}$ & . &  &  &  &  &  & \multicolumn{1}{||l||}{$\left\vert
3\right\rangle =\left\vert 0,\uparrow\right\rangle $} \\ \hline
$\left\langle 4\right\vert $ &  &  &  &  &  & . & $-Y_{6}$ & $-Y_{5}$ & . & .
& $Y_{3}$ &  &  &  &  &  & \multicolumn{1}{||l||}{$\left\vert 4\right\rangle
=\left\vert -,0\right\rangle $} \\ \hline
$\left\langle 5\right\vert $ &  &  &  &  &  & . & . & . & $-Y_{6}$ & $-Y_{5}$
& $Y_{4}$ &  &  &  &  &  & \multicolumn{1}{||l||}{$\left\vert 5\right\rangle
=\left\vert +,0\right\rangle $} \\ \hline\hline
$\left\langle 6\right\vert $ &  &  &  &  &  &  &  &  &  &  &  & $Y_{2}$ & $%
Y_{1}$ & . & . &  & \multicolumn{1}{||l||}{$\left\vert 6\right\rangle
=\left\vert 0,\uparrow\downarrow\right\rangle $} \\ \hline
$\left\langle 7\right\vert $ &  &  &  &  &  &  &  &  &  &  &  & $-Y_{5}$ & .
& $Y_{3}$ & . &  & \multicolumn{1}{||l||}{$\left\vert 7\right\rangle
=\left\vert -,\downarrow\right\rangle $} \\ \hline
$\left\langle 8\right\vert $ &  &  &  &  &  &  &  &  &  &  &  & $Y_{6}$ & .
& . & $Y_{3}$ &  & \multicolumn{1}{||l||}{$\left\vert 8\right\rangle
=\left\vert -,\uparrow\right\rangle $} \\ \hline
$\left\langle 9\right\vert $ &  &  &  &  &  &  &  &  &  &  &  & . & $-Y_{5}$
& $Y_{4}$ & . &  & \multicolumn{1}{||l||}{$\left\vert 9\right\rangle
=\left\vert +,\downarrow\right\rangle $} \\ \hline
$\left\langle 10\right\vert $ &  &  &  &  &  &  &  &  &  &  &  & . & $Y_{6}$
& . & $Y_{4}$ &  & \multicolumn{1}{||l||}{$\left\vert 10\right\rangle
=\left\vert +,\uparrow\right\rangle $} \\ \hline
$\left\langle 11\right\vert $ &  &  &  &  &  &  &  &  &  &  &  & . & . & $%
Y_{6}$ & $Y_{5}$ &  & \multicolumn{1}{||l||}{$\left\vert 11\right\rangle
=\left\vert d,0\right\rangle $} \\ \hline\hline
$\left\langle 12\right\vert $ &  &  &  &  &  &  &  &  &  &  &  &  &  &  &  &
$Y_{3}$ & \multicolumn{1}{||l||}{$\left\vert 12\right\rangle =\left\vert
-,\uparrow\downarrow\right\rangle $} \\ \hline
$\left\langle 13\right\vert $ &  &  &  &  &  &  &  &  &  &  &  &  &  &  &  &
$Y_{4}$ & \multicolumn{1}{||l||}{$\left\vert 13\right\rangle =\left\vert
+,\uparrow\downarrow\right\rangle $} \\ \hline
$\left\langle 14\right\vert $ &  &  &  &  &  &  &  &  &  &  &  &  &  &  &  &
$Y_{5}$ & \multicolumn{1}{||l||}{$\left\vert 14\right\rangle =\left\vert
d,\downarrow\right\rangle $} \\ \hline
$\left\langle 15\right\vert $ &  &  &  &  &  &  &  &  &  &  &  &  &  &  &  &
$-Y_{6}$ & \multicolumn{1}{||l||}{$\left\vert 15\right\rangle =\left\vert
d,\uparrow\right\rangle $} \\ \hline\hline
$\left\langle 16\right\vert $ &  &  &  &  &  &  &  &  &  &  &  &  &  &  &  &
& $\left\vert 16\right\rangle =\left\vert d,\uparrow\downarrow\right\rangle $
\\ \hline\hline
\end{tabular}%
\caption[TABLE Ia]{ \ The table gives the matrix elements of the six
operators $X_{0+}=X_{+}=Y_{1},X_{0-}=X_{-}=Y_{2},X_{\overline{\protect\sigma}%
d}=T_{+}=Y_{3},X_{\protect\sigma d}=T_{-}=Y_{4},C_{\uparrow}=Y_{5}$, and $%
C_{\downarrow}=Y_{6}$ in the basis of the sixteen states defined in the last
column. The matrix is separated into the sub matrices $\left\langle
n\right\vert Y_{j}\left\vert n^{\prime}\right\rangle $ connecting states
with $n=0,1,2,3$ electrons to states with $n^{\prime}=n+1=1,2,3,4$. The
value of the matrix elements is either ${1}$ or ${-1}$, as indicated in the
table. We use $\left\vert d\right\rangle \equiv\left\vert +-\right\rangle $,
as in Table \protect\ref{T2}, to emphasize that $X_\pm\left\vert
d\right\rangle =0$. }
\label{T1a}
\end{table}

\begin{table}[ptb]
\begin{tabular}[t]{||l||l||c|c|c|c||c|c||}
\hline
&  & $Y_{1}$ & $Y_{2}$ & $Y_{3}$ & $Y_{4}$ & $Y_{5}$ & $Y_{6}$ \\
\hline\hline
&  & $X_{0+}=XU$ & $X_{0-}=XD$ & $X_{\overline{\sigma}d}=TU$ & $X_{\sigma
d}=TD$ & $C_{\uparrow}=CU$ & $C_{\downarrow}=CD$ \\ \hline\hline
$1$ & $\left\vert 1\right\rangle =\left\vert 0,0\right\rangle $ & . & . & .
& . & . & . \\ \hline\hline
$1$ & $\left\vert 2\right\rangle =\left\vert 0,\downarrow\right\rangle $ & .
& . & . & . & . & $\left\vert 1\right\rangle $ \\ \hline
$2$ & $\left\vert 3\right\rangle =\left\vert 0,\uparrow\right\rangle $ & . &
. & . & . & $\left\vert 1\right\rangle $ & . \\ \hline
$3$ & $\left\vert 4\right\rangle =\left\vert -,0\right\rangle $ & . & $%
\left\vert 1\right\rangle $ & . & . & . & . \\ \hline
$4$ & $\left\vert 5\right\rangle =\left\vert +,0\right\rangle $ & $%
\left\vert 1\right\rangle $ & . & . & . & . & . \\ \hline\hline
$1$ & $\left\vert 6\right\rangle =\left\vert
0,\uparrow\downarrow\right\rangle $ & . & . & . & . & $\left\vert
2\right\rangle $ & $-\left\vert 3\right\rangle $ \\ \hline
$2$ & $\left\vert 7\right\rangle =\left\vert -,\downarrow\right\rangle $ & .
& $\left\vert 2\right\rangle $ & . & . & . & $-\left\vert 4\right\rangle $
\\ \hline
$3$ & $\left\vert 8\right\rangle =\left\vert -,\uparrow\right\rangle $ & . &
$\left\vert 3\right\rangle $ & . & . & $-\left\vert 4\right\rangle $ & . \\
\hline
$4$ & $\left\vert 9\right\rangle =\left\vert +,\downarrow\right\rangle $ & $%
\left\vert 2\right\rangle $ & . & . & . & . & $-\left\vert 5\right\rangle $
\\ \hline
$5$ & $\left\vert 10\right\rangle =\left\vert +,\uparrow\right\rangle $ & $%
\left\vert 3\right\rangle $ & . & . & . & $-\left\vert 5\right\rangle $ & .
\\ \hline
$6$ & $\left\vert 11\right\rangle =\left\vert d,0\right\rangle $ & . & . & $%
\left\vert 4\right\rangle $ & $\left\vert 5\right\rangle $ & . & . \\
\hline\hline
$1$ & $\left\vert 12\right\rangle =\left\vert -,\uparrow\downarrow
\right\rangle $ & . & $\left\vert 6\right\rangle $ & . & . & $-\left\vert
7\right\rangle $ & $\left\vert 8\right\rangle $ \\ \hline
$2$ & $\left\vert 13\right\rangle =\left\vert +,\uparrow\downarrow
\right\rangle $ & $\left\vert 6\right\rangle $ & . & . & . & $-\left\vert
9\right\rangle $ & $\left\vert 10\right\rangle $ \\ \hline
$3$ & $\left\vert 14\right\rangle =\left\vert d,\downarrow\right\rangle $ & .
& . & $\left\vert 7\right\rangle $ & $\left\vert 9\right\rangle $ & . & $%
\left\vert 11\right\rangle $ \\ \hline
$4$ & $\left\vert 15\right\rangle =\left\vert d,\uparrow\right\rangle $ & .
& . & $\left\vert 8\right\rangle $ & $\left\vert 10\right\rangle $ & $%
\left\vert 11\right\rangle $ & . \\ \hline\hline
$1$ & $\left\vert 16\right\rangle =\left\vert d,\uparrow\downarrow
\right\rangle $ & $.$ & $.$ & $\left\vert 12\right\rangle $ & $\left\vert
13\right\rangle $ & $\left\vert 14\right\rangle $ & $-\left\vert
15\right\rangle $ \\ \hline
\end{tabular}%
\caption[TABLE II]{ \ The elements in the table give the state that is the
result of applying the destruction operators on the top of the table to each
of the sixteen states defined in the second column, where we use $\left\vert
d\right\rangle =\left\vert +-\right\rangle $ to indicate the state with two
local electrons (note that $X_{0+}\left\vert d\right\rangle \equiv
X_{0+}\left\vert +-\right\rangle =0$ and it is neither $\left\vert
-\right\rangle $ nor $\left\vert 0\right\rangle $). The numbers in the first
column gives the ordering of the states in the local subspaces with $%
n=0,1,2,3,4$ electrons }
\label{T2}
\end{table}

\end{document}